\documentclass[sigconf,authorversion,nonacm]{acmart}

\usepackage{caption}
\usepackage{subcaption}
\usepackage{multirow}
\usepackage{booktabs}
\usepackage{threeparttable}
\usepackage{graphicx}
\usepackage{makecell}
\usepackage{siunitx}
\usepackage{enumitem}
\usepackage{lscape}
\usepackage{pdflscape}

\usepackage[acronym]{glossaries}

\newacronym[plural=AOIs]{aoi}{AOI}{Area Of Interest}

\newacronym{adas}{ADAS}{Adaptive Driver Assistance Systems}

\newacronym[plural=ANOVAs]{anova}{ANOVA}{Analysis of Variance}

\newacronym{eeg}{EEG}{Electroencephalogram}

\newacronym[plural=HDDs]{hdd}{HDD}{Head-down Display}

\newacronym[plural=HUDs]{hud}{HUD}{Head-up Display}

\newacronym[plural=MANOVAs]{manova}{MANOVA}{Multivariate Analysis of Variance}

\newacronym{mean}{M}{Mean}

\newacronym{nasa}{NASA-TLX}{NASA Task Load Index}

\newacronym[plural=NDRTs]{ndrt}{NDRT}{Non-Driving Related Task}

\newacronym{pcs}{PCS}{Principal Component Scores}

\newacronym{rmssd}{RMSSD}{Root Mean Square of Successive Differences}

\newacronym{sae}{SAE}{Society of Automotive Engineers}

\newacronym{sge}{SGE}{Stationary Gaze Entropy}

\newacronym{sd}{SD}{Standard Deviation}

\newacronym[plural=ToRs]{tor}{ToR}{Takeover Request}

\glsdisablehyper

\newcommand{\anova}[4]{(F(#1, #2) = #3, p = #4)}

\newcommand{\anovaEta}[5]{(F(#1, #2) = #3, p = #4, \(\eta_p^2\) = #5)}

\newcommand{\chisq}[3]{\(\chi^2_{\text{approx.}}(#1) = #2, p = #3\)}

\newcommand{\manova}[2]{(MATS = #1, \textit{p} = #2)}

\newcommand{\tukey}[4]{(mean difference = #1, p = #2, 95\% CI [#3, #4])}

\AtBeginDocument{%
  \providecommand\BibTeX{{%
    \normalfont B\kern-0.5em{\scshape i\kern-0.25em b}\kern-0.8em\TeX}}}

\setcopyright{acmlicensed}
\copyrightyear{2018}
\acmYear{2018}
\acmDOI{XXXXXXX.XXXXXXX}

\acmConference[Conference acronym 'XX]{Make sure to enter the correct
  conference title from your rights confirmation emai}{June 03--05,
  2018}{Woodstock, NY}
\acmISBN{978-1-4503-XXXX-X/18/06}

\begin{document}

\title[Adapting Takeover Requests for Seamless Handover]{Your Interface, Your Control: Adapting Takeover Requests for Seamless Handover in Semi-Autonomous Vehicles}

\author{Amr Gomaa}
\orcid{0000-0003-0955-3181}
\affiliation{%
  \institution{DFKI}
    \city{Saarbr{\"u}cken}
  \country{Germany}
}

\affiliation{%
  \institution{Saarland Informatics Campus}
    \city{Saarbr{\"u}cken}
  \country{Germany}
}
\email{amr.gomaa@dfki.de}

\author{Simon Engel}
\orcid{0009-0006-3059-8498}
\affiliation{%
  \institution{DFKI}
    \city{Saarbr{\"u}cken}
  \country{Germany}
}
\email{s8sgenge@stud.uni-saarland.de}

\author{Elena Meiser}
\orcid{0000-0003-4842-290X}
\affiliation{%
  \institution{DFKI}
    \city{Saarbr{\"u}cken}
  \country{Germany}
}
\email{elena.meiser@dfki.de}

\author{Abdulrahman Mohamed Selim}
\orcid{0000-0002-4984-6686}
\affiliation{%
  \institution{DFKI}
    \city{Saarbr{\"u}cken}
  \country{Germany}
}
\email{abdulrahman.mohamed@dfki.de}

\author{Tobias Jungbluth}
\orcid{0009-0006-4084-7785}
\affiliation{%
  \institution{DFKI}
    \city{Saarbr{\"u}cken}
  \country{Germany}
}
\email{tobias.jungbluth@dfki.de}

\author{Aeneas Leon Sommer}
\orcid{0009-0004-1009-276X}
\affiliation{%
  \institution{DFKI}
    \city{Saarbr{\"u}cken}
  \country{Germany}
}
\email{aeneas_leon.sommer@dfki.de}

\author{Sarah Kohlmann}
\affiliation{%
  \institution{DFKI}
    \city{Saarbr{\"u}cken}
  \country{Germany}
}
\email{sarah.kohlmann@dfki.de}

\author{Michael Barz}
\orcid{0000-0001-6730-2466}
\affiliation{%
  \institution{DFKI}
    \city{Saarbr{\"u}cken}
  \country{Germany}
}

\affiliation{%
  \institution{University of Oldenburg}
    \city{Oldenburg}
  \country{Germany}
}

\email{michael.barz@dfki.de}

\author{Maurice Rekrut}
\orcid{0000-0002-5829-5409}
\affiliation{%
  \institution{DFKI}
    \city{Saarbr{\"u}cken}
  \country{Germany}
}
\email{maurice.rekrut@dfki.de}

\author{Michael Feld}
\orcid{0000-0001-6755-5287}
\affiliation{%
  \institution{DFKI}
    \city{Saarbr{\"u}cken}
  \country{Germany}
}
\email{michael.feld@dfki.de}

\author{Daniel Sonntag}
\orcid{0000-0002-8857-8709}
\affiliation{%
  \institution{DFKI}
    \city{Saarbr{\"u}cken}
  \country{Germany}
}
\email{sonntag@dfki.de}

\author{Antonio Kr{\"u}ger}
\orcid{0000-0002-8055-8367}
\affiliation{%
  \institution{DFKI}
    \city{Saarbr{\"u}cken}
  \country{Germany}
}

\email{krueger@dfki.de}

\renewcommand{\shortauthors}{Gomaa et al.}

\begin{abstract}

With the automotive industry transitioning towards conditionally automated driving, takeover warning systems are crucial for ensuring safe collaborative driving between users and semi-automated vehicles. However, previous work has focused on static warning systems that do not accommodate different driver states. Therefore, we propose an adaptive takeover warning system that is personalised to drivers, enhancing their experience and safety. 
We conducted two user studies investigating semi-autonomous driving scenarios in rural and urban environments while participants performed non-driving-related tasks such as text entry and visual search. 
We investigated the effects of varying time budgets and head-up versus head-down displays for takeover requests on drivers' situational awareness and mental state. Through our statistical and clustering analyses, we propose strategies for designing adaptable takeover systems, e.g., using longer time budgets and head-up displays for non-hazardous takeover events in high-complexity environments while using shorter time budgets and head-down displays for hazardous events in low-complexity environments.

\end{abstract}




\maketitle

\section{Introduction}
 
Autonomous driving systems, which enable vehicles to operate independently with minimal or no human control or supervision, have the potential to revolutionise transportation as we know it \cite{dimitrakopoulos_chapter_2021, wang_multi-modal_2023}. These systems could allow drivers to focus on \glspl{ndrt} during their journey while the vehicle navigates autonomously \cite{dogan_effects_2019}. The currently deployed level of automation, as classified by the \gls{sae}\footnote{\url{https://www.sae.org/blog/sae-j3016-update}} (N.B., This is an established agency for governing vehicular automation standards), is at the second stage (i.e., L2) out of six. This stage includes features such as adaptive cruise control and lane-keeping assistance but requires continuous driver attention.
However, the automotive industry is currently approaching the transition towards higher automation levels, bringing us closer to fully autonomous driving \cite{khare_exploring_2024}. This presents both opportunities and challenges for industry and academia in conceptualising human interaction within vehicles \cite{detjen_how_2021}. 
Until full vehicle autonomy is achieved, drivers will need to periodically redirect their attention to the driving task (i.e., L3 in the \gls{sae} classification). This necessitates quickly perceiving, processing, and responding to \glspl{tor}.
A \gls{tor} is a notification initiated by an autonomous vehicle upon encountering an unfamiliar or unexpected situation, prompting the transfer of control back to the driver for manual operation \cite{salubre_takeover_2021}. To ensure a safe transition, it is crucial to understand the human interaction with different elements of such systems \cite{riener_automotive_2016}.

Modern vehicles offer multiple locations for visual warning prompts and \glspl{tor}, including the \gls{hud} on the windscreen and the more conventional \gls{hdd} on the centre console. The \gls{hud}, being in the driver's line of sight, is theoretically ideal for capturing the driver's attention, e.g., to show them information~\cite{kim_are_2017, currano_little_2021, gerber_eye_2024}. However, in conditionally autonomous vehicles, drivers may be engrossed in \glspl{ndrt}, such as interacting with their mobile phones, and may not be responsive to warnings displayed on the \gls{hud}.
In addition to the visual warning location, another important factor to consider when it comes to \glspl{tor} is when to trigger them, i.e., the timing of the \glspl{tor} before the actual need for a transfer of control to a human driver. In a meta-analysis of 129 studies on the appropriate timing of \glspl{tor}, \citet{zhang_determinants_2019} demonstrated the volatility of this parameter among drivers. The extent of variability depends on their demographics, current \gls{ndrt}, and surrounding environment, underscoring the necessity for an adaptive interface tailored to specific situations and considering the driver's state and mental workload.
\begin{figure}
    \centering
    \begin{subfigure}{\linewidth}
        \centering
        \includegraphics[width=\linewidth]{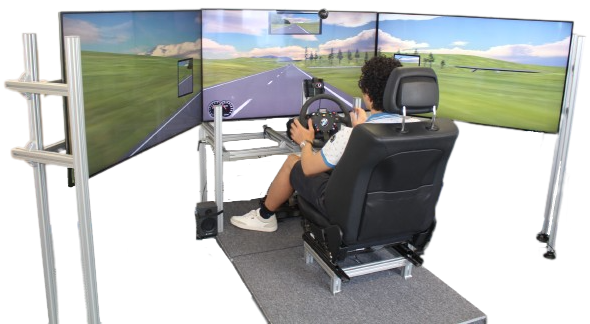}
        \end{subfigure}
    \begin{subfigure}{\linewidth}
        \centering
        \includegraphics[width=\linewidth]{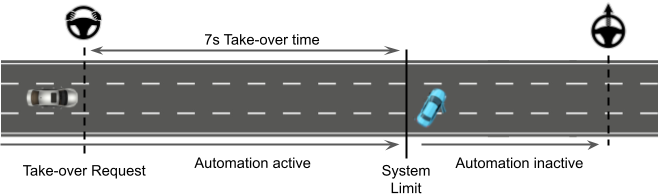}
    \end{subfigure}
    \caption{Top: Our driving simulator setup. Bottom: Schematic of the takeover action in our studies.}
    \Description{Top: Driving simulator setup with three monitors displaying the simulator world and a camera directly above the steering wheel capturing the participant's face. The participant controls the vehicle during a takeover. Bottom: Schematic of the takeover task in our study. \SI{7}{\s} Before arriving at the system boundary, the takeover request is prompted by the hands-on-the-wheel icon. Until the participant responds by pressing the button, or until arrival at the system limit, the automation is active. After that, it is set to inactive, and the user must manually exit the hazard. Once out of the area, the vehicle can regain control, which is signalled by an arrow on the steering wheel.}
    \label{fig:study_setup}
\end{figure}
Moreover, the mental workload imposed on the driver by various \glspl{ndrt} can differ, resulting in varying levels of inattention or distraction. The driving context also plays a crucial role in the \gls{tor} \cite{kim_understanding_2022}. For example, drivers may be more attentive in cities than in rural areas or vice versa. Therefore, it is essential to construct an adaptive interface approach that adapts its warning strategy, e.g., visual warning location and \gls{tor} timing, to the driver's context and behaviour.
Monitoring various factors, such as psychophysiological measures or eye movement patterns, can provide insights into the driver's cognitive state. The human gaze is an interplay of individual cognitive processes, environmental stimuli, and task-related demands. Therefore, monitoring human eye movements using eye tracking can provide information about drivers' internal cognitive state. This use of eye tracking falls under passive gaze-based interaction, i.e., a user does not explicitly interact using their eyes; rather, it monitors and interprets eye movements in the background to potentially adjust the interface \cite{duchowski_gaze-based_2018, qvarfordt_gaze-informed_2017}. One common passive gaze-based interaction method is to monitor users' visual attention and cognitive states \cite{mohamed_selim_review_2024}, which holds true for driving to increase road safety \cite{kotseruba_attention_2022}.
Another important psychophysiological measure to consider is \gls{eeg}, which captures the brain's electrical activity by utilising electrodes positioned on the scalp \cite{nuwer_brain_2014}. 
\gls{eeg} signals can be utilised to identify and classify a driver’s mental workload during both the driving and the \gls{ndrt} \cite{barua_towards_2020}.

Due to the high number of factors, related approaches commonly investigated the impact of individual factors in an isolated manner. However, we aim at understanding the interplay of these factors, considering the potential influence they have on each other. For this, it is crucial to study their collective effects in an adaptive manner.
In this paper, we explore the ergonomics and feasibility of designing an adaptive \gls{tor} system. We focus on \gls{tor} timing and the visual warning location under varying environmental complexities and \glspl{ndrt}. Additionally, we evaluate the driver’s situational awareness throughout the entire semi-automated drive, both before and after taking over control of the vehicle. For this purpose, we conducted two user studies, with 37 and 29 participants, for the \gls{tor} timing and the visual warning location, respectively.
We gathered and analysed reaction times, mental workload (with subjective and objective measures), driving performance measures, and participants' gaze and \gls{eeg} data to propose an adaptive takeover warning system. We suggest several strategies for adapting the warning system based on varying driving contexts and drivers' states. 
Our studies were conducted using a driving simulator, which is shown in~\autoref{fig:study_setup}.

Overall, our contributions are summarised as follows.

\begin{itemize}
    \item \textbf{Investigation of Adaptive \gls{tor} System Elements:} We investigated the key attributes of a \gls{tor} system and its adaptation properties in two user studies. Our focus was on the effects of variable \gls{tor} time budgets and the location of visual warnings on drivers' situational awareness and driving performance throughout the entire driving and takeover procedure.
    
    \item \textbf{Evaluation of Varying Driving Situations:} We assessed our approach under various conditions, including two driving environments (Rural vs Urban), two \glspl{ndrt}, two display types (\gls{hud} vs \gls{hdd}), and three \gls{tor} time budgets (4 seconds vs 8 seconds vs 12 seconds) with different traffic scenarios and takeover reasons (i.e., critical and non-critical situations). We highlight the impact of these conditions on driving performance and quality, revealing interrelated effects and optimal design settings for enhanced takeover performance.
    
    \item \textbf{Guidelines for Designing Adaptive Strategies:} We propose several adaptive strategies for the \gls{tor} system, such as displaying takeover prompts on the \gls{hud} in high-complexity high-speed environments when the driver is engaged in a visual search task or using a longer preceding time for the \gls{tor} in high-traffic situations, regardless of the vehicle speed. We utilised multiple statistical analyses and evaluation metrics such as reaction times, mental workload, driving performance measures, and participants' gaze and \gls{eeg} data. Consequently, these analyses helped us develop and justify these adaptive strategies, enhancing drivers' experience, response, and safety.
\end{itemize}

\section{Background and Related Work}
In our work, we explore an adaptive methodology for presenting \glspl{tor}, which includes multiple facets regarding the system design and the driver's state. Therefore, we begin this section with an overview of the existing literature on adaptive driver assistance systems and different attributes for transferring control. Then, we focus on the various display options available in the vehicle. Finally, we focus on existing evaluation approaches that monitor drivers' situational awareness and workload.

\subsection{Adaptive Driver Assistance Systems} 
\gls{adas} have emerged as a significant area where personalisation can be applied. This is crucial because drivers vary in their preferences, abilities, and requirements.
These preferences can fluctuate based on the driver’s state and the driving situation. Therefore, the goal of personalisation in \gls{adas} is to enhance the driving experience and the driver's performance by tailoring the assistance system to their specific preferences and needs \cite{lilis_personalizing_2017, hasenjager_survey_2020}.
\citet{hasenjager_survey_2020} in their survey paper found that driver models, which predict real-time driving behaviour, serve as the foundation for customising the type, frequency, and quantity of alerts provided by \gls{adas}. However, these models are currently based on static parameters and represent an average driver, limiting their ability to adapt to individual users. This highlights the need for more dynamic and personalised driver models.
\citet{sadeghian_borojeni_feel_2018} conducted a study on \glspl{tor} in highly automated vehicles. They discovered that drivers’ responses to urgent cues vary depending on the road context. For instance, rapid responses are observed when \glspl{tor} are issued on straight roads, while on curved roads, drivers respond more slowly to urgent cues than to non-urgent ones. This finding suggests that the design of \glspl{tor}, and by extension, the design of driver models, should consider the road context to align with natural user responses. It further emphasises the importance of context awareness and personalisation in the design of \gls{adas}.

Therefore, personalisation in \gls{adas} can be implemented in two ways: one approach is explicit personalisation, where drivers are given the option to select their preferred settings from a range of predefined system settings; the other approach is implicit personalisation, which involves estimating drivers’ preferences by observing their behaviour \cite{c_panou_intelligent_2018, hasenjager_personalization_2017, hasenjager_survey_2020}. While both approaches aim to enhance the effectiveness of \gls{adas} and improve the overall driving experience, our work focuses more on the implicit subtle personalisation as a less intrusive approach for seamless improvement to the driver's experience.

\subsection{Transfer of Control} 

According to the \gls{sae}, a key aspect of the third stage of vehicle autonomy (i.e., conditional automation) is the transition of control between the vehicle and the driver. At this stage, the vehicle is in full control of the driving tasks, allowing the driver to engage in \glspl{ndrt}. However, the driver must be ready to take over when needed. This shift in control is facilitated by \glspl{tor}, which are signals that prompt drivers to regain control of the vehicle.
Ensuring the safety of drivers during this transition is a vital and developing concern in human factors research. Researchers, such as \citet{riener_automotive_2016}, have examined how successfully drivers can respond to these requests and how effectively the automated system facilitates them. Many studies have explored the use of different modality warnings. However, as \citet{li_effects_2020} points out, these studies have predominantly suggested a fixed, non-adaptable method for prompting \glspl{tor} based on driving performance.

Driving performance can be assessed in multiple ways, but they can be broadly categorised into two groups: the takeover time, often referred to as reaction time, and the takeover quality. The former is the time between the initiation of a \gls{tor} and the first manual input by the driver. The latter is measured by various features, such as the minimum time to collision (N.B., This is only valid for critical \glspl{tor} use cases), the maximum resulting acceleration after initiating the \gls{tor}, or the correlation to an ideal trajectory \cite{salubre_takeover_2021}.
However, as \citet{eriksson_takeover_2017} and \citet{gold_taking_2016} highlight, all study design aspects, such as the driving scenario or the \gls{ndrt}, significantly influence the performance and quality. Therefore, in this work, we propose studying an adaptive approach for displaying the \glspl{tor}, which could potentially offer a more personalised and effective method for prompting \glspl{tor}. 

\subsection{Head-up versus Head-down Display}

\glspl{hud}, which present information in the user’s line of sight, have a rich history dating back to their initial use by the Royal Air Force in the early 1940s \cite{stokes_aviation_1988}. Since then, \glspl{hud} have evolved and found applications in various fields, including gaming and automotive industries. In vehicles, \glspl{hud} typically display information such as navigation units, speedometers, and tachometers.
Before the advent of \glspl{hud}, \glspl{hdd} were the primary source of in-vehicle infotainment. These traditional displays, mounted on the central console or the dashboard, require drivers to divert their visual attention from the road. In-vehicle \glspl{hud} offer several benefits compared to traditional \glspl{hdd}. 

Research has shown that using a \gls{hud} can lead to shorter reaction times to road occurrences \cite{liu_comparison_2004, horrey_effects_2003}, quicker response times for \glspl{ndrt} \cite{smith_visual_2015, horrey_effects_2003}, better vehicle control \cite{medenica_augmented_2011, liu_comparison_2004}, and reduced levels of mental workload \cite{liu_comparison_2004, horrey_effects_2003} compared to traditional \glspl{hdd}. Additionally, users prefer \glspl{hud} over \glspl{hdd} \cite{smith_visual_2015}. However, it is important to note that using \glspl{hud} poses challenges, such as decreased secondary task success rates \cite{smith_visual_2015}. 
Given these findings, it is clear that a one-size-fits-all approach may not be the most effective. Instead, a dynamic and adaptive approach, as proposed in this work, could potentially offer a more personalised and effective method for displaying information to drivers.

\subsection{Workload and Situational Awareness Measures}

Analysing in-vehicle interfaces often involves assessing the mental effort they impose on the driver. This concept, known as mental workload, has been a topic of broad interest in the academic community since the seminal work of \citet{casali_comparison_1983} on defining and evaluating it. Mental workload, particularly in relation to auditory and visual distractions, significantly impacts the safety and the overall driving experience \cite{gomaa_whats_2022}. The workload imposed on the user is naturally reflected in the user’s driving performance. Therefore, researchers often estimate the workload level based on the primary task performance. In this work, as the car is being driven autonomously, data on driving performance will be collected only during the manual driving phases. Additionally, the reaction time to a transfer of control and the critical event success rate will be monitored.
Mental workload is an essential factor affecting situational awareness \cite{tsang_mental_2006, kotseruba_attention_2022}, and investigating situational awareness might help draw conclusions about mental workload. Drivers’ eye gaze has been used to assess their situational awareness by monitoring their attention and perception \cite{marquart_review_2015, kotseruba_attention_2022}. Moreover, insights about the driver’s cognitive state have successfully been used to predict the following driver’s manoeuvre \cite{martin_dynamics_2018}, crash risk \cite{horrey_-vehicle_2007}, or ability to take over control \cite{gold_take_2013}. 

The importance of visual attention in driving tasks has been emphasised by \citet{baldisserotto_review_2023}, who proposed the integration of \gls{adas} with cognitive state monitoring using eye tracking. They highlighted that over 90\% of driving information is perceived visually, making visual feedback the preferred channel for most driving assistance systems. \citet{baldisserotto_review_2023} argued that this visual information could serve as an early detector of the driver’s intentions, drowsiness, and mental workload when combined with other sensor data. This premise was also explored by \citet{du_predicting_2020}, who conducted a human-in-the-loop experiment to predict drivers' takeover performance in real-time. They collected data from the driving environment and monitored participants' situational awareness and cognitive states using physiological data, including eye movements. Their study aimed to predict takeover performance using physiological data, including eye movements, when drivers are engaged in the same type of \gls{ndrt} but with varying levels of mental workload. Various studies, e.g. \cite{hofbauer_measuring_2020, pakdamanian_deeptake_2021, zhou_using_2022}, have used eye movements to detect drivers' visual attention and monitor their situational awareness to understand their behaviour during conditionally and highly automated driving. These studies highlight the potential of eye tracking in enhancing our understanding of driver behaviour and situational awareness, thereby contributing to the development of safer and more efficient automated driving systems. Therefore, we utilise gaze behaviour in this work as a metric for situational awareness and for decisions regarding system adaptations.

Similarly, \gls{eeg} can be utilised for mental workload estimation \cite{Genvis_workload_1991, holm_estimating_2009, Kartali_workload_2019, raufi_evaluation_2022, kutafina_tracking_2021}. 
\citet{Genvis_workload_1991} used a working memory task with two conditions, a hard and an easy task, to examine the effects of inducing mental workload. They found that the power of the theta band in frontal midline electrodes would rise with the more difficult tasks while the alpha band in central and parietal-central electrodes would attenuate. With these findings in mind, the theta and alpha power can then be exploited to create a mental workload measure \cite{holm_estimating_2009, Kartali_workload_2019, raufi_evaluation_2022}. 
\citet{Kartali_workload_2019} investigated the task of participants playing an aeroplane landing game, where they had to guide aeroplanes onto a landing strip to induce mental workload. The more planes present at the same time, the higher the workload was supposed to be. The mental workload was estimated using the ratio between the theta and alpha bands in certain electrodes (i.e., Fz, Pz, and Cz). They found a statistically significant positive correlation between the number of aeroplanes and the mental workload. This shows that \gls{eeg}-based mental workload measures can be used even in complex environments where participants need to focus on many different tasks at once. Consequently, \gls{eeg}-based mental workload estimation is a fitting method for this work, as a driver has to focus on many different tasks at once, especially when prompted with a \gls{tor}.
While mental workload can be estimated using psychophysiological measures such as heart rate, skin conductance, eye movements, and \gls{eeg}, it can be extremely challenging to replicate and is heavily reliant on interpretation \cite{lohani_review_2019, gomaa_whats_2022}. Therefore, we additionally assess workload based on subjective measures such as the \gls{nasa} \cite{hart_development_1988}. Despite the known disadvantages of subjective measures, the \gls{nasa} has been found effective in automotive studies and consistent with objective measures \cite{carsten_human_2005, naujoks_review_2018}. Therefore, in this work, \gls{nasa} will be used as the guiding metric for the psychophysiological dimensions.

\section{Methods}

We explored adaptive \gls{tor} system elements through two user studies. Study 1 (\textit{Display Type Study}) and Study 2 (\textit{Time Budget Study}) examined the effect of varying \gls{tor} location and \gls{tor} timing, respectively, on driver's performance and situational awareness. 
In this section, we describe the driving simulator, the driving environments, the \glspl{ndrt}, the two user studies detailing their similarities and differences, the evaluation metrics we used, and our hypotheses for the outcomes. 

\subsection{Study Apparatus}

\subsubsection{Driving Simulator}

The driving simulator setup, shown in~\autoref{fig:study_setup}, consists of a driver’s seat surrounded by three 55-inch LCD screens. The left and right monitors are positioned at a 45-degree angle towards the driver to enhance their field of view and perceptual realism. The simulator vehicle is equipped with an automatic transmission, controlled via a steering wheel, gas, and brake pedals. Additionally, the simulator has a virtual rear-view mirror and side mirrors. 
Participants sit in an actual car seat, which has adjustable settings for both the seat and pedals to ensure comfort.

The steering wheel has an embedded red button, which is easily accessible using the thumb, to allow switching between autonomous and manual driving. When a \gls{tor} is triggered (e.g., 250 metres ahead of a construction site), a reaction timer starts, awaiting a button press from the participant. If the button is pressed, the reaction timer stops, and the time is logged. In the \textit{Time Budget Study}, if the participant does not take over, the vehicle stops automatic driving after the allotted time (i.e., 4 seconds, 8 seconds, or 12 seconds) has elapsed. The vehicle resumes automatic driving after the problematic route (i.e., a part where an autonomous vehicle cannot function properly) is finished. However, in the \textit{Display Type Study}, if the participant does not take over and the vehicle reaches the emergency brake trigger (e.g., 30 metres ahead of a construction site), the car automatically halts, and the reaction time is logged as -2. If the participant successfully takes over and leaves the hazardous scene, the trigger to switch back to autonomous control is activated, and the participant switches to autonomous driving by pressing the designated button.

For the \textit{Display Type Study}, the \gls{hud} was positioned virtually, inside the simulation screen, at 0 \textdegree horizontally and vertically in the user’s line of sight to match the setup of \citet{haeuslschmid_recognition_2017} which achieved the shortest reaction time. 
The \gls{hdd} was a mobile tablet securely mounted 30 cm to the right of the steering wheel. It displayed either a black screen or the \gls{tor} warnings during the study. The \gls{hdd} used an Android tablet running a custom application specifically programmed for this study to establish a connection to the simulator. Whenever an event was triggered, the corresponding icon was displayed on the tablet. The \glspl{ndrt} were performed on an additional tablet that the drivers were allowed to leave on the passenger seat (N.B., This was a normal seat placed to the right of the user) when driving. However, for the \textit{Time Budget Study}, the \gls{ndrt} were performed on the \gls{hdd}, and the warning appeared as a countdown timer on the simulation. This separation in warning display modes was intentional to avoid any confounding factors across the two studies and to study each independent variable separately.

\subsubsection{Gaze Tracking}

Gaze behaviour plays a pivotal role in the transition from autonomous to manual driving. By investigating drivers’ gaze, we wanted to gain insight into their visual attention for decision-making. Human gaze behaviour is defined by the dynamics of where and how individuals direct their visual attention. Our eyes constantly move and shift focus, allowing us to explore our surroundings and gather information.
Since the \textit{Time Budget Study} only used the \gls{hud}, we used a 250 Hz Tobii Pro Fusion remote eye tracker\footnote{\url{https://www.tobii.com/products/eye-trackers/screen-based/tobii-pro-fusion}} mounted between the participant and the middle screen. 

On the other hand, the \textit{Display Type Study} used two display modes: the \gls{hud} and the \gls{hdd}. Therefore, it was not possible to use the Tobii Pro Fusion since it would have only captured gaze data within the simulation environment (i.e., only gazes on the \gls{hud} would be captured). Instead, we used a regular RGB camera for gaze estimation. The camera was mounted above the steering wheel; it recorded at 640 × 480 pixels and 30 Hz and was able to capture both the simulation environment and gaze in the real environment (i.e., on the \gls{hdd} tablet). Each frame was timestamped to synchronise gaze angles with the \glspl{ndrt} and driving simulator events.
We used the RT-Gene system \cite{fischer_rt-gene_2018} for the gaze estimation as an alternative to the Tobii Pro Fusion proprietary software. 
Each of the \gls{hud}, \gls{hdd}, road, and \glspl{ndrt} was identified as an \gls{aoi} during the experiment. The final gaze behaviour was represented as a scanpath of timestamps and associated \glspl{aoi}.

\subsubsection{Brain Activity Monitoring}

The \textit{Time Budget Study} used only one display type and had fewer visual aspects for assessing situational awareness as opposed to the \textit{Display Type Study}. This is why we decided to collect additional \gls{eeg} data to estimate participants' mental workload, evaluate situational awareness, and analyse their effect on the different timing and driving experiences. 
We used the 24-channel Dry electrode ANT-Neuro \gls{eeg} device\footnote{\url{https://www.ant-neuro.com/products/eego_24}}. 

\subsection{Driving Environment} 

For our road network design, we used the OpenDRIVE format\footnote{\url{https://www.asam.net/standards/detail/opendrive/}} which can easily be integrated within the OpenDS \cite{math_opends_2013} simulation. 
We used the OpenDS \cite{math_opends_2013} simulation environment as it allows easy development and control over different scenarios and manipulating surrounding objects.
By defining the preferred turnings for a car at each junction, the car follows the desired road. This approach is used to navigate the simulator vehicle during autonomous driving phases. The driving environment and road network were modelled after a real-life road connecting to a European city. 

\begin{figure*}
\centering
     \begin{subfigure}{0.49\linewidth}
         \centering
         \includegraphics[width=\linewidth]{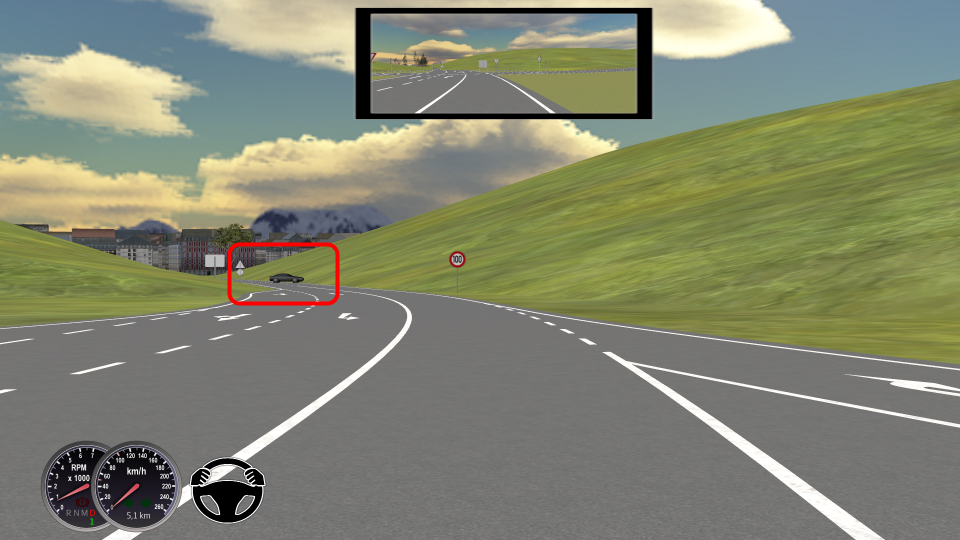}
         \caption{Rural Environment}
         \label{fig:rural_environment}
     \end{subfigure}
     \begin{subfigure}{0.49\linewidth}
         \centering
         \includegraphics[width=\linewidth]{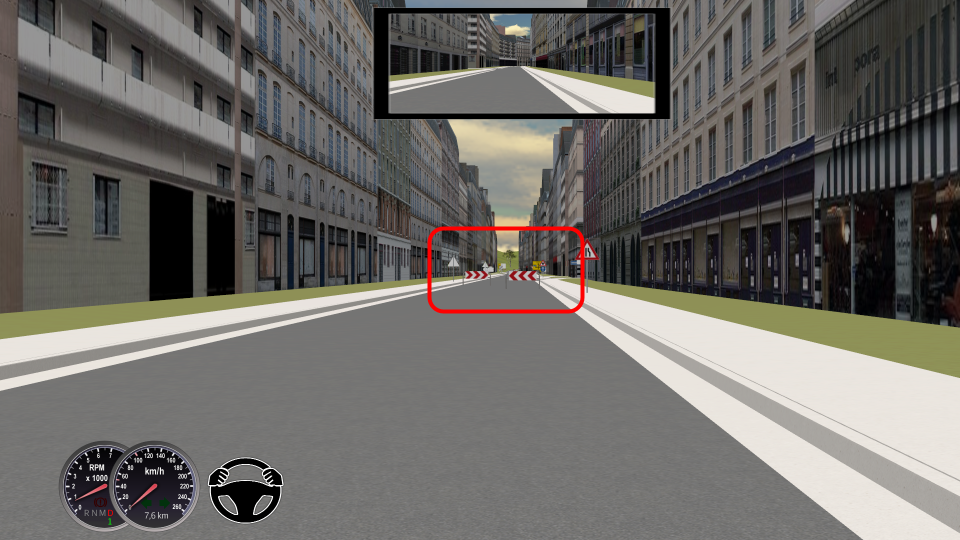}
         \caption{Urban Environment}
         \label{fig:urban_environment}
     \end{subfigure}
    \caption{Examples of the two driving environments.}
  \Description{This figure consists of two images side by side. The left image shows the driving simulator screen while driving in a rural environment. Grassy areas can be seen along the side of the road. The screen also shows the rearview mirror, speedometer, and the hands-on-the-wheel icon. The system boundary on the road ahead is marked by a red frame around it. The right image shows the driving simulator screen while driving in an urban environment. Buildings are visible along the side of the road. The screen also shows the rearview mirror, speedometer, and the hands-on-the-wheel icon. The system limit on the road ahead is marked by a red frame around it.}
  \label{fig:driving_environment}
\end{figure*}

Throughout the driving scenarios in the \textit{Display Type Study}, there were hazardous events where the autonomous car was unable to react and, therefore, requested the driver to take over. The most common event type was collision avoidance. To counteract learning effects, we employed various collision avoidance events, including car crashes, police operations, construction sites, and road blockades, to trigger the \glspl{tor}.
We had two driving scenarios: participants drove either from a rural environment (i.e., countryside or highway) with an average speed limit of 70 km/h to an urban environment (i.e., city) with an average speed limit of 30 km/h, or vice versa. These scenarios simulated a realistic driving experience. This also distinguished the environmental complexity of the rural versus urban environments by the speed limits, making the rural environment a high-complexity one, while the urban environment was a low-complexity environment.

In the \textit{Time Budget Study}, the events were not hazardous; however, the vehicle was simply unable to continue driving once the given time budget expired. The reasoning behind this design choice was to focus on the specific behaviour of drivers at variable time budgets in case of system failures rather than being confounded with nominal behaviour for critical hazardous events. We used similar urban and rural environments; however, we created a loop of six scenarios with high traffic in the urban environment only and a fixed speed limit of 70 km/h for all scenarios. Unlike the previous study, since the speed limit was fixed, this made the urban environment a high-complexity environment (i.e., due to the increased traffic and visual complexity), while the rural environment was a low-complexity environment. These design choices were important to avoid confounding factors in the results of both studies. The order of all scenarios was counterbalanced across participants, and no participant followed the same exact route as another. Additionally, a completely manual driving scenario was added in the \textit{Time Budget Study} to calibrate the \gls{eeg} data and create a baseline for the mental workload estimation.

\textbf{The rural environment}, as shown in~\autoref{fig:rural_environment}, features bland green scenery and two-way roads with a lane width of up to 4 metres. On the rather rectilinear roads with few wide curves spanning hundreds of meters with a large radius.
Traffic in the rural areas is low to none at all. Due to the small number of traffic signs and the monotonous countryside populated by bushes and trees, the visual complexity is considered low. 
\textbf{The urban environment}, as shown in~\autoref{fig:urban_environment}, features narrow city roads with two or more lanes, each with a lane width of up to 3 metres. Due to the variety of buildings, decorations, bus stops, construction sites, and numerous traffic signs that are both relevant (e.g., speed limits and stop signs) and irrelevant (e.g., street names and no parking signs) to the driving task, the visual complexity is considered high.

For the given time budget, in the \textit{Time Budget Study}, the \gls{tor} was issued either 4 seconds, 8 seconds, or 12 seconds before the autonomous car gave up control. However, in the \textit{Display Type Study}, the \gls{tor} was issued at a fixed time of 7 seconds before the autonomous car collided with the object blocking the lane. 

For the manual driving time, in the \textit{Display Type Study}, the hazardous events lasted for a few seconds of manual driving before giving control back to the vehicle. Since the focus was on the reaction time, gaze behaviour, and manoeuvre quality during the takeover time, we did not prolong the driving segment in this study. 
On the other hand, in the \textit{Time Budget Study}, since there were no hazardous events, the manual driving segment lasted for 40 to 60 seconds to assess the driving behaviour along the entire route and to observe the changes in the situational awareness as well as the driving quality for different time budgets.

\subsubsection{Display Type Study Scenarios}

\begin{figure*}
\centering
     \begin{subfigure}{0.49\linewidth}
         \centering
         \includegraphics[width=\linewidth]{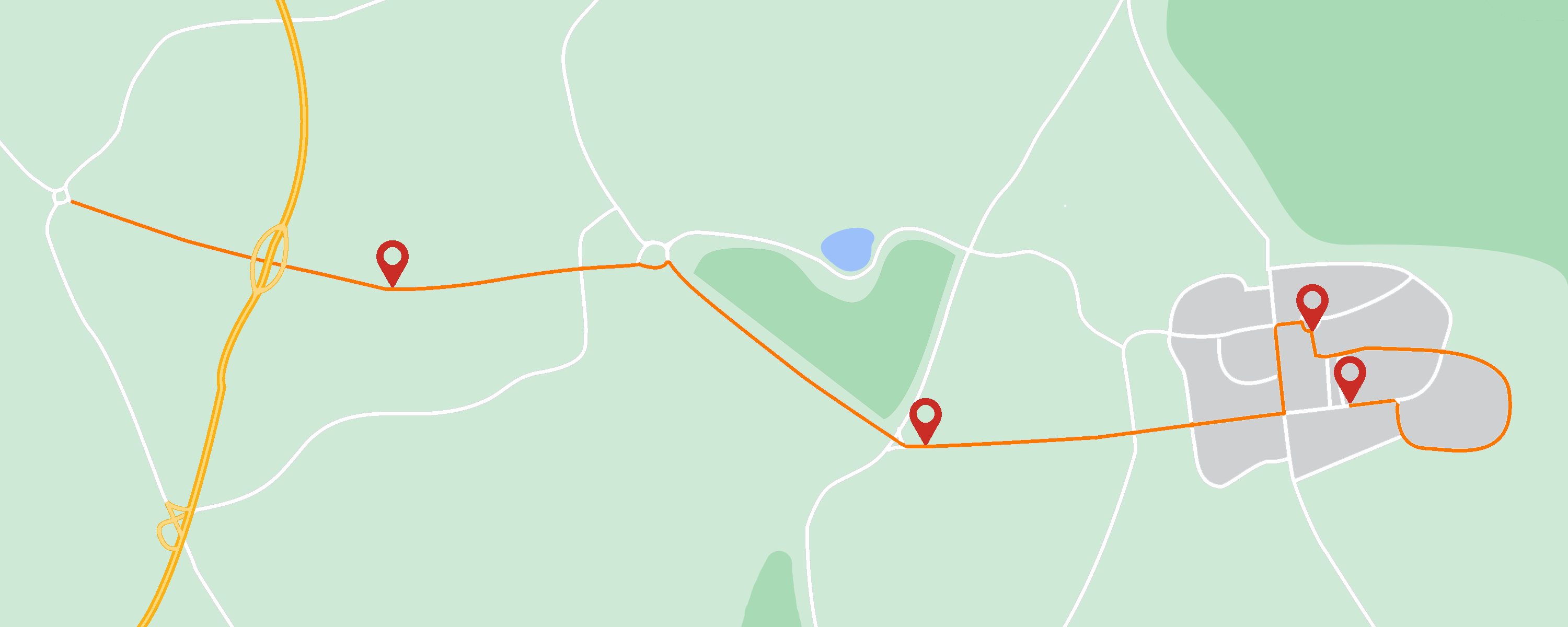}
         \caption{Scenario 1: From rural area to urban}
         \label{fig:scenario_1}
     \end{subfigure}
     \begin{subfigure}{0.49\linewidth}
         \centering
         \includegraphics[width=\linewidth]{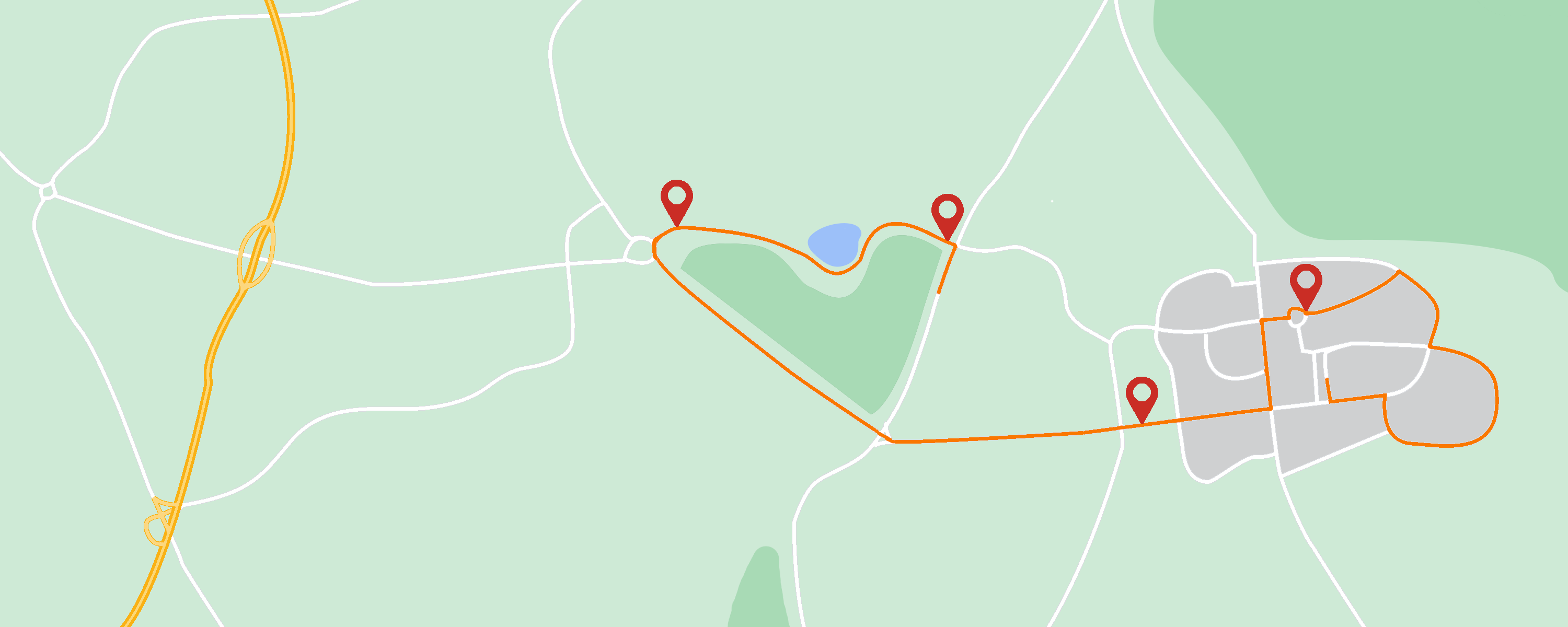}
         \caption{Scenario 2: From urban area to rural}
         \label{fig:scenario_2}
     \end{subfigure}
    \caption{The route and scenarios used in the \textit{Display Type Study}}.
  \Description{This figure consists of two maps side by side for the scenarios in the Display Type Study. The left map shows the driving simulator route for scenario 1. There are four system boundaries on the route, first two in a rural area and then two in an urban area. The right map shows the driving simulator route of scenario 2. There are four system limits on the route, first one in an urban area and then three in a rural area.}
  \label{fig:driving_scenarios}
\end{figure*}

This study presents two driving scenarios with several hazardous events requiring transfer of control from the autonomous vehicle to the user and takes the control back after passing the hazardous (i.e., critical) event.
In Scenario One, the drive starts on a rural road with a speed limit increase from 70 km/h to 100 km/h, encountering a car crash and a police operation requiring transfer of control from the autonomous vehicle to the user. After passing the hazardous events, the vehicle takes control again. The route continues into an urban area with complex traffic and multiple intersections, ending with a chicane and a final transfer of control. The scenario takes about 7 minutes, covering 4 km in rural and 1.8 km in urban areas.
Scenario Two begins where Scenario One ends, with a similar mix of rural and urban driving. However, it is reversed in order. We use these two scenarios for counterbalancing the participants starting order and avoiding confounding factors. This scenario includes a roundabout, a two-vehicle crash, and a broken-down car, all requiring transfer of control from the autonomous vehicle to the user. After each hazardous event is passed, the vehicle takes control again similar to the previous scenario. 
This scenario takes about 6.5 minutes, covering 3.4 km in rural and 1.6 km in urban areas (see~\autoref{fig:driving_scenarios}).

\subsubsection{Time Budget Study Scenarios}

\begin{figure}
    \centering
    \includegraphics[scale=0.4]{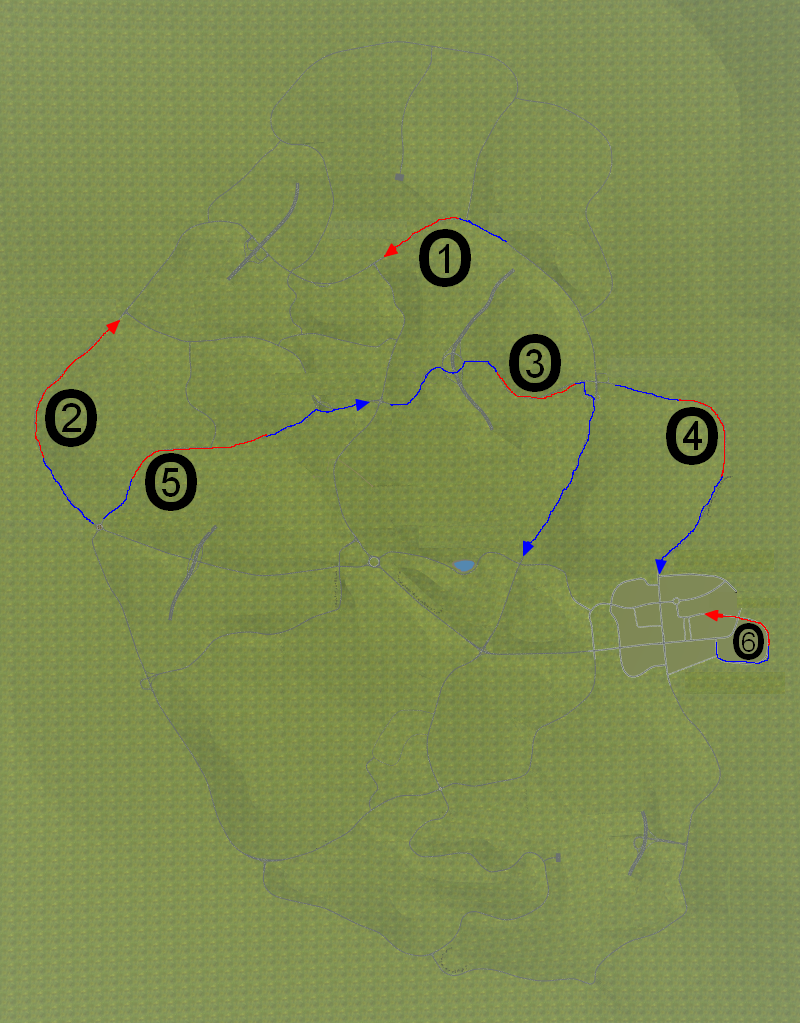}
    \caption{The driving route used in the \textit{Time Budget Study}. The blue lines represent the autonomous driving segments of the road, and the red lines represent the manual driving segments. Note that there is always an automatic autonomous driving segment after each manual segment, even if it is not shown in the graph.}
    \Description{This figure contains a map showing the six routes of takeover scenarios from an aerial view used in the Time Budget Study. Each scenario is represented by a two-coloured arrow. The arrowheads and ends are blue, indicating manual driving. The middle of the arrow is red, indicating autonomous driving.}
    \label{fig:driving_scenarios_time_budget_study}
\end{figure}

The \textit{Time Budget Study} used similar routes to the \textit{Display Type Study}. 
However, as the number of independent variables is larger in this study, additional scenarios were created to avoid any confounding factors.
We extended the entire driving route to include a total of six scenarios.~\autoref{fig:driving_scenarios_time_budget_study} shows a map of the six scenarios. This image is a high-level abstract view taken directly from the simulation, rather than a detailed schematic. Each scenario highlights the segment where the exact takeover procedure occurred, as illustrated in the bottom view of~\autoref{fig:study_setup}.
As mentioned earlier, the order of the scenarios was counterbalanced. Each participant went through an entirely different order of scenarios to avoid any confounding factors and learning effects. Each participant went through the six scenarios twice, but not consecutively, with different visual complexities in the surrounding environment.

\subsection{Non-driving Related Tasks}

\begin{figure*}
\centering
     \begin{subfigure}{0.49\linewidth}
         \centering
         \includegraphics[width=\linewidth]{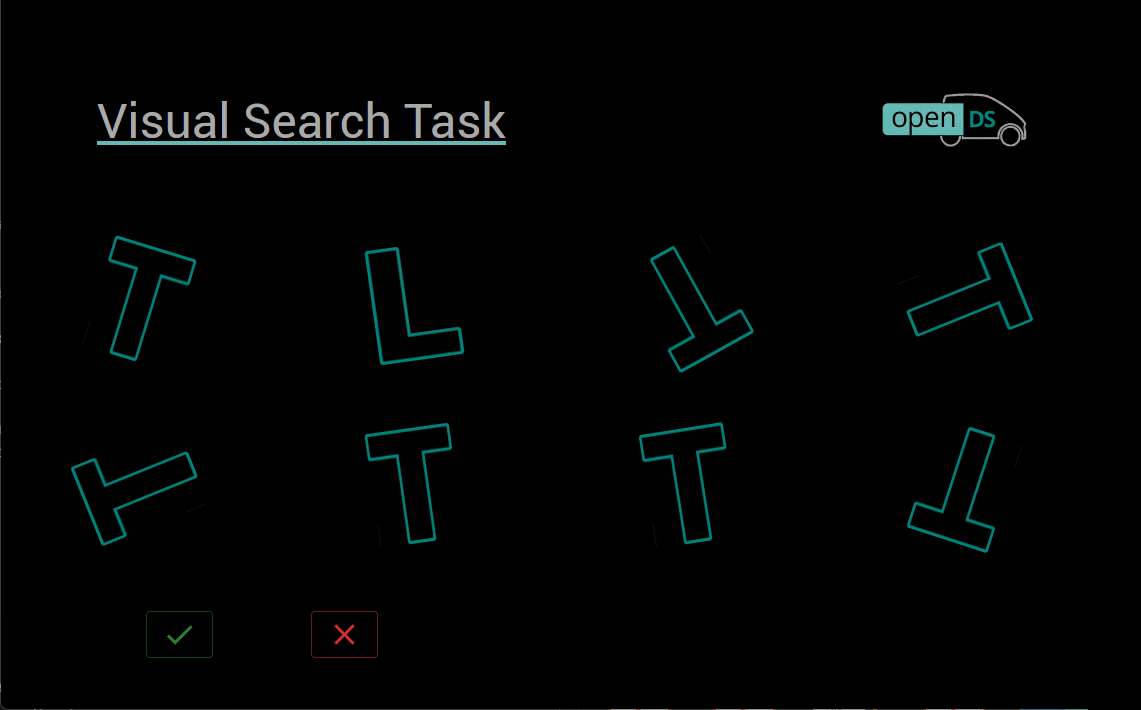}
         \caption{Visual Search Task}
         \label{fig:visual_search_task}
     \end{subfigure}
     \begin{subfigure}{0.49\linewidth}
         \centering
         \includegraphics[width=\linewidth]{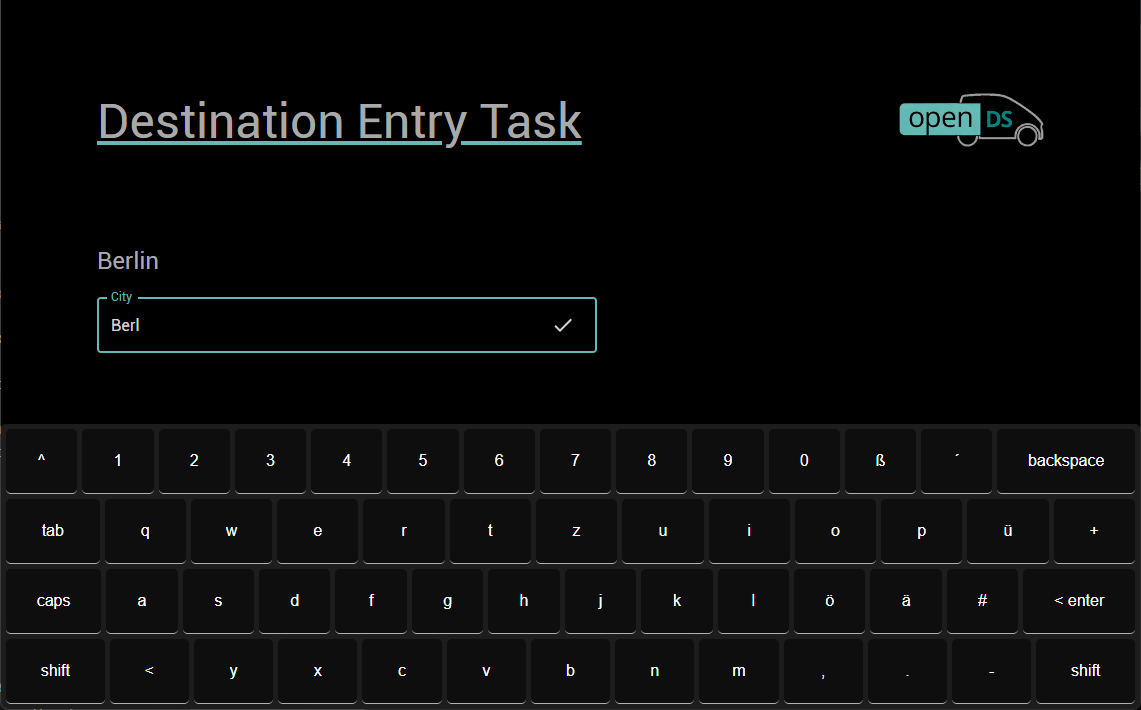}
         \caption{Destination Entry Task}
         \label{fig:destination_entry_task}
     \end{subfigure}
     \begin{subfigure}{0.6\linewidth}
         \centering
         \includegraphics[width=\linewidth]{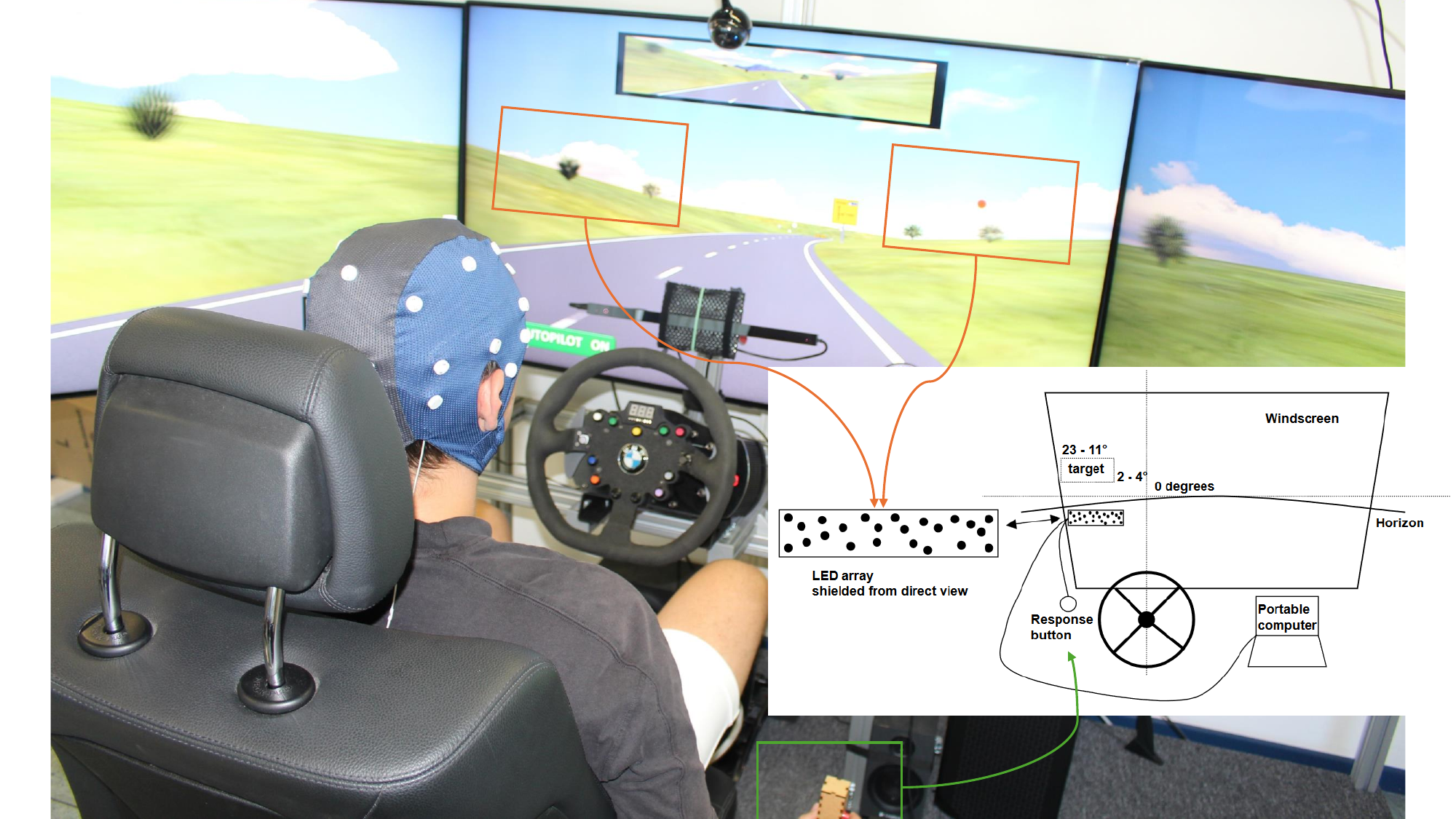}
         \caption{Peripheral Detection Task: The schematic is from \cite{olsson_measuring_2000}.}
         \label{fig:peripehral_detection_task}
     \end{subfigure}
    \caption{The \glspl{ndrt} used in both user studies.}
  \Description{This figure contains three images: two side by side at the top, and one at the bottom. The top left image shows a screen displaying the Visual Search Task. There are eight green letters, either “T” (n=7) or “L” (n=1), on a black background. There is also a checkmark and an “X” that you need to press to complete the task. The top right image shows a screen displaying the Destination Entry Task. There is a text field where you need to enter a location. The keyboard is visible on the screen for this purpose. The bottom image shows the driving simulator setup with the Peripheral Detection Task. On the middle simulator screen, two areas are marked with a red frame. A red dot is visible in one of these areas. The picture shows a schematic illustrating how the Peripheral Detection Task should be structured.}
  \label{fig:ndrts}
\end{figure*}

In the \textit{Display Type Study}, Participants were given an additional tablet to perform the \glspl{ndrt}. 
During the manual driving phases, participants were instructed to either place the tablet in their lap or on the chair next to them, which acted as a passenger seat. Overall, there were three different \glspl{ndrt}, but each study only had two. A visual search task was common in both studies. \textit{Time Budget Study} had a peripheral detection task, while \textit{Display Type Study} had a destination entry task. In each task, the expected and actual results were logged, as well as the start and end timestamps for every single submission for manipulation checks during the statistical analysis phase. 
All \glspl{ndrt} were utilised in previous work for inducing or estimating certain levels of mental workload in dual-task scenarios. An overview of all three tasks can be seen in~\autoref{fig:ndrts}.

\subsubsection{Visual Search Task}

This \gls{ndrt} was adapted from \cite{lavie_distracted_2005, gomaa_whats_2022}, and simulates the distraction of interacting with in-vehicle infotainment systems. Participants had to identify the existence of an L-shaped target among multiple T-shaped distracting items, as shown in~\autoref{fig:visual_search_task}, and respond by pressing on the L-shaped item. All items were randomly rotated between 0\textdegree and 360\textdegree.

\subsubsection{Destination Entry Task}

This \gls{ndrt} simulates typical entry tasks such as texting or destination entry for navigation. In this task, the application presented a city name that had to be typed correctly into the text input field, as shown in~\autoref{fig:destination_entry_task}. The entry could be confirmed either by clicking the confirmation button located to the right of the input field or by pressing the enter key on the keyboard. As soon as an entry was submitted, the next city name was displayed. The selection of city names was derived from the list of major cities of the country where the study took place, excluding names that contained white spaces.

\subsubsection{Peripheral Detection Task}

This \gls{ndrt} was adapted from \cite{olsson_measuring_2000, martens_measuring_2000, caber_driver_2024, bengler_assessment_2012}, where participants needed to detect red dots appearing on the centre screen of the simulator, as shown in~\autoref{fig:peripehral_detection_task}. They had to respond by pressing a button on the right side of the driver’s seat. This task simulates the lapses in attention that occur due to environmental distractions.

\subsection{User Studies}

We designed two user studies to evaluate our adaptation approach under multiple conditions (e.g., different traffic levels and environmental visual complexity) for different elements of a standard \gls{tor} warning system (e.g., different time budgets and different warning display locations). Both studies utilised the same driving simulator and similar environments across different scenarios and various situations for the takeover (i.e., critical and non-critical situations). 

Both studies were designed as within-participant counterbalanced driving experiments, where each participant experienced all conditions during the study. Participants had to complete a takeover upon a \gls{tor} initiated by the autonomous simulator. The \glspl{tor} occurred at nearly fixed time intervals. Between the takeover scenarios, the car drove autonomously, and the participants completed one of the two secondary \glspl{ndrt}. Both studies were piloted multiple times to mitigate any design issues.

To minimise the learning effect, each takeover situation was designed with unique characteristics, including varying hazard types, speed limits, and lane curvature. Furthermore, by alternating the starting scenario between \textit{Scenario One} and \textit{Scenario Two} in the \textit{Display Type Study}, and counter-balancing the six scenarios in the \textit{Time Budget Study}, the learning effect was evenly distributed across the scenarios and, consequently, the transfer of control situations. 

All participants were fluent in English and had a valid driver’s licence. Each participant provided informed consent prior to their participation in the study and received monetary compensation. They were given a detailed information sheet explaining the purpose and procedures of the research and were briefed on the camera recording, with assurances that all footage would be deleted once the gaze data had been extracted. To ensure anonymity and confidentiality, each participant was assigned a unique identification code, and any identifying information was kept separate from the research data. The study was conducted in accordance with ethical guidelines and received approval from the ethics committee. Participants were also informed that they could withdraw from the study at any time.

In both studies, participants were introduced to the complete setup and given time to read the instructions for the \glspl{ndrt} and practice them. They were also given the opportunity to freely drive around in the simulator in a tutorial scenario to familiarise themselves with the vehicle. Afterwards, they completed the driving scenarios. Only in the \textit{Time Budget Study}, the additional fully manual driving task was conducted, and participants were given a break in the middle due to the long duration of this study compared to the \textit{Display Type Study}. For both studies, while doing the \gls{ndrt}, participants were instructed to focus deeply on the task. The data from the participants were manually verified to ensure no data gaps and correct synchronisation. The participants filled a \gls{nasa} questionnaire verbally after each takeover while the vehicle was driving autonomously to the next scenario destination. 

To compute the required sample size for each study, we conducted an a priori analysis using G*Power \cite{faul_statistical_2009} for a medium effect (Cohen’s \textit{d} = $0.25$) according to conventions, with an alpha of \textit{p} = $0.05$ and a test power of $0.80$. The results indicated a required sample size of approximately 34 participants for the \textit{Time Budget Study} and approximately 20 participants for the \textit{Display Type Study}. We collected slightly more participants to account for participant data exclusion. A summary of the collected participants' data from both studies is shown in Table~\ref{tab:participant_demographics}.

\begin{table*}
    \centering
    \begin{tabular}{@{}lll@{}}
        \toprule
        & \textbf{Time Budget Study} & \textbf{Display Type Study} \\ 
        \midrule
        \textbf{Retained Participants} 
            & \textbf{N = 37} & \textbf{N = 29} \\ 
        \textbf{Participant Gender} 
            & M = 54.8\%, F = 45.2\% & M = 72.4\%, F = 27.6\% \\ 
        \textbf{Average Age}
            & 24.58 years & 26.8 years \\ 
        \textbf{Average Driving Experience} 
            & 5.79 years & 7.5 years \\ 
        \textbf{Automatic Transmission Experience} 
            & 59.5\% & 65.5\% \\ 
        \textbf{Simulator Experience}
            & 47.6\% & 27.6\% \\ 
        \bottomrule
    \end{tabular}
    \caption{Participant demographics and experience.}
    \Description{The table has two columns for each of our two studies: the Time Budget and the Display Type study. Each column is presented with six rows of information regarding participants' demographics. The Retained participants were 37 and 29 for the Time Budget study and Display Type study, respectively. Participants were 54.8\% and 72.4\% Male for the Time Budget study and the Display Type study, respectively. The remaining participants were female, with no participants identified as non-binary or choosing the self-describe option. The table also shows the average age as 24.58 and 26.8 years, while the average driving experience is 5.79 and 7.5 years for the Time Budget and Display Type study, respectively. The fifth and sixth rows show the automatic transmission experience as 59.5\% and 65.5\% while the simulator experience as 47.6\% and 27.6\% for the Time Budget and the Display Type study, respectively.}
    \label{tab:participant_demographics}
\end{table*}

\subsubsection{Display Type Study}

This study focused on varying the display types (\gls{hud} vs \gls{hdd}) for showing the \gls{tor}. The secondary \glspl{ndrt} for this study were the visual search task and the destination entry task. We used a regular RGB camera for gaze estimation. Before the actual study, the \gls{aoi} gaze tracking was calibrated individually for each participant. The study instructor directed participants to look at specific areas, such as the four corners of the three monitors, the \gls{hdd}, one area for the \gls{hud}, and the tablet in their hand. Each gaze at an area resulted in 18 frames per participant.

We collected data from 33 participants; however, four participants were excluded due to technical problems in video capturing (two participants), improper task execution (one participant), and inability to detect eye gaze due to reflections on glasses (one participant). The final sample size of 29 participants provided sufficient test power for the statistical analyses.

During the experiments, participants performed a \gls{ndrt} on the tablet until a warning appeared on either the \gls{hud} or the \gls{hdd}. Upon noticing the warning, they pressed the red button on the steering wheel, exited the hazardous area, and pressed the red button again to relinquish control back to the car. The time budget for the \gls{tor} occurred at fixed intervals of 7 seconds before collision. Participants were given a 3-minute break between the two scenarios.

\subsubsection{Time Budget Study}

This study focused on varying the \gls{tor} time budget, where the request was issued 4 seconds, 8 seconds, or 12 seconds before the autonomous car stopped driving. Participants could take control anytime during this period by pressing a button on the driving wheel. The secondary \glspl{ndrt} for this study were the visual search task and the peripheral detection task. As mentioned earlier, we used an \gls{eeg} headset and the Tobii eye tracker to collect data for situational awareness monitoring during this study. 

We collected data from 42 participants; however, five participants were excluded due to autopilot issues (two participants), motion sickness (one participant), and incorrect task execution (one participant). The final sample size of 37 participants provided sufficient test power for the statistical analyses.

During the experiments, participants were instructed to keep driving straight until the autopilot took over. About 60 seconds before the \gls{tor}, the experiment conductor instructed the participant to start a specific \gls{ndrt}. Participants were told to focus on this task until the next \gls{tor}. The \gls{tor} was communicated via audio and visual messages. The time budget for the \gls{tor} varied (4 seconds vs 8 seconds vs 12 seconds), but it was counterbalanced, and displayed by a countdown. Participants switched to manual driving by pressing the red button on the steering wheel.
In each driving scenario, a slow vehicle (e.g., a tractor) would appear in front of the simulator car, and participants were free to overtake if they wished. At a certain checkpoint, the autopilot would resume control. Participants were given a 5-minute break in the middle between scenarios.

\subsection{Evaluation Metrics}

We used various metrics to evaluate both user studies. We categorised the metrics into common ones used across both studies and study-specific ones used across only one study. 

\subsubsection{Common Metrics (i.e., Dependant Variables)}

\begin{itemize}
    \item \textbf{Reaction Time:} This refers to the time between a \gls{tor} and the actual takeover. This metric is crucial for assessing the responsiveness of participants, with lower values indicating better performance.
    
    \item \textbf{Mental Workload:} Measured using the \gls{nasa} score, which assesses the subjectively experienced workload for each transfer of control. This score provides insight into the cognitive state experienced by participants. We also considered the six rating sub-scales of \gls{nasa}, which are the Mental Demand, Physical Demand, Temporal Demand, Performance, Effort, and Frustration Level. Higher values indicate that the driver is more strained.

    \item \textbf{Takeover Quality:} This is analysed differently in the two studies.
    \begin{itemize}
        \item \textit{Time Budget Study} used the remaining time budget (sometimes referred to as \textit{Time Left till System Boundary}) as a metric to evaluate the takeover quality. This metric indicates how much time the driver decided to leave as a buffer before taking control. Since the time budget for taking control is variable in this study, the remaining time budget provides a better indicator of the driver's decision-making process. It is important to note that in this study, drivers do not observe the \gls{tor} reason as a hazardous event, allowing them to wait until the entire time budget expires, as some participants did.
        
        \item \textit{Display Type Study} used the reaction time as a metric to evaluate the takeover quality. Unlike the \textit{Time Budget Study}, the reaction time can be consistently measured in this context, providing a clear indicator of the driver's responsiveness to the \gls{tor}.
    \end{itemize}

    \item \textbf{Driving Performance:} This is measured using the commonly used metrics of lane deviation, speed deviation, and steering wheel position using their \gls{mean}, \gls{sd} and \gls{rmssd}~\cite{kozlowski_assessment_2016, gomaa_whats_2022, wei_state_2023}. Additionally, in the \textit{Display Type Study}, lane deviation is compared against an ideal manoeuvre trajectory using a person correlation coefficient. Higher lane deviation coefficient values indicate better manoeuvre quality.

    \item \textbf{Gaze Data:} This is analysed differently in the two studies.
    \begin{itemize}
        \item \textit{Time Budget Study} used the \gls{sge}, which applies Shannon’s entropy equation to the probability distribution of fixation coordinates. A higher entropy, which indicates a wider distribution of fixations (i.e., events where the eyes remain relatively still for a period lasting between a few tens of milliseconds up to a few seconds \cite{Holmqvist_Nystrom_Andersson_Dewhurst_Jarodzka_Weijer_2011}) across the visual field and suggests greater gaze dispersion, serves as an indicator of situational awareness and scanning efficiency as indicated by \citet{shiferaw_stationary_2018}.
        
        \item \textit{Display Type Study} analysed the scanpaths, which refer to the sequential eye movement pattern across space over a period of time~\cite{Holmqvist_Nystrom_Andersson_Dewhurst_Jarodzka_Weijer_2011}.
    \end{itemize}

\end{itemize}

\subsubsection{Common Independent Variables (IVs)}

\begin{itemize}    
    
    \item \textbf{Environmental Complexity:} This covers metrics related to the within factors of environmental complexity. This helps in understanding how different environmental conditions affect participant performance.
    \begin{itemize}
        \item In the \textit{Time Budget Study}, the \textbf{Urban environments} represent high-complexity environments, and the \textbf{Rural environments} represent low-complexity environments. This is because of the fixed speed limit in both environments and the increased traffic and visual complexity in the urban environment.
        
        \item In the \textit{Display Type Study}, the \textbf{Urban environments} represent low-complexity environments, and the \textbf{Rural environments} represent high-complexity environments.  This is because of the higher speeds in the rural environment.
    \end{itemize}
    
    \item \textbf{Driving Mode:} This refers to the mode of driving, whether manual or takeover semi-autonomous driving (i.e. when the driver is prompted to take control from the autonomous vehicle).
    
    \item \textbf{\gls{ndrt} Type:} This was considered in both studies for their respective \glspl{ndrt}.
\end{itemize}

\subsubsection{Display Type Study Specific Metrics and IVs}

\begin{itemize}
    \item \textbf{\gls{ndrt} Performance:} This refers to the assessed performance of participants in the \gls{ndrt} with respect to the number of tasks solved, the average time taken (in milliseconds), and the error rate.
    
    \item \textbf{Perceived Driving Difficulty:} This is assessed in rural and urban areas through a manipulation check by quantifying the mean derivative of filtered lateral and longitudinal acceleration and speed exhibited by participants during the manual driving phases.
    
    \item \textbf{Display Type:} This includes the different display types for showing the \glspl{tor} (\gls{hdd} vs \gls{hud}).
\end{itemize}

\subsubsection{Time Budget Study Specific Metrics and IVs}

\begin{itemize}
    \item \textbf{\gls{eeg} Data:} The \gls{eeg} data from each participant was collected to analyse their brain activity during the driving tasks. We processed the \gls{eeg} signals similar to \citet{Kartali_workload_2019} by computing the ratio between the theta and alpha bands in certain electrodes.
    \item \textbf{\gls{tor} Time Budget:} This includes the within factors of the different time budgets (4 seconds vs 8 seconds vs 12 seconds).
\end{itemize}

\subsection{Statistical Analysis Prerequisites and Metrics Preprocessing }

We used the Shapiro-Wilk Test \cite{shapiro_analysis_1965} to initially assess if each metric data follows a normal distribution. To investigate the differences between groups and conditions, we employed both \gls{anova} and \gls{manova}. \gls{anova} tests are generally robust to violations of the normality assumption \cite{blanca_non-normal_2017, schmider_is_2010, salkind_encyclopedia_2010, lumley_importance_2002, harwell_summarizing_1992, glass_consequences_1972}, as they compare the means of different groups rather than examine individual data points.

The F-test in \gls{anova} analyses compares the between-group variability to the within-group variability. This ratio tends to be robust to non-normality because it relies on the differences between group means rather than the distributional properties of individual observations. In traditional statistical literature, a commonly cited rule of thumb is to consider a sample size of around 30 (±5) as sufficient for the central limit theorem to approximate the distribution of means to approximately normal \cite{blanca_non-normal_2017, herzog_understanding_2019}, regardless of the underlying population distribution, assuming the data is not extremely skewed. To reduce skewness further and bring the distribution closer to normality, a logarithmic transformation was applied to the non-normal data \cite{berry_logarithmic_1987}. If significant differences were found, pair-wise Tukey’s HSD (honestly significant difference) test \cite{abdi_tukeys_2010} was used to confirm the statistical significance.

The study design ensured that any within factors for the planned \gls{manova} and \gls{anova} were independent and nominally scaled. All dependent variables were assumed to be interval-scaled. Univariate outliers of dependent variables were winsorised during data preparation.
The additional time budget in the \textit{Time Budget Study} was calculated from the winsorised reaction time.


\subsection{Hypotheses}
Our hypotheses can be categorised into \textit{Common Hypotheses}, which include hypotheses of interest to both user studies, and \textit{Time Budget Study} and \textit{Display Type Study}, which include study-specific hypotheses. 

\subsubsection{Common Hypotheses (CH)}

\begin{enumerate}[label=CH\arabic*:] 
    \item \textbf{Non-driving Related Tasks Engagement:} Higher \gls{ndrt} engagement (i.e., Disengagement from the driving environment during the "Car in Control" phase) results in increased reaction time, reduced takeover quality and driving performance, lower situation awareness, and higher mental workload.
    
    \item \textbf{Environmental Complexity:} Higher environmental complexity leads to increased reaction time, reduced takeover quality and driving performance, lower situation awareness, and higher mental workload.

\end{enumerate}

\subsubsection{Display Type Study}

\paragraph{Main Effects (DM)}

\begin{enumerate}[label=DM\arabic*:]
    \item \textbf{Non-driving Related Tasks:} The destination entry task leads to increased reaction time, reduced takeover quality and driving performance, and higher mental workload compared to the visual search task.
    
    \item \textbf{Non-driving Related Tasks Interruption Frequency:} Lower \gls{ndrt} interruption frequency (e.g., checking \gls{hud}, \gls{hdd}, or road) leads to increased reaction time, reduced takeover quality and driving performance, and higher mental workload.
    
    \item \textbf{Display:} Warnings presented on the \gls{hdd} result in increased reaction time, reduced takeover quality and driving performance, and higher mental workload compared to the \gls{hud}.
\end{enumerate}

\paragraph{Interaction Effects (DI)}

\begin{enumerate}[label=DI\arabic*:]
    \item \textbf{Environmental Complexity:} \glspl{tor} on the \gls{hud} reduce the effects of driving environment difficulty compared to the \gls{hdd}.
    \begin{itemize}
        \item In difficult driving environments, the increase in reaction time, reduction in takeover quality and driving performance, and increase in mental workload are less severe when \glspl{tor} are presented on the \gls{hud} compared to the \gls{hdd}.
    \end{itemize}

    \item \textbf{Non-driving Related Tasks:} \glspl{tor} on the \gls{hud} (compared to the \gls{hdd}) reduce the effects of the destination entry task (compared to the visual search task).
    \begin{itemize}
        \item While performing the visual-manual destination entry task, the increase in reaction time, reduction in takeover quality and driving performance, and increase in mental workload are less severe when \glspl{tor} are presented on the \gls{hud} compared to the \gls{hdd}.
    \end{itemize}

    \item \textbf{Non-driving Related Tasks Engagement:} \glspl{tor} on the \gls{hud} reduce the effects of high \gls{ndrt} engagement compared to the \gls{hdd}.
    \begin{itemize}
        \item When \gls{ndrt} engagement is high, the increase in reaction time, reduction in takeover quality and driving performance, and increase in mental workload are less severe when \glspl{tor} are presented on the \gls{hud} compared to the \gls{hdd}.
    \end{itemize}

    \item \textbf{Non-driving Related Tasks Interruption Frequency:} \glspl{tor} on the \gls{hud} reduce the effects of low \gls{ndrt} interruption frequency compared to the \gls{hdd}.
    \begin{itemize}
        \item When \gls{ndrt} interruption frequency is low, the increase in reaction time, reduction in takeover quality and driving performance, and increase in mental workload are less severe when \glspl{tor} are presented on the \gls{hud} compared to the \gls{hdd}.
    \end{itemize}
\end{enumerate}

\subsubsection{Time Budget Study}

\paragraph{Main Effects (TM)}

\begin{enumerate}[label=TM\arabic*:]
    \item \textbf{Driving Mode:} Autonomous driving results in reduced takeover quality and driving performance, lower situation awareness, and higher mental workload compared to manual driving in similar situations.
    
    \item \textbf{Time Budget:} A low time budget leads to reduced takeover quality and driving performance, lower situation awareness, and higher mental workload compared to medium or high time budgets.
\end{enumerate}

\paragraph{Interaction Effects (TI)}

\begin{enumerate}[label=TI\arabic*:]
    \item \textbf{Environmental Complexity:} A high time budget mitigates the effects of environmental complexity compared to low or medium time budgets.
    \begin{itemize}
        \item In highly complex environments, the increase in reaction time, reduction in takeover quality and driving performance, decrease in situation awareness, and increase in mental workload are less pronounced with a high time budget compared to a medium or low time budget.
    \end{itemize}

    \item \textbf{Non-driving Related Tasks:} A high time budget mitigates the effects of environmental disengagement compared to low or medium time budgets.
    \begin{itemize}
        \item During disengagement from the environment (e.g., visual search task), the increase in reaction time, reduction in takeover quality and driving performance, decrease in situation awareness, and increase in mental workload are less severe with a high time budget compared to a medium or low time budget.
    \end{itemize}

    \item \textbf{Driving Mode:} Lower environmental complexity results in a smaller difference between manual and autonomous driving in similar situations.
    \begin{itemize}
        \item In low-complexity environments, the reduction in takeover quality and driving performance, decrease in situation awareness, and increase in mental workload during autonomous driving are less severe compared to high-complexity environments.
    \end{itemize}
\end{enumerate}

\section{Results}

In this section, we present our results for each user study separately. 
Due to the significance of the gaze data analysis in both studies, we present its related findings in a separate section before addressing each hypothesis individually; this also holds true for the \gls{eeg} data analysis in the \textit{Time Budget Study}.
In both user studies, all our statistical tests were conducted on within-subject factors.

In order to avoid any confusion in this section, we briefly restate the following points: \textbf{(1)} For the \gls{nasa} results, we explored the subscales in addition to the overall computed workload, based on prior research findings that highlighted the benefits of this approach \cite{mckendrick_deeper_2018, sugiono_investigating_2017}; 
\textbf{(2)} For the driving mode, takeover driving refers to when the driver is prompted to take control of the semi-autonomous vehicle; \textbf{(3)} In the \textit{Time Budget Study} the urban environment represents high-complexity and the rural environment represents low-complexity, but this is reversed in the \textit{Display Type Study} with urban representing low-complexity and rural representing high-complexity environments.

Finally, we summarize all the results and hypotheses in~\autoref{tab:DT_summary} and~\autoref{tab:TB_summary} which are located at the end of each study subsection.

\subsection{Display Type Study}

\subsubsection{Gaze Analysis}

\begin{figure}
    \centering
    \begin{subfigure}{0.48\textwidth}
        \includegraphics[width=\linewidth]{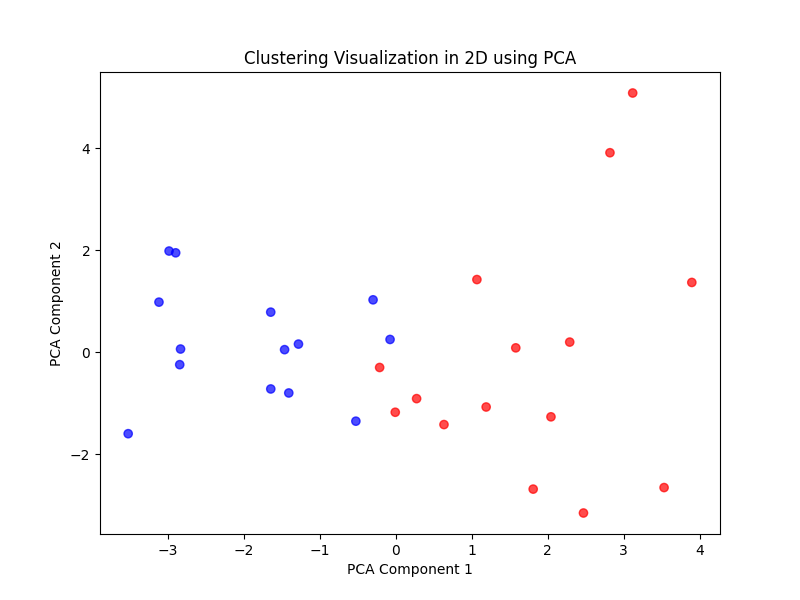}
        \caption{Clustering for $K=2$ visualized in 2D space.}
        \label{fig:cluster_K=2}
    \end{subfigure}
    \hspace{0.2cm}
    \begin{subfigure}{0.48\textwidth}
        \includegraphics[width=\linewidth]{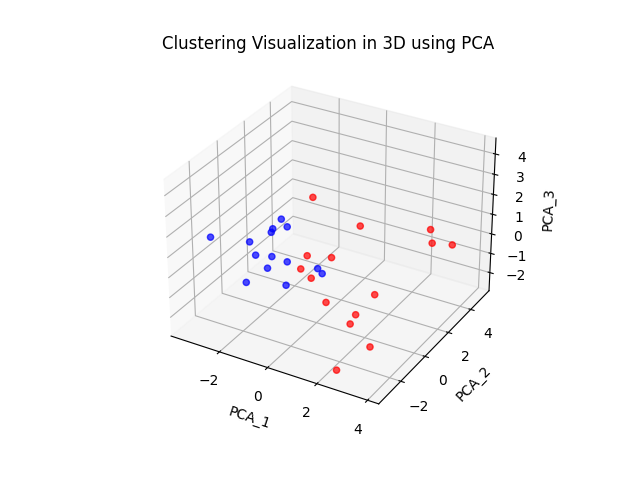}
        \caption{Clustering for $K=2$ visualized in 3D space.}
        \label{fig:cluster_K=2_3d}
    \end{subfigure}
    \caption{Clustering visualisation showing distinct participant groups based on their eye gaze patterns for $K=2$ with a silhouette score of 0.2240. \textit{Cluster 1}: 14 members, \textit{Cluster 2}: 15 members.
    PCA was used to reduce the high-dimensional data to (a) two and (b) three dimensions. Similar data points are close together, while dissimilar points are farther apart.}
    \Description{This figure contains two images side by side. The left image shows a scatter plot in 2D with two distinct clusters of points plotted on a two-dimensional plane. The horizontal axis is labelled "PCA Component 1", and the vertical axis is labelled "PCA Component 2". One cluster is represented by blue points (i.e., the left cluster) and the other by red points (i.e., the right cluster). The right image shows a 3D scatter plot with two distinct clusters of data points in a three-dimensional space. The axes are labelled PCA_1, PCA_2, and PCA_3. One cluster is represented by blue points (i.e., the left cluster) and the other by red points (i.e., the right cluster). This figure illustrates the result of PCA providing a visual representation of how data can be clustered into groups based on similarities in their features.}
    \label{fig:cluster_visualization}
\end{figure}

We transformed the collected raw gaze data (i.e., the vertical and horizontal head pose angles, as well as the vertical and horizontal eye gaze angles) into scanpaths by assigning them to \glspl{aoi} based on the data obtained during the calibration process. We categorised the scanpaths based on the phase, i.e., the \gls{ndrt} phase, the warning phase, and the manual driving phase. 
We found that the \gls{ndrt} phase contained the longest fixation duration, with an average duration of \SI{3332}{\ms}. Despite detecting the \gls{tor} on the \gls{hud} or the \gls{hdd} being the primary task, the road had an average fixation duration of \SI{1129}{\ms}. 
On average, participants fixated on the \gls{hud} for longer and more often than on the \gls{hdd}.
However, during the destination entry \gls{ndrt}, participants spent less time fixating on the \gls{ndrt}, with the remaining time distributed almost equally across the \gls{hud}, \gls{hdd}, and road. In high-complexity environments, more attention was paid to the road, \gls{hud}, and \gls{hdd}.
Most \gls{aoi} transitions were toward the \gls{ndrt} task during takeover driving. 
Participants often interrupted their \gls{ndrt} to fixate on the \gls{hud} or transitioned from the road to the \gls{hud}. In high-complexity environments, more transitions were observed between the road and \gls{hud}, as well as between the road and \gls{hdd}, compared to low-complexity environments. 

To identify different patterns among our 29 participants, we applied the K-means clustering algorithm \cite{macqueen_methods_1967, lloyd_least_1982}. Selecting an appropriate value of K is crucial as it directly impacts the resulting clustering. For each participant, we used the mean duration per \gls{aoi} (4), the mean number of fixations per \gls{aoi} (4), and the mean number of transitions from one \gls{aoi} to another (12), totalling 20 features. After standardising our data, we ran the clustering algorithm multiple times with varying numbers of clusters.
To evaluate the quality of clustering results, we used the silhouette score \cite{rousseeuw_silhouettes_1987}, which measures how well-separated the clusters are and indicates the appropriateness of the number of clusters for the given data. We decided on two clusters with a silhouette score of 0.2240, which are shown in~\autoref{fig:cluster_K=2}. Participants were equally distributed across the two clusters, with 14 members in \textit{Cluster 1} and 15 in \textit{Cluster 2}. 
In~\autoref{fig:cluster_visualization}, we used Principal Component Analysis (PCA) to reduce the high-dimensional data (i.e., 20 features) to two and three dimensions; the visualisation shows some data points close to each other despite being in different clusters.

The main differences between the two clusters are the fixation duration on the road and the \glspl{ndrt}. \textit{Cluster 1} participants spent, on average, \SI{1241}{\ms} more fixating on the road and \SI{1316}{\ms} less on the \glspl{ndrt}. Additionally, the average number of fixations on the road in \textit{Cluster 2} exceeds that of \textit{Cluster 1}. 
Although the number of fixations on the \gls{hud} and \gls{hdd} is similar in both clusters, the mean duration spent fixating on the \gls{hud} or \gls{hdd} is on average \SI{43}{\ms} longer for members of \textit{Cluster 1}.
When comparing reaction times, participants from \textit{Cluster 1} missed 4 \glspl{tor}, while participants from \textit{Cluster 2} missed 14, with 12 of the missed \glspl{tor} issued on the \gls{hdd}. The mean reaction time for \textit{Cluster 2} is \SI{223}{\ms} longer compared to \textit{Cluster 1}. For \textit{Cluster 1}, there is no difference in the reaction times between \gls{hud} and \gls{hdd} (\SI{2748}{\ms} vs \SI{2750}{\ms}). For \textit{Cluster 2}, participants took on average \SI{2446}{\ms} to react to warnings on the \gls{hud} and \SI{2812}{\ms} to warnings on the \gls{hdd}.

Participants from \textit{Cluster 2} generally took longer to react to warnings when performing the destination entry task (\textit{M} = \SI{3487}{\ms} vs \textit{M} = \SI{2961}{\ms}). There is also a difference between reaction times in the destination entry task and the visual search task for \textit{Cluster 2}, with faster reactions during the visual search task (\SI{2760}{\ms} vs \SI{3487}{\ms}). For members of \textit{Cluster 1}, no such differences were found.
Members of \textit{Cluster 1} performed better in low-complexity environments compared to high-complexity ones (\SI{2570}{\ms} vs \SI{3239}{\ms}). For \textit{Cluster 2}, no such difference was found between the driving environments.

Regarding \glspl{ndrt} performance, participants in \textit{Cluster 1} accomplished, on average, 6.68 more tasks than those in \textit{Cluster 2}, and with a lower failure rate in both \glspl{ndrt} compared to \textit{Cluster 2}. 
Members of \textit{Cluster 2} solved more tasks in low-complexity environments (26.19) compared to high-complexity environments (22.87).
The mean number of interruptions of the \gls{ndrt} during the takeover driving phases is slightly higher in both clusters in low-complexity environments.

In terms of driving quality, participants in \textit{Cluster 2} had a higher lateral offset correlation coefficient. For both clusters, there is a difference in the lateral offset correlation coefficient when comparing the two display conditions; on average, \gls{tor} on the \gls{hud} adds 0.2 to the coefficient, indicating better quality. 
The \gls{hud} also reduces steering wheel activity in both clusters, indicating better quality. The differences in driving environment conditions and \gls{ndrt} conditions are similar in both clusters.

We conducted one-way \gls{anova} tests to investigate the impact of the two clusters on reaction time, where missed reactions were imputed with the maximum takeover time of \SI{7000}{\ms}, yielding significant differences \anova{1}{228}{4.1}{0.04}. Post hoc results suggested that these differences were only significant for \textit{Cluster 2}. For \textit{Cluster 2}, reaction times were significantly lower when presenting \glspl{tor} on the \gls{hud} \tukey{1062}{0.003}{258}{1845}. For workload \anova{1}{228}{6.6}{0.95} and the \gls{nasa} subscales, no significant differences were found. Additionally, no differences in driving quality were observed.

\subsubsection{Common Hypotheses (CH)}

\paragraph{CH1: Non-driving Related Tasks Engagement} 

To investigate the impact of \gls{ndrt} engagement on our dependent variables, we conducted \gls{anova} tests. The analysis revealed no significant differences in reaction time \anova{1}{230}{1.01}{0.31} or workload \anova{1}{230}{0.03}{0.85}, including the \gls{nasa} subscale ratings, between high and low \gls{ndrt} engagement. However, a significant difference was observed in the mean derivative of the steering wheel position \anova{1}{212}{8.15}{0.004}. Tukey's HSD post hoc test indicated that the mean derivative of the steering wheel position was significantly higher when \gls{ndrt} engagement was high compared to when it was low \tukey{0.0001}{0.004}{0.0}{0.0001}.

Overall, \gls{ndrt} engagement does not significantly affect the reaction time or workload. However, it does affect the driving quality, as indicated by the steering wheel position. Therefore, hypothesis CH1 is partially supported, although the evidence based on the steering wheel position is relatively weak.

\paragraph{CH2: Environmental Complexity}

Comparing the mean reaction times for the two driving environments revealed that the takeover time was higher in high-complexity ones. Participants took, on average, \SI{2431}{\ms} in low-complexity environments, as opposed to \SI{2921}{\ms} in high-complexity environments. 
However, 11 takeover requests were missed in low-complexity environments, compared to only 7 in high-complexity environments. When imputing the missed reactions with the maximum takeover time of \SI{7000}{\ms}, the high-complexity takeover situations still took longer with \SI{3167}{\ms} compared to \SI{2864}{\ms}. 


We performed separate one-way \gls{anova} tests to investigate the effect of driving difficulties on reaction times, driving performance measures, and mental workload. The analysis revealed a statistically significant effect of driving difficulty on reaction time \anova{1}{228}{8.95}{0.003}. Tukey's HSD post hoc test indicated a statistically significant difference between low-complexity and high-complexity takeover situations \tukey{489.66}{0.0031}{167.18}{812.13}. 
These findings suggest that high-complexity takeover situations have significantly higher values compared to low-complexity takeover situations. However, when imputing the missed reactions with the maximum takeover time of \SI{7000}{\ms}, no significant difference was found \anova{1}{228}{1.81}{0.17}.

For the effect of driving difficulty on subjective mental workload, no significant difference was found \anova{1}{228}{0.68}{0.40}. Nevertheless, \gls{anova} tests for the \gls{nasa} subscales showed differences in physical demand \anova{1}{230}{4.95}{0.02}, temporal demand \anova{1}{230}{6.19}{0.01}, frustration level \anova{1}{230}{5.73}{0.01}, and performance \anova{1}{230}{7.74}{0.005}. Post hoc tests revealed that physical demand \tukey{0.68}{0.02}{0.0788}{1.2919}, temporal demand \tukey{0.81}{0.01}{0.1708}{1.4671}, and frustration level \tukey{0.87}{0.01}{0.1544}{1.587} were rated higher in the \gls{hdd} condition, while performance \tukey{-0.87}{0.01}{-1.5017}{-0.2569} was rated higher in the \gls{hud} condition.

For driving quality, a significant difference was found in the mean lateral offset correlation coefficient \anova{1}{212}{11.06}{0.001}. Tukey's HSD post hoc test indicated a statistically significant difference between the two groups \tukey{0.16}{0.001}{0.03}{0.15}, suggesting a higher correlation coefficient in high-complexity takeover situations. For the mean derivative of the steering wheel position, the \gls{anova} revealed a significant difference \anova{1}{212}{11.04}{0.001}. Tukey's HSD post hoc test revealed that in low-complexity takeover situations, the mean derivative steering wheel position is significantly higher than in high-complexity situations \tukey{0.16}{0.001}{0.06}{0.26}.

Further \gls{anova} tests for two single takeover situations, one in the low-complexity environment and one in the high-complexity environment, both with warnings issued on the \gls{hud} and the same reason for the transfer of control (i.e., a car crash ahead), showed a significant difference in reaction time \anova{1}{56}{24.61}{0.0001}. The post hoc test revealed that reaction time is significantly lower in low-complexity environments compared to high-complexity ones \tukey{1583.06}{0.00001}{943.90}{2222.23}. For mental workload, no difference was found \anova{1}{56}{0.47}{0.49}. Significant effects on driving quality were found for the mean lateral offset correlation coefficient \anova{1}{56}{31.48}{0.00001} and mean derivative of the steering wheel position \anova{1}{56}{31.48}{0.00001}. Both the mean correlation coefficient \tukey{-0.14}{0.00001}{-0.2112}{-0.0879} and mean derivative of the steering wheel position \tukey{-0.0001}{0.00001}{-0.0001}{0.0} were higher in the low-complexity environment.

Overall, no significant difference in mental workload was found, but the higher driving difficulty negatively affects the reaction times and the driving quality in terms of lateral offset correlation coefficients. Therefore, hypothesis CH2 is partially supported.

\subsubsection{Main Effects Hypotheses (DM)}
\paragraph{DM1: Non-driving Related Tasks}


Comparing the mean reaction times for the two \glspl{ndrt} revealed that the takeover time was higher for the destination entry task. Participants took, on average, \SI{2839}{\ms} during the destination entry task, as opposed to \SI{2528}{\ms} for the visual search task. 
Furthermore, during the destination entry task, participants missed 11 \glspl{tor} compared to 7 during the visual search task.
Regarding \gls{ndrt} performance, more tasks were solved in the visual search task (33.35) since one visual search task requires less time to solve compared to the destination entry task (8.81). The failure rate was higher in the destination entry task at 6.5\% compared to the visual search task at 2.9\%. However, the number of interruptions was not affected by the \gls{ndrt}.

We conducted further one-way \gls{anova} tests to investigate whether the \glspl{ndrt} had a significant effect on reaction time, driving quality, and mental workload. The performed \gls{ndrt} did not yield a significant effect on reaction time \anova{1}{230}{3.51}{0.06}, workload \anova{1}{230}{0.58}{0.44}, or any of the driving performance measures. However, a significant difference in physical demand was found \anova{1}{230}{4.34}{0.038}, and Tukey's HSD test indicated that physical demand in the destination entry task was significantly higher compared to the visual search task \tukey{0.6422}{0.03}{0.0349}{1.2496}.

Even though physical demand was higher in the destination entry task condition, the overall workload was not affected. Therefore, hypothesis DM1 could not be supported because the \glspl{ndrt} did not have any significant effect on reaction time, driving quality, or mental workload in our study setup.

\paragraph{DM2: Non-driving Related Tasks Interruption Frequency}    

Each glance from the \gls{ndrt} tablet towards any of the monitors or the \gls{hdd} was considered a \gls{ndrt} interruption. The number of \gls{ndrt} interruptions did not seem to be affected by the driving environment (28.7 in the low-complexity environment vs 27.8 in the high-complexity environment). Nevertheless, more interruptions were observed in low-complexity environments when performing the visual search task.

We conducted further one-way \gls{anova} tests to investigate the impact of \gls{ndrt} interruption frequency during the takeover driving phases on the dependent variables. The interruption frequency was considered high when it exceeded $0.3/\unit{\s}$; otherwise, it was considered low. The interruption frequency did not yield a significant effect on reaction time \anova{1}{230}{5.15}{0.07} or workload \anova{1}{230}{0.14}{0.70}. However, a significant effect on the frustration level was found \anova{1}{230}{6.70}{0.01}. The post hoc test results indicated that a high interruption frequency led to a significantly higher frustration level among participants \tukey{0.94}{0.01}{0.2270}{1.6643}.
Additionally, an effect on the mean derivative of the steering wheel position was found \anova{1}{212}{5.15}{0.02}. The post hoc Tukey's HSD test revealed that the mean derivative of the steering wheel position was significantly lower \tukey{-0.05}{0.024}{-0.090}{-0.007} when there were higher \gls{ndrt} interruptions during the takeover driving phases. 
Therefore, we can conclude that DM2 was met because the frequency of \gls{ndrt} interruptions affected the driving quality, with a higher frequency resulting in better driving quality after taking over.

\paragraph{DM3: Display}   

The slowest reaction time was observed when the warning was displayed on the \gls{hud} in high-complexity environments during the destination entry task (\SI{3421}{\ms}). However, the fastest reaction times were recorded in low-complexity environments when the warning was displayed on the \gls{hud} during the visual search task (\SI{1986}{\ms}).
Comparing the mean reaction times for the display conditions, warnings on the \gls{hud} generally resulted in shorter reaction times (\SI{2593}{\ms}) compared to the \gls{hdd} (\SI{2779}{\ms}). The highest correlation was found between reaction time and display type, with the \gls{hud} negatively correlated to reaction time, indicating that showing warnings on it reduced the time, while showing warnings on the \gls{hdd} increased it. Additionally, there were only 3 missed takeover requests on the \gls{hud} compared to 15 on the \gls{hdd}. When imputing the missed reactions with the maximum takeover time of \SI{7000}{\ms}, the difference in mean reaction times between \gls{hud} and \gls{hdd} increased to \SI{618}{\ms}. The standard deviations in both conditions were high.
%
Moreover, the mean lateral offset correlation coefficient in the \gls{hud} condition was 0.66, which is 0.20 higher than in the \gls{hdd} condition. The mean derivative of the steering wheel position was also higher in the \gls{hdd} condition (0.14 vs 0.06), though large variances were found for the lateral offset correlation coefficient (up to 0.40) and steering wheel position (up to 0.21).

We performed separate one-way \gls{anova} tests to investigate the effect of the warning display on reaction time, takeover quality, and mental workload. The analysis revealed a significant difference when imputing the reaction times for missed takeovers with the maximum takeover time of \SI{7000}{\ms} \anova{1}{230}{7.73}{0.005}. Tukey's HSD post hoc test indicated that requests on the \gls{hdd} resulted in significantly higher reaction times than requests on the \gls{hud} \tukey{617.59}{0.0059}{-1055.10}{-180.08}. 
The \gls{anova} test on mental workload showed no significant difference \anova{1}{230}{0.61}{0.43}. However, differences were found in physical demand \anova{1}{230}{4.9}{0.02}, temporal demand \anova{1}{230}{6.1}{0.01}, performance \anova{1}{230}{5.7}{0.01}, and frustration level \anova{1}{230}{5.2}{0.01}. Post hoc Tukey's HSD tests revealed that physical demand was significantly higher when displaying the takeover warning on the \gls{hdd} \tukey{0.68}{0.07}{0.0788}{1.2919}. The same held for temporal demand and frustration level, both higher in the \gls{hdd} conditions (\tukey{0.819}{0.0135}{0.1708}{1.4671}, \tukey{0.8707}{0.01}{0.1544}{1.5870}). Performance, in contrast, was significantly higher in \gls{hud} conditions \tukey{-0.8793}{0.005}{-1.5017}{-0.2569}.

Driving quality, in terms of lateral offset correlation coefficient \anova{1}{212}{18.69}{0.00001} and mean derivative of the steering wheel position \anova{1}{212}{11.96}{0.0006}, was significantly different between \gls{hdd} and \gls{hud}. Post hoc Tukey's HSD test revealed that the correlation coefficient was significantly higher \tukey{-0.21}{0.0001}{-0.30}{-0.11} in the \gls{hud} condition, and the mean derivative of the steering wheel position was significantly lower \tukey{0.0794}{0.0007}{0.03}{0.12} in the \gls{hud} condition compared to \gls{hdd}.

Therefore, we can conclude that hypothesis DM3 is only partially supported due to the absence of a significant difference in overall mental workload. However, since three out of the six workload subscales were rated significantly lower in \gls{hud} conditions, the workload seems to be affected by the display, though this is not reflected in the overall workload. Additionally, reaction time is indeed lower when issuing \glspl{tor} on the \gls{hud}, and driving quality is improved compared to warnings on the \gls{hdd}.

\subsubsection{Interaction Effect Hypotheses (DI)}

\paragraph{DI1: Environmental Complexity}
When considering the display type, reaction times for the low-complexity environment were lower when issuing the \gls{tor} on the \gls{hud} as opposed to the \gls{hdd}. This is also reflected in the mean lateral offset correlation coefficient and the mean derivative of the steering wheel position. However, reaction times for high-complexity environments slightly increased when the \gls{tor} was prompted on the \gls{hud} compared to the \gls{hdd}. The same effect on driving quality was observed in low-complexity environments. 

We conducted multiple separate \gls{anova} tests to test the interaction effect of showing \glspl{tor} warnings on the \gls{hud} instead of the \gls{hdd} on driving difficulty. When imputing the missed reactions with the maximum takeover time of \SI{7000}{\ms}, a significant effect of the display type on reaction time was found \anova{1}{228}{13.53}{0.0002}. Tukey's HSD post hoc tests revealed that \glspl{tor} on the \gls{hud} compared to the \gls{hdd} in high-complexity environments did not significantly reduce reaction time \tukey{-420}{0.04}{-1003}{163}. However, the usage of the \gls{hud} in low-complexity environments significantly reduced reaction time \tukey{752}{0.0054}{169}{1335}.

According to the analysis results, the workload was not significantly impacted by the usage of the \gls{hud} \anova{1}{228}{2.79}{0.09}. However, the analysis of variance found a significant difference in mental workload \anova{1}{228}{5.2}{0.02}, and the post hoc test showed that, at least for low-complexity environments, warnings on the \gls{hud} significantly reduced physical demand and temporal demand (\tukey{1.3879}{0.082}{0.2693}{2.5065}, \tukey{1.5172}{0.0063}{0.3251}{2.7094}). In contrast, a significantly higher performance value was detected when issuing warnings on the \gls{hud} in low-complexity environments \tukey{-1.4741}{0.005}{-2.6254}{-0.3228}.

\gls{anova} tests for driving quality revealed that the choice of the display had no significant effect on the lateral offset correlation coefficient \anova{1}{210}{0.64}{0.42}, but had a significant effect on the mean derivative of the steering wheel position \anova{1}{210}{10.30}{0.001}. Tukey's HSD post hoc test reported that it had no significant effect on high-complexity environments, but significantly lowered the mean derivative of the steering wheel position in low-complexity environments \tukey{0.1529}{0.00001}{0.07}{0.23}.

Therefore, we can conclude that hypothesis DI1 is partially supported because it does not affect the overall mental workload and only affects reaction time and driving quality in low-complexity environments.

\paragraph{DI2: Non-driving Related Tasks} 

When considering the display type, reaction times for \glspl{ndrt} were lower when issuing the request on the \gls{hud}. This is also reflected in the mean lateral offset correlation coefficient and the mean derivative of the steering wheel position. However, the workload did not seem to be affected.

We conducted separate \gls{anova} tests to assess the interaction effect of display type on the destination entry task. Using a \gls{hud} instead of an \gls{hdd} did not yield a significant difference in reaction time, regardless of the imputation method \anova{1}{228}{0.07}{0.79}. Additionally, no significant difference in workload \anova{1}{228}{0.03}{0.85} or any of the subscale ratings and driving measures was found.
Therefore, hypothesis DI2 could not be supported.

\paragraph{DI3: Non-driving Related Tasks Engagement} 

Further analyses were conducted to determine whether \gls{hud} usage could mitigate the effects of \gls{ndrt} engagement. In the context of CH1, only significant effects were found on driving quality. Different display types did not yield significant differences in reaction time \anova{1}{228}{0.02}{0.086}, workload \anova{1}{228}{0.006}{0.93}, or any \gls{nasa} subscale ratings. Nonetheless, a significant difference was found in the mean derivative of the steering wheel position \anova{1}{210}{8.20}{0.004}. The post hoc test revealed that issuing the \gls{tor} on the \gls{hdd} while engagement is high results in a mean difference of 0.0391, but due to a p-value of 0.24, the null hypothesis cannot be rejected. However, when engagement is low, displaying the TOR on the \gls{hud} significantly decreases the mean derivative of the steering wheel position \tukey{0.018}{0.0002}{0.07}{0.30}.

Regarding driving quality measures, the mean lateral offset correlation coefficient was marginally higher when participants were performing the visual search task before taking over. Additionally, the mean of the first-order derivative of the steering wheel position was lower (0.09) compared to the destination entry task condition (0.11).
Therefore, hypothesis DI3 is only partially met by decreasing the effects of low engagement on driving quality.

\paragraph{DI4: Non-driving Related Tasks Interruption Frequency} 

\glspl{tor} on the \gls{hud} reduce the effects of low \gls{ndrt} interruption frequency, compared to the \gls{hdd}. When \gls{ndrt} interruption frequency is low, the increase in takeover time, reduction in takeover quality, and increase in mental workload are less severe when \glspl{tor} are presented on the \gls{hud} compared to the \gls{hdd}.

We investigated whether the effects of interruption frequency could be reduced by using the \gls{hud}. \gls{anova} tests regarding reaction time did not yield significant differences, regardless of the imputation method \anova{1}{228}{1.38}{0.99}. Similarly, analyses of variance did not find convincing differences in participants' experienced mental workload \anova{1}{228}{0.81}{0.36} or any of the \gls{nasa} subscale ratings. Additionally, the increased mean derivative of the steering wheel position could not be reduced when using the \gls{hud}, as no significant difference was found \anova{1}{210}{0.86}{0.35}. Therefore, we could not find enough evidence in the data to support DI4.

Overall, our results for the \textit{Display Type Study} show that hypotheses \textbf{CH1, CH2, DM2, DM3, DI1, DI3} were partially retained. However, we could not find enough evidence to support hypotheses \textbf{DM1, DI2, DI4} based on the data we have collected and analysed.

\begin{table*}[!tbp]
\centering
\resizebox{\textwidth}{!}{
\begin{threeparttable}
\centering
\caption{Summarized Results Display Type Study}
\label{tab:DT_summary}

\begin{tabular}{@{}llllll@{}}
\textbf{Hypothesis} & \textbf{Variable} & \textbf{Analysis type} & \textbf{F-score} & \textbf{p-value} & \textbf{Interpretation}
\\ \midrule
\multirow{3}{*}{CH1} & Reaction time & \multirow{3}{*}{ANOVA} & F(1, 230) = 1.01 & p = 0.310 (n.s.) & \multirow{3}{*}{Hypothesis partially met} \\
& Workload & & F(1, 230) = 0.03 & p = 0.850 (n.s.) & \\
& Driving quality & & F(1, 212) = 8.15 & p = 0.004 (**) &
\\ \midrule
\multirow{3}{*}{CH2} & Reaction time & \multirow{3}{*}{ANOVA} & F(1, 228) = 8.95 & p = 0.003 (**) & \multirow{3}{*}{Hypothesis partially met} \\
& Workload & & F(1, 228) = 0.68 & p = 0.400 (n.s.) & \\
& Driving quality & & F(1, 212) = 11.06 & p = 0.001 (**) & 
\\ \midrule
\multirow{2}{*}{DM1} & Reaction time & \multirow{2}{*}{ANOVA} & F(1, 230) = 3.51 & p = 0.060 (n.s.) & \multirow{2}{*}{Hypothesis not met} \\
& Workload & & F(1, 230) = 0.58 & p = 0.440 (n.s.) 
\\ \midrule
\multirow{3}{*}{DM2} & Reaction time & \multirow{3}{*}{ANOVA} & F(1, 230) = 5.15 & p = 0.070 (n.s.) & \multirow{3}{*}{Hypothesis partially met} \\
& Workload & & F(1, 230) = 0.14 & p = 0.700 (n.s.) & \\
& Driving quality & & F(1, 212) = 5.15 & p = 0.020 (*) & 
\\ \midrule
\multirow{3}{*}{DM3} & Reaction time & \multirow{3}{*}{ANOVA} & F(1, 230) = 7.73 & p = 0.005 (**) & \multirow{3}{*}{Hypothesis partially met} \\
& Workload & & F(1, 230) = 0.61 & p = 0.430 (n.s.) & \\
& Driving quality & & F(1, 212) = 18.69 & p < 0.001 (***) & 
\\ \midrule 
\multirow{3}{*}{DI1} & Reaction time & \multirow{3}{*}{ANOVA} & F(1, 228) = 13.53 & p < 0.001 (***) & \multirow{3}{*}{Hypothesis partially met} \\
& Workload & & F(1, 228) = 2.79 & p = 0.090 (n.s.) & \\
& Driving quality & & F(1, 210) = 10.30 & p = 0.001 (**) & 
\\ \midrule
\multirow{2}{*}{DI2} & Reaction time & \multirow{2}{*}{ANOVA} & F(1, 228) = 0.07 & p = 0.790 (n.s.) & \multirow{2}{*}{Hypothesis not met} \\
& Workload & & F(1, 228) = 0.03 & p = 0.850 (n.s.) &
\\ \midrule
\multirow{3}{*}{DI3} & Reaction time & \multirow{3}{*}{ANOVA} & F(1, 228) = 0.02 & p = 0.086 (n.s.) & \multirow{3}{*}{Hypothesis partially met} \\ 
& Workload & & F(1, 228) = 0.006 & p = 0.930 (n.s.) & \\
& Driving quality & & F(1, 210) = 8.20 & p = 0.004 (**) & 
\\ \midrule
\multirow{3}{*}{DI4} & Reaction time & \multirow{3}{*}{ANOVA} & F(1, 228) = 1.38 & p = 0.990 (n.s.) & \multirow{3}{*}{Hypothesis not met} \\
& Workload & & F(1, 228) = 0.81 & p = 0.350 (n.s.) & \\
& Driving quality & & F(1, 210) = 0.86 & p = 0.350 (n.s.) & \\ 
\bottomrule 
\end{tabular}

    \begin{tablenotes}
          \item Note: n.s. = not significant, * = \textit{p} <.05, ** = \textit{p} <.01, *** = \textit{p} <.001
        \end{tablenotes}
    \end{threeparttable}
        }

\end{table*}

\subsection{Time Budget Study}


\subsubsection{EEG Analysis}
 
We computed a single value representing mental workload using \gls{eeg}, following the method by \citet{Kartali_workload_2019}. Levene's test \cite{levene_robust_1960} was applied to check for homoscedasticity, confirming that the variances were homogeneous across all conditions (\( p > 0.05 \)).
 
We conducted an \gls{anova} test with the factors of driving mode and environmental complexity. The results showed no significant effects for environmental complexity \anovaEta{1}{36}{2.24}{0.143}{0.059}, driving mode \anovaEta{1}{36}{0.45}{0.507}{0.012}, or their interaction \anovaEta{1}{36}{1.78}{0.191}{0.047}.
 
An additional \gls{anova} test with the factors of environmental complexity, time budget, and \gls{ndrt} also revealed no significant effects: \anovaEta{1}{36}{0.37}{0.550}{0.010} for environmental complexity, \anovaEta{1.01}{16.35}{1.06}{0.311}{0.029}\footnote{\label{fn:note1}Sphericity corrections after Greenhouse-Geisser \cite{greenhouse_methods_1959}.} for time budget, \anovaEta{1}{36}{0.96}{0.334}{0.026} for \gls{ndrt}, \anovaEta{1.05}{37.75}{1.18}{0.287}{0.032}\footref{fn:note1} for the interaction between environmental complexity and time budget, \anovaEta{1}{36}{0.89}{0.352}{0.024} for the interaction between environmental complexity and \gls{ndrt}, \anovaEta{1.01}{36.27}{0.98}{0.329}{0.027}\footref{fn:note1} for the interaction between time budget and \gls{ndrt}, and \anovaEta{1.12}{40.40}{1.37}{0.253}{0.037}\footref{fn:note1} for the three-way interaction. These results suggest that our hypotheses in the \textit{Time Budget Study} could not be supported using \gls{eeg}-based workload measures.
 
\subsubsection{Gaze Analysis}
 
For the gaze analysis, we employed two approaches. The first approach, similar to previous work on situational awareness and our \textit{Display Type Study}, involved analysing different gaze behaviours on \glspl{aoi} such as the traffic signs, speedometer, and mirrors. The second approach utilised the \gls{sge} as an indicator of generic situational awareness, based on the extent of visual exploration.
 
\paragraph{Speedometer Fixations}
 
We performed an \gls{anova} test for the speedometer fixations, considering the driving mode and environmental complexity as factors. Since homoscedasticity was not met across all conditions, we applied an aligned rank transformation. The results did not indicate any significant effects: \anovaEta{1}{36}{3.60}{0.066}{0.165} for environmental complexity, \anovaEta{1}{36}{1.67}{0.204}{0.068} for driving mode, and \anovaEta{1}{36}{3.25}{0.080}{0.106} for their interaction.
 
We performed a further \gls{anova} test considering the environmental complexity, time budget, and \gls{ndrt} as factors. We used Levene's test to check for homoscedasticity, and for all of the conditions, the variances were homogeneous.
However, none of the results were significant: \anovaEta{1}{36}{3.90}{0.056}{0.098} for environmental complexity, \anovaEta{2}{72}{1.08}{0.347}{0.029} for time budget, \anovaEta{1}{36}{0.06}{0.802}{0.002} for \gls{ndrt}, \anovaEta{1.56}{55.99}{1.03}{0.347}{0.028} for the interaction between environmental complexity and time budget, \anovaEta{1}{36}{0.03}{0.863}{<0.001} for the interaction between environmental complexity and \gls{ndrt}, \anovaEta{1.64}{59.11}{0.64}{0.502}{0.017} for the interaction between time budget and \gls{ndrt}, and \anovaEta{2}{72}{0.334}{0.717}{0.009} for the three-way interaction.
 
\paragraph{Traffic Sign Fixations}
 
We performed an \gls{anova} test for traffic sign fixations, considering the driving mode and environmental complexity as factors. Homoscedasticity was confirmed using Levene's test, and the results indicated a significant interaction between environmental complexity and driving mode \anovaEta{1}{36}{11.32}{0.002}{0.239}. Simple main effects showed that environmental complexity had a significant impact on fixation percentages in both manual \anovaEta{1}{36}{6.98}{0.012}{0.162} and takeover driving \anovaEta{1}{36}{21.08}{<0.001}{0.369}, with participants glancing more frequently at signs in low-complexity environments.
In addition, participants showed a higher overall percentage of fixating on signs during takeover driving compared to manual driving, indicating an interaction effect between environmental complexity and driver situational awareness across different driving modes.
 
We performed a further \gls{anova} test considering the environmental complexity, time budget, and \gls{ndrt} as factors. We used Levene's test to check for homoscedasticity, and for all of the conditions, the variances were homogeneous.
However, none of the results were significant: \anovaEta{1}{36}{3.08}{0.088}{0.079} for environmental complexity, \anovaEta{2}{72}{0.216}{0.806}{0.006} for time budget, \anovaEta{1}{36}{0.75}{0.393}{0.020} for \gls{ndrt}, \anovaEta{2}{72}{1.77}{0.178}{0.047} for the interaction between environmental complexity and time budget, \anovaEta{1}{36}{0.44}{0.513}{0.012} for the interaction between environmental complexity and \gls{ndrt}, \anovaEta{1.7}{61.16}{0.61}{0.532}{0.017} for the interaction between time budget and \gls{ndrt}, and \anovaEta{2}{72}{0.68}{0.512}{0.018} for the three-way interaction.
  
\paragraph{Rear-view and Side Mirrors Fixations}

We performed a bootstrapped corrected \gls{manova} test for fixations on the three different mirrors, considering the driving mode and environmental complexity as factors. 
Box’s M test \citet{box_general_1949} indicated homogeneity of the variance-covariance matrix \chisq{6}{13.61}{0.034}, and no multivariate outliers were found using Mahalanobis distances \cite{mahalanobis_generalised_1936}. The results were not significant for environmental complexity \manova{EnvCom}{0.001}{1.000}, driving mode \manova{2.00}{0.264}, or their interaction \manova{0.001}{1.000}.

We performed a further \gls{manova} test considering the environmental complexity, time budget, and \gls{ndrt} as factors. Box’s M test indicated heterogeneity of the variance-covariance matrix \chisq{66}{150.39}{<0.001}, which was compensated by bootstrapping.
Significant effects were found for environmental complexity \manova{28.39}{<0.001}, with fewer fixations on the right mirror in high-complexity environments. 
However, the other effects were not significant: \manova{2.24}{0.356} for time budget, \manova{1.98}{0.350} for \gls{ndrt}, \manova{3.39}{0.278} for the interaction between environmental complexity and time budget, \manova{3.65}{0.136} for the interaction between environmental complexity and \gls{ndrt}, \manova{2.29}{0.421} for the interaction between time budget and \gls{ndrt}, and \manova{0.91}{0.870} for the three-way interaction.

\paragraph{All \gls{aoi} Glances Combined}

We performed a bootstrapped corrected \gls{manova} test for the percentage of fixations on the mirrors, speedometer, and traffic signs, considering driving mode and environmental complexity as factors. Box’s M test indicated heterogeneity of the variance-covariance matrix \chisq{18}{70.72}{<0.001}, but no multivariate outliers were found using Mahalanobis distances. A significant interaction between environmental complexity and driving mode \manova{4.78}{0.031} was identified. Simple main effects showed significant differences for driving mode in both low-complexity \manova{19.48}{<0.001} and high-complexity environments \manova{20.08}{<0.001}. 
Univariate comparisons indicated more mirror fixations in manual (\textit{M} = 7.66, \textit{SD} = 4.76) versus takeover driving (\textit{M} = 4.05, \textit{SD} = 2.96) in low-complexity environments, and more in manual (\textit{M} = 7.39, \textit{SD} = 4.87) versus takeover (\textit{M} = 3.45, \textit{SD} = 3.58) in high-complexity environments.

We performed a further \gls{manova} test considering the environmental complexity, time budget, and \gls{ndrt} as factors. Box’s M test indicated heterogeneity of the variance-covariance matrix \chisq{66}{180.13}{< 0.001}, and no multivariate outliers were found using Mahalanobis distances.
Significant effects were found for environmental complexity \manova{14.39}{< 0.001}. Univariate comparisons indicated more mirror fixations in low-complexity environments (\textit{M} = 4.60, \textit{SD} = 3.39) compared to the high-complexity environments (\textit{M} = 3.59, \textit{SD} = 3.33). 
Other effects were not significant: \manova{1.99}{0.623} for time budget, \manova{1.80}{0.340} for \gls{ndrt}, \manova{4.76}{0.317} for the interaction between environmental complexity and time budget, \manova{2.62}{0.184} for the interaction between environmental complexity and \gls{ndrt}, \manova{2.32}{0.672} for the interaction between time budget and \gls{ndrt}, and \manova{2.19}{0.799} for the three-way interaction.

\paragraph{Stationary Gaze Entropy (\gls{sge})}

We performed an \gls{anova} test for the \gls{sge} measurement, considering the driving mode and environmental complexity as factors.
Homoscedasticity was confirmed using Levene's test, and the results indicated a significant main effect for environmental complexity \anovaEta{1}{36}{4.42}{0.043}{0.109}, with high-complexity environments (\textit{M} = 1.77, \textit{SD} = 0.38) causing more gaze entropy than low-complexity environments (\textit{M} = 1.70, \textit{SD} = 0.35).
However, the results were not significant for driving mode \anovaEta{1}{36}{1.06}{0.310}{0.029}, or for the interaction between environmental complexity and driving mode \anovaEta{1}{36}{0.05}{0.823}{0.001}.

We performed a further \gls{anova} test considering the environmental complexity, time budget, and \gls{ndrt} as factors. We used Levene's test to check for homoscedasticity, and for all of the conditions, the variances were homogeneous.
However, none of the results were significant: \anovaEta{1}{36}{0.55}{0.465}{0.015} for environmental complexity, \anovaEta{2}{72}{0.40}{0.675}{0.011} for time budget, \anovaEta{1}{36}{3.68}{0.063}{0.093} for \gls{ndrt}, \anovaEta{2}{72}{0.82}{0.445}{0.022} for the interaction between environmental complexity and time budget, \anovaEta{1}{36}{0.06}{0.804}{0.002} for the interaction between environmental complexity and \gls{ndrt}, \anovaEta{2}{72}{1.36}{0.263}{0.036} for the interaction between time budget and \gls{ndrt}, and \anovaEta{2}{72}{0.20}{0.820}{0.005} for the three-way interaction.

\subsubsection{Common Hypotheses (CH)}

\paragraph{CH1: Non-driving Related Tasks Engagement}

We conducted an \gls{anova} for each \gls{ndrt}, considering the environmental complexity and time budget as factors. These analyses revealed no significant main effects, indicating that both \glspl{ndrt} were equally engaging across all factor combinations using \gls{eeg} and gaze metrics. Additional analysis using the subjective \gls{nasa} also indicated no significant differences for the \glspl{ndrt}. 

We further tested this using a \gls{manova} analysis with the factors of environmental complexity, time budget, and \gls{ndrt}. Box's M test indicated homogeneity of the variance-covariance matrix \chisq{231}{237.46}{0.371}. To check for multivariate outliers, we calculated 36 \gls{pcs} due to the sample size limitations. The Mahalanobis distances were then calculated and showed no multivariate outliers. 
Significant main effects were found for environmental complexity \manova{45.71}{< 0.001}, time budget \manova{17.04}{0.023}, and \gls{ndrt} \manova{18.94}{0.01} in the \gls{nasa} scores. 
However, the univariate comparisons for the main effect of the \glspl{ndrt} showed no significant differences for the \gls{nasa} subscales. 
The insignificant tendencies on each scale imply that the visual search task, which led to more disengagement from the driving scene, made takeover driving overall more demanding.
In addition, as previously shown, neither the gaze analysis nor the \gls{eeg} analysis provided any support for CH1, which means that CH1 could not be supported.

\paragraph{CH2: Environmental Complexity}

To assess the takeover quality using the remaining time budget metric, we conducted an \gls{anova} similar to previous analyses using the within factors environmental complexity, time budget, and \gls{ndrt}. Homoscedasticity was checked with Levene's test, and for all conditions, homogeneity of variances was confirmed. While this analysis was significant for the time budget, it was not significant for the other factors: \anovaEta{1}{36}{3.33}{0.076}{0.085} for environmental complexity, \anovaEta{1}{36}{0.24}{0.629}{0.007} for \gls{ndrt}, \anovaEta{2}{72}{0.85}{0.432}{0.023} for the interaction between environmental complexity and time budget, \anovaEta{1}{36}{1.25}{0.272}{0.033} for the interaction between environmental complexity and \gls{ndrt}, \anovaEta{1.7}{61.16}{0.28}{0.759}{0.008} for the interaction between time budget and \gls{ndrt}, and \anovaEta{1.45}{52.09}{2.91}{0.079}{0.075} for the three-way interaction.

For Lane Deviation, we conducted a \gls{manova} with the driving mode and environmental complexity as factors. Box’s M test showed homogeneity of the variance-covariance matrix \chisq{6}{10.41}{0.108}. Mahalanobis distances were calculated and showed no multivariate outliers. The results showed a main effect for environmental complexity \manova{13.63}{0.001}, for which univariate comparisons showed a significant difference in lane deviation mean and lane deviation RMSSD, indicating a higher deviation as well as higher entropy for low-complexity environments. The remaining effects were not significant \manova{0}{1}, \manova{0}{1}, showing no support for hypothesis TM1 or TI3.

For Speed Deviation, we conducted two bootstrapped corrected \gls{manova} tests with the driving mode and environmental complexity as factors. Box’s M test showed homogeneity of the variance-covariance matrix \chisq{6}{21.72}{0.001}. Mahalanobis distances were calculated and showed no multivariate outliers. The \gls{manova} showed a main effect for environmental complexity \manova{25.04}{<0.001}, for which univariate comparisons showed significant differences in speed deviation mean with higher deviation for high-complexity environments. The other effects were not significant: \manova{0.001}{1.000} for environmental complexity, and \manova{0.001}{1.000} for the interaction between environmental complexity and driving mode.
We further tested this using a \gls{manova} analysis with the factors of environmental complexity, time budget, and \gls{ndrt}. Box’s M test showed heterogeneity of the variance-covariance matrix \chisq{66}{206.57}{<0.001}, which is already compensated by the bootstrapping in the \gls{manova}. Mahalanobis distances were calculated and showed no multivariate outliers. The calculation showed a main effect for environmental complexity \manova{52.92}{<0.001}, for which univariate comparisons showed significant differences in speed deviation mean and RMSSD with higher deviation mean for high-complexity environments. 
However, the remaining effects were not significant: \manova{2.03}{0.765} for time budget, \manova{1.94}{0.479} for \gls{ndrt}, \manova{4.76}{0.449} for the interaction between environmental complexity and time budget, \manova{1.15}{0.759} for the interaction between environmental complexity and \gls{ndrt}, \manova{6.42}{0.295} for the interaction between time budget and \gls{ndrt}, and \manova{3.46}{0.627} for the three-way interaction.

For Steering Wheel Position, we again conducted two bootstrapped corrected \gls{manova} tests for the three steering wheel measures with the driving mode and environmental complexity as factors. Box’s M test showed heterogeneity of the variance-covariance matrix \chisq{6}{186.99}{<0.001}, which is already compensated by the bootstrapping in the \gls{manova}. Mahalanobis distances were calculated and showed no multivariate outliers. However, the \gls{manova} showed no significant effects: \manova{4.94}{0.171} for environmental complexity, \manova{0.001}{1.000} for driving mode, and \manova{0.001}{1.000} for the interaction between environmental complexity and driving mode.
We further tested this using a \gls{manova} analysis with the factors of environmental complexity, time budget, and \gls{ndrt}. Box’s M test showed heterogeneity of the variance-covariance matrix \chisq{66}{420.91}{<0.001}, which is already compensated by the bootstrapping in the \gls{manova}. Mahalanobis distances were calculated and showed no multivariate outliers. While the result showed a significant main effect for environmental complexity \manova{6.43}{<0.001}, univariate comparisons showed no significant differences for the individual variables. The remaining effects were not significant: \manova{9.88}{0.223} for time budget, \manova{3.75}{0.298} for \gls{ndrt}, \manova{7.65}{0.272} for the interaction between environmental complexity and time budget, \manova{3.55}{0.332} for the interaction between environmental complexity and \gls{ndrt}, \manova{11.20}{0.231} for the interaction between time budget and \gls{ndrt}, and \manova{5.54}{0.370} for the three-way interaction.

Overall, there are strong indicators for the effects of environmental complexity on multiple metrics. As previously indicated, gaze analysis showed significant effects for environmental complexity using the traffic signs fixations, side mirrors fixations, and the \gls{sge} measurement.
This indicates that CH2 is partially supported with a statistical significance in gaze analysis and the majority of the driving performance metrics. Finally, the univariate comparisons of the previous \gls{nasa} \gls{manova} for the main effect of environmental complexity showed significant differences for mental demand, physical demand, temporal demand, and effort, with the high-complexity environment rated as more demanding.

\subsubsection{Main Effects Hypotheses (TM)}
\paragraph{TM1: Driving Mode}

The previous gaze analysis partially supports TM1. While speedometer fixations were not statistically significant, the overall fixations at all \glspl{aoi} and traffic signs showed statistical significance, with the driver making more fixations during the baseline manual driving compared to the takeover driving. This suggests higher situational awareness in scenarios where the driver is performing complete manual driving compared to taking over control from a semi-automated vehicle. However, this effect is not reflected in driving performance, as all the previously conducted \gls{manova} tests showed no significant effect for TM1.

From a mental workload perspective, we conducted a \gls{manova} on the \gls{nasa} results with the driving mode and environmental complexity as factors. Box's M test indicated homogeneity of the variance-covariance matrix \chisq{63}{71.02}{0.228}. Mahalanobis distances were calculated and showed no multivariate outliers. The \gls{manova} revealed main effects for environmental complexity \manova{23.83}{0.001} and driving mode \manova{30.22}{0.007} in the \gls{nasa} scores. For the main effect of driving mode, univariate comparisons indicated significant differences in temporal demand, with autonomous driving rated as more temporally demanding than manual driving. Insignificant tendencies in the other \gls{nasa} subscales suggest an overall higher demand in the takeover condition. Therefore, TM1 is partially supported.

\paragraph{TM2: Time Budget}  
The gaze analyses showed no statistical support for TM2. However, subjective mental workload estimation using the \gls{nasa} questionnaire indicated an effect for the time budget. Univariate comparisons from the previous \gls{manova} analysis for the main effect of the time budget showed significant differences for temporal load \manova{10.85}{0.030}. Pairwise comparisons indicate that the short time budget was significantly more temporally demanding than the long time budget (CI [0.071;1.61], \textit{p} = 0.028). Furthermore, statistically insignificant tendencies in the remaining ratings suggest the same conclusion: the short time budget is more demanding than the medium and high time budgets, supporting hypothesis TM2.

The previously mentioned \gls{anova} tests conducted for the time till system boundary (i.e., remaining time budget) showed a significant effect for time budget \anovaEta{1.3}{46.86}{511.41}{<0.001}{0.934}. Pairwise comparisons reached significance for all comparisons (i.e., all \textit{p} < 0.001). Time till system boundary increased with an increase in time budget, with the short time budget (\textit{M} = 1274.35, \textit{SD} = 803.17) being significantly different from the medium (\textit{M} = 4612.86, \textit{SD} = 2443.87) and the long time budget (\textit{M} = 8268.34, \textit{SD} = 2443.87). Therefore, TM2 is partially supported.

\begin{table*}[!tbp]
\centering
\resizebox{\textwidth}{!}{ 
\begin{threeparttable}
\caption{Summarized Results Time Budget Study}
\label{tab:TB_summary}
\begin{tabular}{@{}lllp{1.5cm}lllp{2cm}@{}}
\textbf{Hypothesis} & \textbf{Variable} & \textbf{Measurement} & \textbf{Analysis type} & \textbf{MATS} & \textbf{F-score} & \textbf{p-value} & \textbf{Interpretation} \\ 
\midrule
CH1 & Workload & EEG & ANOVA & - & F(1.05, 37.75) = 1.18 & p = 0.287 (n.s.) & Hypothesis not met \\ \midrule
\multirow{8}{*}{CH2} & \multirow{5}{*}{Gaze} & Speedometer fixations & \multirow{3}{*}{ANOVA} & - & F(1, 36) = 3.25 & p = 0.080 (n.s.) & \multirow{8}{2cm}{Hypothesis partially met} \\
& & Traffic sign fixations & & - & F(1, 36) = 11.32 & p = 0.002 (**) & \\
& & Stationary gaze entropy & & - & F(1, 36) = 0.05 & p = 0.823 (n.s.) & \\
\cline{3-7}
& & Rear-view \& side mirrors fixations & \multirow{2}{*}{MANOVA} & MATS = 0.001 & - & p = 1.000 (n.s.) & \\
& & All AOI glances & & MATS = 4.78 & - & p = 0.031 (*)& \\
\cline{2-7}
& \multirow{3}{*}{Driving quality} & Lane deviation & \multirow{3}{*}{MANOVA} & MATS = 13.63 & - & p = 0.001 (**) & \\
& & Speed deviation & & MATS = 25.04 & - & p < 0.001 (***) & \\
& & Steering wheel position & & MATS = 4.94 & - & p = 0.171 (n.s.) & \\ 
\midrule 
\multirow{6}{*}{TM1} & \multirow{5}{*}{Gaze} & Speedometer fixations & \multirow{3}{*}{ANOVA} & - & F(1, 36) = 1.67 & p = 0.204 (n.s.) & \multirow{6}{2cm}{Hypothesis partially met} \\
& & Traffic sign fixations & & - & F(1, 36) = 11.32 & p = 0.002 (**) & \\
& & Stationary gaze entropy & & - & F(1, 36) = 1.06 & p = 0.310 (n.s.) & \\
\cline{3-7}
& & rear-view \& side mirrors fixations & \multirow{2}{*}{MANOVA} & MATS = 2.00 & - & p = 0.264 (n.s.) & \\
& & All AOI glances combined & & MATS = 4.78 & - & p = 0.031 (*) & \\ 
\cline{2-7}
& Workload & Temporal load & MANOVA & MATS = 30.22 & - & p = 0.007 (**) & \\
\midrule
\multirow{7}{*}{TM2} & \multirow{5}{*}{Gaze} & Speedometer fixations & \multirow{3}{*}{ANOVA} & - & F(2,72) = 1.08 & p = 0.347 (n.s.) & \multirow{7}{2cm}{Hypothesis partially met} \\
& & Traffic sign fixations & & - & F(2,72) = 0.216 & p = 0.806 (n.s.) & \\
& & Stationary gaze entropy & & - & F(2,72) = 0.40 & p = 0.675 (n.s.) & \\
\cline{3-7}
& & Rear-view \& side mirrors fixations & \multirow{2}{*}{MANOVA} & MATS = 2.24 & - & p = 0.356 (n.s.) & \\
& & All AOI glances combined & & MATS = 1.99 & - & p = 0.623 (n.s.) & \\ 
\cline{2-7}
& Workload & Temporal load & MANOVA & MATS = 10.85 & - & p = 0.030 (*) & \\
\cline{2-7}
& Takeover quality & Time till system boundary & ANOVA & - & F(1.3, 46.86) = 511.41 & p = <0.001 (***) & \\
\midrule
TI1 & - & - & MANOVA & MATS = 45.71 & - & p = <0.001 & Hypothesis not met \\
\midrule
TI2 & - & - & MANOVA & MATS = 45.71 & - & p = <0.001 & Hypothesis not met \\
\midrule
TI3 & - & - & MANOVA & MATS = 0.63 & - & p = 0.881 (n.s.) & Hypothesis not met \\

\bottomrule
\end{tabular}

\begin{tablenotes}
    \item Note: MATS = modified ANOVA-Type statistic, n.s. = not significant, * = \textit{p} < .05, ** = \textit{p} < .01, *** = \textit{p} < .001
\end{tablenotes}
\end{threeparttable}
} 
\end{table*}

\subsubsection{Interaction Effect Hypotheses (TI)}

All three interaction effect hypotheses could not be supported. The previous \gls{nasa} and \gls{manova} tests showed no significant interaction between time budget and environmental complexity \manova{45.71}{<0.001} or \gls{ndrt} \manova{45.71}{<0.001}, providing no support for hypotheses TI1 and TI2. Additionally, there was no significant interaction effect for driving mode and environmental complexity \manova{0.63}{0.881}, providing no support for TI3. The only effect partially supported for TI3 was found during the gaze analysis for fixations at all \glspl{aoi} and traffic signs. However, this is a weak correlation for supporting TI3. Therefore, future studies should explore interaction effects further.

Overall, our results for the \textit{Time Budget Study} show that hypotheses \textbf{CH2, TM1, TM2} were partially supported. However, we could not find enough evidence to support hypotheses \textbf{CH1, TI1, TI2, TI3} based on the data we have collected and analysed.

\section{Discussion}

The discussion section is divided into four parts. The first part discusses the interpretation of the previously reported results from the \textit{Display Type Study} and how they relate to the proposed hypotheses, while the second part focuses on the \textit{Time Budget Study}. The third part focuses on proposing guidelines for designing adaptive interfaces based on the results obtained from both studies. Finally, the fourth part highlights the limitations of the studies and proposes suggestions for improvements in future work.

\subsection{Display Type Study Discussion}

We begin our discussion by positioning our work within the context of previous research to verify and validate our results and interpretations. 
In a meta-analysis study by \citet{zhang_determinants_2019}, the mean reaction time across 129 studies included in the analysis was reported as \SI{2720}{\ms}. The mean reaction time across all conditions in our study was \SI{2687}{\ms}, indicating a consistent pattern across the studies.
Furthermore, our findings on lower reaction times for takeover warnings on \glspl{hud} align with prior research \cite{kim_exploring_2013, horrey_effects_2003, liu_comparison_2004}. Similar to the studies by \citet{kim_exploring_2013} and \citet{smith_head-up_2016}, participants in our study generally ranked their preference for \glspl{hud} higher than for conventional \gls{hdd}. Additionally, the increased driving quality associated with \glspl{hud}, as mentioned in \cite{horrey_driving_2004, medenica_augmented_2011}, is supported by our study as well (i.e., hypothesis DM3).

Our hypothesis that driving context difficulty negatively affects workload, reaction time, and driving quality (i.e., hypothesis CH2) is only partially supported. The driving quality, in terms of the lateral offset correlation coefficient, is higher in urban takeovers. However, the steering wheel activity, one of the main measurements for driving quality, is also higher. This may be due to the takeover situations involving roundabouts. Prior research on taking over control primarily focused on highway situations and incorporated different levels of difficulty based solely on traffic density. 
Nevertheless, \citet{gold_taking_2016} and \citet{radlmayr_how_2014} made similar observations when comparing highways with (high-complexity) and without (low-complexity) traffic, as we did when comparing rural (high-complexity) and urban (low-complexity) environments. Therefore, we conclude that greater driving context difficulty, either through increased traffic or an increased number of manoeuvres, leads to increased reaction time and worse driving quality.

Regarding the impact of the type of \gls{ndrt} (i.e., hypothesis DM1), our results show no discernible difference between the destination entry task and the visual search task concerning takeover time, mental workload, or other performance measures used to assess driver behaviour. 
Prior research has presented conflicting findings, with some studies reporting an effect of the \gls{ndrt} type on certain measures, while others found no such effect. Our current study aligns with the latter category, corroborating existing research that indicates the type of \gls{ndrt} does not significantly influence the quality of takeover performance \cite{dogan_effects_2019, radlmayr_influence_2019, radlmayr_how_2014, zeeb_why_2017, vogelpohl_transitioning_2018}. However, these preceding studies did not investigate the effect of \gls{ndrt} engagement (i.e., hypothesis CH1) and \gls{ndrt} interruption frequency (i.e., hypothesis DM2) before taking over control, as we did in this work. 
One of our findings is that both a low interruption frequency and high \gls{ndrt} engagement each independently lead to lower driving quality. 
Higher \gls{ndrt} engagement is usually highly related to lower \gls{ndrt} interruption frequency, though they do not have to occur simultaneously. These insights could be related to the findings of \citet{dogan_effects_2019}, which state that an \gls{ndrt} should induce a medium amount of task demands, i.e., sufficiently engaging but not too captivating to deplete the driver's mental capabilities. Even though we found no differences in mental workload for the \gls{ndrt} conditions or \gls{ndrt} engagement, there seems to be a relation between engagement and driving performance that might be related to increased workload.

We found evidence for the positive interaction effect of \gls{hud} usage on takeover performance only in low-complexity environments (i.e., hypothesis DI1). Moreover, this interaction effect positively affected high \gls{ndrt} engagement (i.e., hypothesis DI3). When comparing the different 2 $\times$ 2 $\times$ 2 conditions, we observed that the combination of \gls{hud}, visual search task, and low-complexity environment yielded the lowest workload, lowest reaction times, and high driving quality. Conversely, the lowest takeover performance quality was found in the condition combining \gls{hdd}, destination entry task, and high-complexity environment.

Regarding mental workload using subjective measures (e.g., \gls{nasa}), previous work presents conflicting verdicts on their accuracy in capturing the actual mental workload of the driver. Some driving simulator studies, including our \textit{Time Budget Study}, found significant differences in workload among different study conditions using \gls{nasa}; however, in this study, we did not find any significant differences, similar to the findings of \citet{kim_exploring_2013}.

When analysing the previously discussed participant clusters observed in this study, we found that participants who spent a lot of time fixating on the road (i.e., \textit{Cluster 2}) exhibited worse takeover quality. This is supported by the findings of \citet{radlmayr_how_2014}, which state that having the eyes on the road for a considerably longer time during autonomous driving does not guarantee situation awareness. To our knowledge, we are the first to cluster drivers by their gaze behaviour to gain insights into their takeover quality in semi-autonomous cars. Additionally, we found that participants in \textit{Cluster 2} took longer to react to a warning when performing the destination entry task compared to \textit{Cluster 1}. 
On the other hand, \textit{Cluster 1} was negatively affected by the high-complexity environments in terms of reaction time. However, driving quality cannot be tied to reaction time regarding the two clusters. 
In addition, the \gls{hud} significantly increased the driving quality for both clusters (i.e., hypothesis DM3). Furthermore, we examined the drivers of each cluster separately based on their demographic data and found no significant difference between them in terms of age, acquisition of driver's licences, or driving experiences. Consequently, we suppose that the dissimilarities in takeover quality are indeed a result of variations in gaze behaviour.

Regarding the preferences expressed by participants in the post-study questionnaire, we observed a direct correlation with our findings in terms of the warning display location. 86\% of the participants preferred the \gls{hud}, which is the display that we found to have the overall better takeover quality. For the \glspl{ndrt}, 79\% of the participants preferred the visual search task, though we did not find any significant differences in takeover quality between the two tasks. Lastly, only 55\% of the participants preferred takeover situations in the low-complexity environments, even though we found sufficient evidence that takeovers in the low-complexity environments yield lower reaction times and better driving quality. Therefore, adapting the warning strategy solely based on user preference might not result in the most efficient and safe system. 
We discuss this and other detailed guidelines for adapting takeover systems in Section~\ref{sec:guidelines}.

\subsection{Time Budget Study Discussion}

As previously reported, we included additional metrics in this study as opposed to the \textit{Display Type Study}. 
Since this study focuses on the time budget rather than visual takeover warnings, there are fewer constraints on visual and sensor burdens. Therefore, we utilised a dry electrode \gls{eeg} headset to assess mental workload, as in previous work by \citet{Kartali_workload_2019}. However, our hypotheses were not supported using this \gls{eeg}-based metric. 
However, this metric still provides an important finding regarding the use of \gls{eeg}-based analysis in automotive applications. After further analysis of the \gls{eeg} data on a sample level per timestamp and assessing the preprocessing techniques provided by \citet{Kartali_workload_2019}, we concluded that this method is not mature enough for highly mechanical tasks such as driving. While \citet{Kartali_workload_2019} tested their method on an aviation task, the participants in their study made minimal physical movements. In our study, and in the driving context in general, drivers perform strenuous physical activities that significantly affect \gls{eeg} data, as confirmed by our analysis, rendering the mental workload values inconclusive. Further studies should assess mental workload in similarly complex tasks to evaluate the maturity of the proposed methods.

In contrast, the gaze analysis provided more informative results. The significant interaction between environmental complexity and driving mode for traffic sign fixations revealed that participants paid more attention to traffic signs in high-complexity environments. For instance, in these complex settings with a dense array of signs and signals, drivers glanced more frequently at traffic signs during takeover situations. This suggests that in high-complexity environments, drivers need more time to process the visual information around them before resuming control. Therefore, a longer time budget (i.e., 12 seconds) in these environments is recommended, as suggested by the partially supported hypothesis CH2.

The analysis of \glspl{ndrt} did not indicate significant effects on either mental workload or situational awareness (i.e., hypothesis CH1). While our non-significant results cannot conclude whether the \glspl{ndrt} has an effect on our conditions, the performance in the \glspl{ndrt} indicates a differentiating factor between them. 
The peripheral detection task reached a mean correct rate of 86.33\%, while the visual search task rate was 96.80\%. Thus, the manipulation check of the autonomous driving phase with different \glspl{ndrt} can be seen as successful. This finding is crucial, as it suggests that the type of secondary task performed during autonomous driving might not influence the driver's ability to take over control. Whether the driver is engaged in a simple or complex task, our study suggests that there is no impact on mental workload or situational awareness. Therefore, decisions about the time budget for the transfer of control should focus more on environmental complexity rather than the type of \gls{ndrt} in future studies and for adaptive systems.

CH2, which hypothesised that environmental complexity would influence takeover quality, was partially supported. The study found that environmental complexity significantly affected driving performance metrics, including lane deviation and speed deviation. For instance, drivers experienced higher lane and speed deviation in rural, low-complexity environments, suggesting that they may struggle more to regain control in these settings. 
Interestingly, while urban environments impose higher visual and cognitive demands, drivers seemed to perform worse in simpler, rural settings, possibly due to reduced vigilance. These findings support the recommendation for an 8-second time budget in low-complexity environments instead of 4 seconds, allowing drivers enough time to stabilise their control without the additional pressure of heightened visual demands.

The interaction hypotheses (i.e., hypotheses TI1, TI2, and TI3) were not supported by the data. There were no significant interactions between environmental complexity, driving mode, or \glspl{ndrt}, suggesting that these factors do not interact to significantly affect mental workload or situational awareness. This absence of interaction effects might suggest that these variables should be considered independently when designing transfer of control systems. For instance, environmental complexity should dictate the time budget, regardless of the specific \gls{ndrt} being performed or the driving mode, as there is no substantial evidence that these variables influence each other in a meaningful way. However, future research should aim to further explore these potential interactions, as they could reveal subtle but important effects that were not captured in our study.

\subsection{Guidelines for an Adaptive Takeover Interface}\label{sec:guidelines}

For the \textit{Display Type Study}, we generally found shorter reaction times and fewer missed takeovers when issuing the request on the \gls{hud}, confirming \textbf{DM3}. Therefore, takeover systems should start with displaying warning messages on \gls{hud} over \gls{hdd} as a baseline default.

When comparing surrounding environment complexity, we were not able to find enough evidence to support \textbf{DI1} for high-complexity driving environments (i.e., it is not statistically significant). However, we found that using the \gls{hud} in low-complexity conditions not only shows decrements in physical demand, temporal demand, and higher subjective performance but also shows a reduction in reaction time and improvements in driving quality, suggesting that the \gls{hud} is the more suitable display type in low-complexity environments. 

As for the \glspl{ndrt}, we did not find enough evidence to support our hypothesis \textbf{DM1} that there is a significant difference between the destination entry and the visual search tasks. 
We also did not find enough evidence for hypothesis \textbf{DI2} that the usage of the \gls{hud} can reduce the found effects of \glspl{ndrt}. However, we were able to show evidence that the usage of \gls{hud} can improve the takeover in terms of driving quality. Surprisingly, it does not enhance the takeover for those drivers that are highly engaged in the \gls{ndrt} task but for those with low engagement (\textbf{DI3}). 
When the engagement is low, we suggest prompting warnings on the \gls{hud}, though when engagement is high, not enough evidence was found for a superior display type with respect to the driving quality metric.

When analysing the drivers' gaze behaviour, we found that longer times spent fixating on the road (i.e., \textit{Cluster 2}) do not ultimately yield shorter reaction times and decrease the awareness of warnings on the \gls{hdd}. As for participants highly engaged in the \gls{ndrt} (\textit{Cluster 1}), they benefit from the \gls{hud} in terms of driving quality but not in terms of reaction time. Those participants are also highly influenced by the high-complexity environments, whereas members fixating primarily on the road are rather influenced by the task they are performing. This shows the need for an adaptive system based on the situational awareness of the drivers (i.e., \textit{Cluster 1} versus \textit{Cluster 2}) rather than based on driving performance and surroundings-related conditions.

In conclusion, an example of an adaptive warning system based on these findings would be to use the \gls{hdd} as a default for faster reaction (i.e., lower reaction time) and then switch to \gls{hud} when it is noticed that the participant is fixating on the road or when the driving quality deteriorates. Another example would be to use \gls{hdd} when reaction time is prioritised (i.e., in critical hazardous situations) and use \gls{hud} when driving quality is prioritised (i.e., in non-critical takeover requests).

As for the \textit{Time Budget Study}, the results of our study provide clear guidelines for optimising time budgets in different driving environments. In high-complexity environments, a longer time budget of 12 seconds is recommended to give drivers enough time to process the complex visual and cognitive demands. In contrast, a moderate time budget of 8 seconds is recommended for low-complexity environments, where drivers face less visual demand but may require sufficient time to regain control without becoming complacent. These recommendations strike a balance between driver workload, situational awareness, and performance, ultimately enhancing safety during the transfer of control in autonomous driving.

Supporting this recommendation, we found confirming trends in gaze behaviour, particularly with the speedometer and mirror glances, even though these effects were not statistically significant. For example, in high-complexity environments, participants tended to glance more at the speedometer. Additionally, the significant effect of environmental complexity on mirror fixations, with fewer fixations in low-complexity environments, further underscores the difference in visual demand across environments. Surprisingly, drivers exhibited higher gaze entropy (i.e., more random visual exploration) in low-complexity environments, indicating that even though they fixated less on mirrors, they engaged in more exploratory behaviour. To accommodate this, a moderate time budget of 8 seconds is recommended for low-complexity environments instead of using the lowest value of 4 seconds, which is close to the average value used in critical situations takeover, as seen in the \textit{Display Type Study}. This provides drivers ample time to transition while not overwhelming them with unnecessary delays and enhancing driving quality.

Finally, when comparing completely manual driving to takeover driving, we found that drivers made more frequent fixations while manually controlling the vehicle, which indicates a more active engagement with their surroundings, as indicated by the partially met hypothesis \textbf{TM1}. However, this increased situational awareness did not translate into better driving performance; the \gls{nasa} scores revealed that autonomous driving was perceived as more temporally demanding than manual driving, implying that the transition from autonomous to manual control may increase the temporal burden on drivers. These findings support the previous recommendations to adjust the time budget to account for this increased demand (i.e., use an 8-second time budget instead of 4 seconds), particularly in high-complexity environments (i.e., by increasing the time budget further to 12 seconds), as indicated by the partially supported \textbf{TM2}.


\subsection{Limitation and Future Work}

Despite conducting pilot studies to assess all aspects of our research and aligning our findings with earlier studies, some limitations of our work should be considered.

In general, both experiments were conducted within a driving simulator. While this increases internal validity and allows control over the study, it lacks external validity. Future studies could attempt to reproduce our work in a real environment. However, safety and technical conditions are hard to fulfil in critical hazardous situations. Failure to react or even crashing does not really affect the driver in the simulator. However, the speed of their reactions suggested that they treated the takeover situation seriously.

A greater significance for interpretation is that participants knew they would need to take over control. This awareness likely influenced their response times. Consequently, the reaction times observed in our study might be shorter than those encountered in real-world driving, where longer periods of autonomous driving and infrequently required interventions are more common. We were able to mitigate this partially by having longer periods of autonomous driving in the \textit{Time Budget Study}. However, some participants were still able to predict roughly when the takeover would occur. This consideration highlights the importance of contextualising our findings and recognising that the results could be influenced by the anticipation of \glspl{tor}.

In the \textit{Display Type Study}, although we conducted our study as a within-subject study to ensure no learning effect for the different critical situations, there could be a slight influence of the scenarios over the display type used in them. Therefore, future work could reproduce the same experiment as a between-subjects study so that both display types (i.e., \gls{hud} and \gls{hdd}) are compared against the exact same critical scenarios.

Finally, other attributes could be studied in future work, such as environmental changes that are easily controlled by a simulator. Evaluating drivers' reactions in adverse weather conditions, such as rain, snow, or fog, may provide crucial insights into the adaptability and reliability of takeover systems in these challenging situations and highlight the adaptation attributes further. However, we argue that this can only be achieved by understanding the trade-offs and implications of different design approaches as presented in our work.

\section{Conclusion}

In this work, we conducted two complementary studies aimed at optimising takeover performance in semi-autonomous vehicles by focusing on the role of display type and time budget during \glspl{tor}. These studies provide essential insights into how automotive interfaces should adapt to varying driving conditions and user states to ensure safe and efficient takeover scenarios. Although the time budget and display type are among the most crucial attributes for takeover warning systems, other attributes should be similarly investigated for adaptation and personalisation capability. We see our approach and design guidelines as the foundation block for future studies on truly adaptive takeover systems.

In our \textit{Display Type Study}, we explored the impact of different display types on takeover performance. While our findings show statistically  superiority for using \gls{hud} over \gls{hdd} in several conditions, they also indicate the importance of an adaptive system and not using \gls{hud} exclusively as in previous work. For example, using \gls{hdd} for faster reaction situations (i.e., in critical situations) and using \gls{hud} for prioritised driving quality (i.e., in non-critical situations when the time budget is higher).

In our \textit{Time Budget Study}, we investigated how the time available for a takeover, or time budget, affects driver performance under varying environmental complexities. The results showed that a longer time budget (i.e., 12 seconds in our study) is essential in high-complexity environments where cognitive and visual demands are higher. In contrast, a moderate time budget of 8 seconds was sufficient in low-complexity environments. 
Our study also found that the type of \gls{ndrt} did not significantly influence mental workload or situational awareness, indicating that time budget considerations should primarily be guided by environmental factors. The driving performance analysis showed that drivers demonstrated better lane-keeping and fewer deviations with longer time budgets in high-complexity conditions. These results align with our recommendation for an adaptive time budget system that tailors the time provided for takeover based on environmental complexity. Gaze patterns also support a moderate time budget, as opposed to a short one, as drivers still exhibited more exploratory gaze behaviour in low-complexity environments, indicating a need for additional time to stabilise control.

By combining the findings of both studies, our research provides clear design guidelines for adaptive takeover systems in semi-autonomous vehicles. For example, in highly complex non-hazardous environments with high visual and cognitive distractions, we recommend using a \gls{hud} combined with a longer time budget (e.g., 12 seconds) to allow drivers sufficient time to regain situational awareness and control. Conversely, in less complex, yet hazardous environments, using an \gls{hdd} with a moderate time budget (e.g., 8 seconds) may suffice, given the lower visual demands and faster reaction requirements. Moreover, the system should dynamically adjust its warning strategy based on driver engagement with the \gls{ndrt} and real-time gaze behaviour, leveraging \gls{hud} when drivers show low engagement with the \gls{ndrt} or exhibit signs of distraction, and switching to \gls{hdd} when the situation demands rapid responses in critical scenarios.
In conclusion, by aligning these factors with environmental complexity and driver states, we provide a pathway for improving the efficiency, safety, and adaptability of \glspl{tor} in semi-autonomous vehicles. We believe that our guidelines are a cornerstone in the development of safer, user-centred autonomous driving technologies, enhancing overall driving safety and comfort.

\bibliographystyle{ACM-Reference-Format}
\bibliography{references}


\begin{thebibliography}{91}


\ifx \showCODEN    \undefined \def \showCODEN     #1{\unskip}     \fi
\ifx \showISBNx    \undefined \def \showISBNx     #1{\unskip}     \fi
\ifx \showISBNxiii \undefined \def \showISBNxiii  #1{\unskip}     \fi
\ifx \showISSN     \undefined \def \showISSN      #1{\unskip}     \fi
\ifx \showLCCN     \undefined \def \showLCCN      #1{\unskip}     \fi
\ifx \shownote     \undefined \def \shownote      #1{#1}          \fi
\ifx \showarticletitle \undefined \def \showarticletitle #1{#1}   \fi
\ifx \showURL      \undefined \def \showURL       {\relax}        \fi
\providecommand\bibfield[2]{#2}
\providecommand\bibinfo[2]{#2}
\providecommand\natexlab[1]{#1}
\providecommand\showeprint[2][]{arXiv:#2}

\bibitem[Abdi and Williams(2010)]%
        {abdi_tukeys_2010}
\bibfield{author}{\bibinfo{person}{Herve Abdi} {and} \bibinfo{person}{Lynne~J. Williams}.} \bibinfo{year}{2010}\natexlab{}.
\newblock \showarticletitle{Tukey's {Honestly} {Significant} {Difference} ({HSD}) {Test}}.
\newblock In \bibinfo{booktitle}{\emph{Encyclopedia of {Research} {Design}}}, \bibfield{editor}{\bibinfo{person}{Neil~J. Salkind}} (Ed.). \bibinfo{publisher}{SAGE Publications, Inc.}, \bibinfo{address}{Thousand Oaks}, \bibinfo{pages}{1566--1570}.
\newblock
\href{https://doi.org/10.4135/9781412961288}{doi:\nolinkurl{10.4135/9781412961288}}


\bibitem[Baldisserotto et~al\mbox{.}(2023)]%
        {baldisserotto_review_2023}
\bibfield{author}{\bibinfo{person}{Filippo Baldisserotto}, \bibinfo{person}{Krzysztof Krejtz}, {and} \bibinfo{person}{Izabela Krejtz}.} \bibinfo{year}{2023}\natexlab{}.
\newblock \showarticletitle{A {Review} of {Eye} {Tracking} in {Advanced} {Driver} {Assistance} {Systems}: {An} {Adaptive} {Multi}-{Modal} {Eye} {Tracking} {Interface} {Solution}}. In \bibinfo{booktitle}{\emph{Proceedings of the 2023 {Symposium} on {Eye} {Tracking} {Research} and {Applications}}} \emph{(\bibinfo{series}{{ETRA} '23})}. \bibinfo{publisher}{Association for Computing Machinery}, \bibinfo{address}{New York, NY, USA}, \bibinfo{pages}{1--3}.
\newblock
\showISBNx{9798400701504}
\href{https://doi.org/10.1145/3588015.3589512}{doi:\nolinkurl{10.1145/3588015.3589512}}


\bibitem[Barua et~al\mbox{.}(2020)]%
        {barua_towards_2020}
\bibfield{author}{\bibinfo{person}{Shaibal Barua}, \bibinfo{person}{Mobyen~Uddin Ahmed}, {and} \bibinfo{person}{Shahina Begum}.} \bibinfo{year}{2020}\natexlab{}.
\newblock \showarticletitle{Towards {Intelligent} {Data} {Analytics}: {A} {Case} {Study} in {Driver} {Cognitive} {Load} {Classification}}.
\newblock \bibinfo{journal}{\emph{Brain Sciences}} \bibinfo{volume}{10}, \bibinfo{number}{8} (\bibinfo{date}{Aug.} \bibinfo{year}{2020}), \bibinfo{pages}{526}.
\newblock
\showISSN{2076-3425}
\href{https://doi.org/10.3390/brainsci10080526}{doi:\nolinkurl{10.3390/brainsci10080526}}
\newblock
\shownote{Number: 8 Publisher: Multidisciplinary Digital Publishing Institute}.


\bibitem[Bengler et~al\mbox{.}(2012)]%
        {bengler_assessment_2012}
\bibfield{author}{\bibinfo{person}{Klaus Bengler}, \bibinfo{person}{Martin Kohlmann}, {and} \bibinfo{person}{Christian Lange}.} \bibinfo{year}{2012}\natexlab{}.
\newblock \showarticletitle{Assessment of cognitive workload of in-vehicle systems using a visual peripheral and tactile detection task setting}.
\newblock \bibinfo{journal}{\emph{Work}} \bibinfo{volume}{41}, \bibinfo{number}{Supplement 1} (\bibinfo{date}{Jan.} \bibinfo{year}{2012}), \bibinfo{pages}{4919--4923}.
\newblock
\showISSN{1051-9815}
\href{https://doi.org/10.3233/WOR-2012-0786-4919}{doi:\nolinkurl{10.3233/WOR-2012-0786-4919}}
\newblock
\shownote{Publisher: IOS Press}.


\bibitem[Berry(1987)]%
        {berry_logarithmic_1987}
\bibfield{author}{\bibinfo{person}{Donald~A. Berry}.} \bibinfo{year}{1987}\natexlab{}.
\newblock \showarticletitle{Logarithmic {Transformations} in {ANOVA}}.
\newblock \bibinfo{journal}{\emph{Biometrics}} \bibinfo{volume}{43}, \bibinfo{number}{2} (\bibinfo{year}{1987}), \bibinfo{pages}{439--456}.
\newblock
\showISSN{0006-341X}
\href{https://doi.org/10.2307/2531826}{doi:\nolinkurl{10.2307/2531826}}
\newblock
\shownote{Publisher: [Wiley, International Biometric Society]}.


\bibitem[Blanca et~al\mbox{.}(2017)]%
        {blanca_non-normal_2017}
\bibfield{author}{\bibinfo{person}{María~J. Blanca}, \bibinfo{person}{Rafael Alarcón}, \bibinfo{person}{Jaume Arnau}, \bibinfo{person}{Roser Bono}, {and} \bibinfo{person}{Rebecca Bendayan}.} \bibinfo{year}{2017}\natexlab{}.
\newblock \showarticletitle{Non-normal data: {Is} {ANOVA} still a valid option?}
\newblock \bibinfo{journal}{\emph{Psicothema}} \bibinfo{volume}{29}, \bibinfo{number}{4} (\bibinfo{date}{Nov.} \bibinfo{year}{2017}), \bibinfo{pages}{552--557}.
\newblock
\showISSN{1886-144X}
\href{https://doi.org/10.7334/psicothema2016.383}{doi:\nolinkurl{10.7334/psicothema2016.383}}


\bibitem[BOX(1949)]%
        {box_general_1949}
\bibfield{author}{\bibinfo{person}{G.~E.~P. BOX}.} \bibinfo{year}{1949}\natexlab{}.
\newblock \showarticletitle{A {GENERAL} {DISTRIBUTION} {THEORY} {FOR} {A} {CLASS} {OF} {LIKELIHOOD} {CRITERIA}}.
\newblock \bibinfo{journal}{\emph{Biometrika}} \bibinfo{volume}{36}, \bibinfo{number}{3-4} (\bibinfo{date}{Dec.} \bibinfo{year}{1949}), \bibinfo{pages}{317--346}.
\newblock
\showISSN{0006-3444}
\href{https://doi.org/10.1093/biomet/36.3-4.317}{doi:\nolinkurl{10.1093/biomet/36.3-4.317}}


\bibitem[C.~Panou(2018)]%
        {c_panou_intelligent_2018}
\bibfield{author}{\bibinfo{person}{Maria C.~Panou}.} \bibinfo{year}{2018}\natexlab{}.
\newblock \showarticletitle{Intelligent personalized {ADAS} warnings}.
\newblock \bibinfo{journal}{\emph{European Transport Research Review}} \bibinfo{volume}{10}, \bibinfo{number}{2} (\bibinfo{date}{Dec.} \bibinfo{year}{2018}), \bibinfo{pages}{59}.
\newblock
\showISSN{1866-8887}
\href{https://doi.org/10.1186/s12544-018-0324-6}{doi:\nolinkurl{10.1186/s12544-018-0324-6}}


\bibitem[Caber et~al\mbox{.}(2024)]%
        {caber_driver_2024}
\bibfield{author}{\bibinfo{person}{Nermin Caber}, \bibinfo{person}{Bashar~I. Ahmad}, \bibinfo{person}{Jiaming Liang}, \bibinfo{person}{Simon Godsill}, \bibinfo{person}{Alexandra Bremers}, \bibinfo{person}{Philip Thomas}, \bibinfo{person}{David Oxtoby}, {and} \bibinfo{person}{Lee Skrypchuk}.} \bibinfo{year}{2024}\natexlab{}.
\newblock \showarticletitle{Driver {Profiling} and {Bayesian} {Workload} {Estimation} {Using} {Naturalistic} {Peripheral} {Detection} {Study} {Data}}.
\newblock \bibinfo{journal}{\emph{IEEE Transactions on Intelligent Vehicles}} \bibinfo{volume}{9}, \bibinfo{number}{1} (\bibinfo{date}{Jan.} \bibinfo{year}{2024}), \bibinfo{pages}{3047--3060}.
\newblock
\showISSN{2379-8904}
\href{https://doi.org/10.1109/TIV.2023.3313419}{doi:\nolinkurl{10.1109/TIV.2023.3313419}}


\bibitem[Carsten et~al\mbox{.}(2005)]%
        {carsten_human_2005}
\bibfield{author}{\bibinfo{person}{Oliver. Carsten}, \bibinfo{person}{Natasha. Merat}, \bibinfo{person}{Wiel Janssen}, \bibinfo{person}{Emma Johansson}, \bibinfo{person}{Mark. Fowkes}, {and} \bibinfo{person}{Karel Brookhuis}.} \bibinfo{year}{2005}\natexlab{}.
\newblock \bibinfo{booktitle}{\emph{Human {Machine} {Interaction} and the {Safety} of {Traffic} in {Europe}}}.
\newblock \bibinfo{type}{Final {Publishable} {Report}}. \bibinfo{institution}{Project and European Commission}. \bibinfo{pages}{1--63} pages.
\newblock


\bibitem[Casali and Wierwille(1983)]%
        {casali_comparison_1983}
\bibfield{author}{\bibinfo{person}{John~G. Casali} {and} \bibinfo{person}{Walter~W. Wierwille}.} \bibinfo{year}{1983}\natexlab{}.
\newblock \showarticletitle{A {Comparison} of {Rating} {Scale}, {Secondary}-{Task}, {Physiological}, and {Primary}-{Task} {Workload} {Estimation} {Techniques} in a {Simulated} {Flight} {Task} {Emphasizing} {Communications} {Load}}.
\newblock \bibinfo{journal}{\emph{Human Factors}} \bibinfo{volume}{25}, \bibinfo{number}{6} (\bibinfo{date}{Dec.} \bibinfo{year}{1983}), \bibinfo{pages}{623--641}.
\newblock
\showISSN{0018-7208}
\href{https://doi.org/10.1177/001872088302500602}{doi:\nolinkurl{10.1177/001872088302500602}}
\newblock
\shownote{Publisher: SAGE Publications Inc}.


\bibitem[Currano et~al\mbox{.}(2021)]%
        {currano_little_2021}
\bibfield{author}{\bibinfo{person}{Rebecca Currano}, \bibinfo{person}{So~Yeon Park}, \bibinfo{person}{Dylan~James Moore}, \bibinfo{person}{Kent Lyons}, {and} \bibinfo{person}{David Sirkin}.} \bibinfo{year}{2021}\natexlab{}.
\newblock \showarticletitle{Little {Road} {Driving} {HUD}: {Heads}-{Up} {Display} {Complexity} {Influences} {Drivers}’ {Perceptions} of {Automated} {Vehicles}}. In \bibinfo{booktitle}{\emph{Proceedings of the 2021 {CHI} {Conference} on {Human} {Factors} in {Computing} {Systems}}} \emph{(\bibinfo{series}{{CHI} '21})}. \bibinfo{publisher}{Association for Computing Machinery}, \bibinfo{address}{New York, NY, USA}, \bibinfo{pages}{1--15}.
\newblock
\showISBNx{978-1-4503-8096-6}
\href{https://doi.org/10.1145/3411764.3445575}{doi:\nolinkurl{10.1145/3411764.3445575}}


\bibitem[Detjen et~al\mbox{.}(2021)]%
        {detjen_how_2021}
\bibfield{author}{\bibinfo{person}{Henrik Detjen}, \bibinfo{person}{Sarah Faltaous}, \bibinfo{person}{Bastian Pfleging}, \bibinfo{person}{Stefan Geisler}, {and} \bibinfo{person}{Stefan Schneegass}.} \bibinfo{year}{2021}\natexlab{}.
\newblock \showarticletitle{How to {Increase} {Automated} {Vehicles}’ {Acceptance} through {In}-{Vehicle} {Interaction} {Design}: {A} {Review}}.
\newblock \bibinfo{journal}{\emph{International Journal of Human–Computer Interaction}} \bibinfo{volume}{37}, \bibinfo{number}{4} (\bibinfo{date}{Feb.} \bibinfo{year}{2021}), \bibinfo{pages}{308--330}.
\newblock
\showISSN{1044-7318}
\href{https://doi.org/10.1080/10447318.2020.1860517}{doi:\nolinkurl{10.1080/10447318.2020.1860517}}
\newblock
\shownote{Publisher: Taylor \& Francis \_eprint: https://doi.org/10.1080/10447318.2020.1860517}.


\bibitem[Dimitrakopoulos et~al\mbox{.}(2021)]%
        {dimitrakopoulos_chapter_2021}
\bibfield{author}{\bibinfo{person}{George Dimitrakopoulos}, \bibinfo{person}{Aggelos Tsakanikas}, {and} \bibinfo{person}{Elias Panagiotopoulos}.} \bibinfo{year}{2021}\natexlab{}.
\newblock \showarticletitle{Chapter 6 - {A} path of structural transformation for the automotive and insurance industries toward autonomous vehicles}.
\newblock In \bibinfo{booktitle}{\emph{Autonomous {Vehicles}}}, \bibfield{editor}{\bibinfo{person}{George Dimitrakopoulos}, \bibinfo{person}{Aggelos Tsakanikas}, {and} \bibinfo{person}{Elias Panagiotopoulos}} (Eds.). \bibinfo{publisher}{Elsevier}, \bibinfo{pages}{69--83}.
\newblock
\showISBNx{978-0-323-90137-6}
\href{https://doi.org/10.1016/B978-0-323-90137-6.00003-5}{doi:\nolinkurl{10.1016/B978-0-323-90137-6.00003-5}}


\bibitem[Dogan et~al\mbox{.}(2019)]%
        {dogan_effects_2019}
\bibfield{author}{\bibinfo{person}{Ebru Dogan}, \bibinfo{person}{Vincent Honnêt}, \bibinfo{person}{Stéphan Masfrand}, {and} \bibinfo{person}{Anne Guillaume}.} \bibinfo{year}{2019}\natexlab{}.
\newblock \showarticletitle{Effects of non-driving-related tasks on takeover performance in different takeover situations in conditionally automated driving}.
\newblock \bibinfo{journal}{\emph{Transportation Research Part F: Traffic Psychology and Behaviour}}  \bibinfo{volume}{62} (\bibinfo{date}{April} \bibinfo{year}{2019}), \bibinfo{pages}{494--504}.
\newblock
\showISSN{1369-8478}
\href{https://doi.org/10.1016/j.trf.2019.02.010}{doi:\nolinkurl{10.1016/j.trf.2019.02.010}}


\bibitem[Du et~al\mbox{.}(2020)]%
        {du_predicting_2020}
\bibfield{author}{\bibinfo{person}{Na Du}, \bibinfo{person}{Feng Zhou}, \bibinfo{person}{Elizabeth Pulver}, \bibinfo{person}{Dawn Tilbury}, \bibinfo{person}{Lionel~P. Robert}, \bibinfo{person}{Anuj~K. Pradhan}, {and} \bibinfo{person}{X.~Jessie Yang}.} \bibinfo{year}{2020}\natexlab{}.
\newblock \showarticletitle{Predicting {Takeover} {Performance} in {Conditionally} {Automated} {Driving}}. In \bibinfo{booktitle}{\emph{Extended {Abstracts} of the 2020 {CHI} {Conference} on {Human} {Factors} in {Computing} {Systems}}} \emph{(\bibinfo{series}{{CHI} {EA} '20})}. \bibinfo{publisher}{Association for Computing Machinery}, \bibinfo{address}{New York, NY, USA}, \bibinfo{pages}{1--8}.
\newblock
\showISBNx{978-1-4503-6819-3}
\href{https://doi.org/10.1145/3334480.3382963}{doi:\nolinkurl{10.1145/3334480.3382963}}


\bibitem[Duchowski(2018)]%
        {duchowski_gaze-based_2018}
\bibfield{author}{\bibinfo{person}{Andrew~T. Duchowski}.} \bibinfo{year}{2018}\natexlab{}.
\newblock \showarticletitle{Gaze-based interaction: {A} 30 year retrospective}.
\newblock \bibinfo{journal}{\emph{Computers \& Graphics}}  \bibinfo{volume}{73} (\bibinfo{date}{June} \bibinfo{year}{2018}), \bibinfo{pages}{59--69}.
\newblock
\showISSN{0097-8493}
\href{https://doi.org/10.1016/j.cag.2018.04.002}{doi:\nolinkurl{10.1016/j.cag.2018.04.002}}


\bibitem[Eriksson and Stanton(2017)]%
        {eriksson_takeover_2017}
\bibfield{author}{\bibinfo{person}{Alexander Eriksson} {and} \bibinfo{person}{Neville~A. Stanton}.} \bibinfo{year}{2017}\natexlab{}.
\newblock \showarticletitle{Takeover {Time} in {Highly} {Automated} {Vehicles}: {Noncritical} {Transitions} to and {From} {Manual} {Control}}.
\newblock \bibinfo{journal}{\emph{Human Factors}} \bibinfo{volume}{59}, \bibinfo{number}{4} (\bibinfo{date}{June} \bibinfo{year}{2017}), \bibinfo{pages}{689--705}.
\newblock
\showISSN{0018-7208}
\href{https://doi.org/10.1177/0018720816685832}{doi:\nolinkurl{10.1177/0018720816685832}}
\newblock
\shownote{Publisher: SAGE Publications Inc}.


\bibitem[Faul et~al\mbox{.}(2009)]%
        {faul_statistical_2009}
\bibfield{author}{\bibinfo{person}{Franz Faul}, \bibinfo{person}{Edgar Erdfelder}, \bibinfo{person}{Axel Buchner}, {and} \bibinfo{person}{Albert-Georg Lang}.} \bibinfo{year}{2009}\natexlab{}.
\newblock \showarticletitle{Statistical power analyses using {G}*{Power} 3.1: {Tests} for correlation and regression analyses}.
\newblock \bibinfo{journal}{\emph{Behavior Research Methods}} \bibinfo{volume}{41}, \bibinfo{number}{4} (\bibinfo{date}{Nov.} \bibinfo{year}{2009}), \bibinfo{pages}{1149--1160}.
\newblock
\showISSN{1554-3528}
\href{https://doi.org/10.3758/BRM.41.4.1149}{doi:\nolinkurl{10.3758/BRM.41.4.1149}}


\bibitem[Fischer et~al\mbox{.}(2018)]%
        {fischer_rt-gene_2018}
\bibfield{author}{\bibinfo{person}{Tobias Fischer}, \bibinfo{person}{Hyung~Jin Chang}, {and} \bibinfo{person}{Yiannis Demiris}.} \bibinfo{year}{2018}\natexlab{}.
\newblock \showarticletitle{{RT}-{GENE}: {Real}-{Time} {Eye} {Gaze} {Estimation} in {Natural} {Environments}}. In \bibinfo{booktitle}{\emph{Proceedings of the {European} {Conference} on {Computer} {Vision} ({ECCV})}}. \bibinfo{pages}{334--352}.
\newblock
\urldef\tempurl%
\url{https://openaccess.thecvf.com/content_ECCV_2018/html/Tobias_Fischer_RT-GENE_Real-Time_Eye_ECCV_2018_paper.html}
\showURL{%
\tempurl}


\bibitem[Gerber et~al\mbox{.}(2024)]%
        {gerber_eye_2024}
\bibfield{author}{\bibinfo{person}{Michael~A. Gerber}, \bibinfo{person}{Ronald Schroeter}, \bibinfo{person}{Daniel Johnson}, \bibinfo{person}{Christian~P. Janssen}, \bibinfo{person}{Andry Rakotonirainy}, \bibinfo{person}{Jonny Kuo}, {and} \bibinfo{person}{Mike Lenné}.} \bibinfo{year}{2024}\natexlab{}.
\newblock \showarticletitle{An {Eye} {Gaze} {Heatmap} {Analysis} of {Uncertainty} {Head}-{Up} {Display} {Designs} for {Conditional} {Automated} {Driving}}. In \bibinfo{booktitle}{\emph{Proceedings of the {CHI} {Conference} on {Human} {Factors} in {Computing} {Systems}}} \emph{(\bibinfo{series}{{CHI} '24})}. \bibinfo{publisher}{Association for Computing Machinery}, \bibinfo{address}{New York, NY, USA}, \bibinfo{pages}{1--16}.
\newblock
\showISBNx{9798400703300}
\href{https://doi.org/10.1145/3613904.3642219}{doi:\nolinkurl{10.1145/3613904.3642219}}


\bibitem[Gevins et~al\mbox{.}(1997)]%
        {Genvis_workload_1991}
\bibfield{author}{\bibinfo{person}{A Gevins}, \bibinfo{person}{M~E Smith}, \bibinfo{person}{L McEvoy}, {and} \bibinfo{person}{D Yu}.} \bibinfo{year}{1997}\natexlab{}.
\newblock \showarticletitle{High-resolution {EEG} mapping of cortical activation related to working memory: effects of task difficulty, type of processing, and practice.}
\newblock \bibinfo{journal}{\emph{Cerebral Cortex}} \bibinfo{volume}{7}, \bibinfo{number}{4} (\bibinfo{date}{June} \bibinfo{year}{1997}), \bibinfo{pages}{374--385}.
\newblock
\showISSN{1047-3211}
\href{https://doi.org/10.1093/cercor/7.4.374}{doi:\nolinkurl{10.1093/cercor/7.4.374}}


\bibitem[Glass et~al\mbox{.}(1972)]%
        {glass_consequences_1972}
\bibfield{author}{\bibinfo{person}{Gene~V Glass}, \bibinfo{person}{Percy~D. Peckham}, {and} \bibinfo{person}{James~R. Sanders}.} \bibinfo{year}{1972}\natexlab{}.
\newblock \showarticletitle{Consequences of {Failure} to {Meet} {Assumptions} {Underlying} the {Fixed} {Effects} {Analyses} of {Variance} and {Covariance}}.
\newblock \bibinfo{journal}{\emph{Review of Educational Research}} \bibinfo{volume}{42}, \bibinfo{number}{3} (\bibinfo{date}{Sept.} \bibinfo{year}{1972}), \bibinfo{pages}{237--288}.
\newblock
\showISSN{0034-6543}
\href{https://doi.org/10.3102/00346543042003237}{doi:\nolinkurl{10.3102/00346543042003237}}
\newblock
\shownote{Publisher: American Educational Research Association}.


\bibitem[Gold et~al\mbox{.}(2013)]%
        {gold_take_2013}
\bibfield{author}{\bibinfo{person}{Christian Gold}, \bibinfo{person}{Daniel Damböck}, \bibinfo{person}{Lutz Lorenz}, {and} \bibinfo{person}{Klaus Bengler}.} \bibinfo{year}{2013}\natexlab{}.
\newblock \showarticletitle{“{Take} over!” {How} long does it take to get the driver back into the loop?}
\newblock \bibinfo{journal}{\emph{Proceedings of the Human Factors and Ergonomics Society Annual Meeting}} \bibinfo{volume}{57}, \bibinfo{number}{1} (\bibinfo{date}{Sept.} \bibinfo{year}{2013}), \bibinfo{pages}{1938--1942}.
\newblock
\showISSN{1071-1813}
\href{https://doi.org/10.1177/1541931213571433}{doi:\nolinkurl{10.1177/1541931213571433}}
\newblock
\shownote{Publisher: SAGE Publications Inc}.


\bibitem[Gold et~al\mbox{.}(2016)]%
        {gold_taking_2016}
\bibfield{author}{\bibinfo{person}{Christian Gold}, \bibinfo{person}{Moritz Körber}, \bibinfo{person}{David Lechner}, {and} \bibinfo{person}{Klaus Bengler}.} \bibinfo{year}{2016}\natexlab{}.
\newblock \showarticletitle{Taking {Over} {Control} {From} {Highly} {Automated} {Vehicles} in {Complex} {Traffic} {Situations}: {The} {Role} of {Traffic} {Density}}.
\newblock \bibinfo{journal}{\emph{Human Factors}} \bibinfo{volume}{58}, \bibinfo{number}{4} (\bibinfo{date}{June} \bibinfo{year}{2016}), \bibinfo{pages}{642--652}.
\newblock
\showISSN{0018-7208}
\href{https://doi.org/10.1177/0018720816634226}{doi:\nolinkurl{10.1177/0018720816634226}}
\newblock
\shownote{Publisher: SAGE Publications Inc}.


\bibitem[Gomaa et~al\mbox{.}(2022)]%
        {gomaa_whats_2022}
\bibfield{author}{\bibinfo{person}{Amr Gomaa}, \bibinfo{person}{Alexandra Alles}, \bibinfo{person}{Elena Meiser}, \bibinfo{person}{Lydia~Helene Rupp}, \bibinfo{person}{Marco Molz}, {and} \bibinfo{person}{Guillermo Reyes}.} \bibinfo{year}{2022}\natexlab{}.
\newblock \showarticletitle{What’s on your mind? {A} {Mental} and {Perceptual} {Load} {Estimation} {Framework} towards {Adaptive} {In}-vehicle {Interaction} while {Driving}}. In \bibinfo{booktitle}{\emph{Proceedings of the 14th {International} {Conference} on {Automotive} {User} {Interfaces} and {Interactive} {Vehicular} {Applications}}} \emph{(\bibinfo{series}{{AutomotiveUI} '22})}. \bibinfo{publisher}{Association for Computing Machinery}, \bibinfo{address}{New York, NY, USA}, \bibinfo{pages}{215--225}.
\newblock
\showISBNx{978-1-4503-9415-4}
\href{https://doi.org/10.1145/3543174.3546840}{doi:\nolinkurl{10.1145/3543174.3546840}}


\bibitem[Greenhouse and Geisser(1959)]%
        {greenhouse_methods_1959}
\bibfield{author}{\bibinfo{person}{Samuel~W. Greenhouse} {and} \bibinfo{person}{Seymour Geisser}.} \bibinfo{year}{1959}\natexlab{}.
\newblock \showarticletitle{On methods in the analysis of profile data}.
\newblock \bibinfo{journal}{\emph{Psychometrika}} \bibinfo{volume}{24}, \bibinfo{number}{2} (\bibinfo{date}{June} \bibinfo{year}{1959}), \bibinfo{pages}{95--112}.
\newblock
\showISSN{1860-0980}
\href{https://doi.org/10.1007/BF02289823}{doi:\nolinkurl{10.1007/BF02289823}}


\bibitem[Haeuslschmid et~al\mbox{.}(2017)]%
        {haeuslschmid_recognition_2017}
\bibfield{author}{\bibinfo{person}{Renate Haeuslschmid}, \bibinfo{person}{Susanne Forster}, \bibinfo{person}{Katharina Vierheilig}, \bibinfo{person}{Daniel Buschek}, {and} \bibinfo{person}{Andreas Butz}.} \bibinfo{year}{2017}\natexlab{}.
\newblock \showarticletitle{Recognition of {Text} and {Shapes} on a {Large}-{Sized} {Head}-{Up} {Display}}. In \bibinfo{booktitle}{\emph{Proceedings of the 2017 {Conference} on {Designing} {Interactive} {Systems}}} \emph{(\bibinfo{series}{{DIS} '17})}. \bibinfo{publisher}{Association for Computing Machinery}, \bibinfo{address}{New York, NY, USA}, \bibinfo{pages}{821--831}.
\newblock
\showISBNx{978-1-4503-4922-2}
\href{https://doi.org/10.1145/3064663.3064736}{doi:\nolinkurl{10.1145/3064663.3064736}}


\bibitem[Hart and Staveland(1988)]%
        {hart_development_1988}
\bibfield{author}{\bibinfo{person}{Sandra~G. Hart} {and} \bibinfo{person}{Lowell~E. Staveland}.} \bibinfo{year}{1988}\natexlab{}.
\newblock \showarticletitle{Development of {NASA}-{TLX} ({Task} {Load} {Index}): {Results} of {Empirical} and {Theoretical} {Research}}.
\newblock In \bibinfo{booktitle}{\emph{Advances in {Psychology}}}, \bibfield{editor}{\bibinfo{person}{Peter~A. Hancock} {and} \bibinfo{person}{Najmedin Meshkati}} (Eds.). \bibinfo{series}{Human {Mental} {Workload}}, Vol.~\bibinfo{volume}{52}. \bibinfo{publisher}{North-Holland}, \bibinfo{pages}{139--183}.
\newblock
\href{https://doi.org/10.1016/S0166-4115(08)62386-9}{doi:\nolinkurl{10.1016/S0166-4115(08)62386-9}}


\bibitem[Harwell(1992)]%
        {harwell_summarizing_1992}
\bibfield{author}{\bibinfo{person}{Michael~R. Harwell}.} \bibinfo{year}{1992}\natexlab{}.
\newblock \showarticletitle{Summarizing {Monte} {Carlo} {Results} in {Methodological} {Research}}.
\newblock \bibinfo{journal}{\emph{Journal of Educational Statistics}} \bibinfo{volume}{17}, \bibinfo{number}{4} (\bibinfo{date}{Dec.} \bibinfo{year}{1992}), \bibinfo{pages}{297--313}.
\newblock
\showISSN{0362-9791}
\href{https://doi.org/10.3102/10769986017004297}{doi:\nolinkurl{10.3102/10769986017004297}}
\newblock
\shownote{Publisher: American Educational Research Association}.


\bibitem[Hasenjäger et~al\mbox{.}(2020)]%
        {hasenjager_survey_2020}
\bibfield{author}{\bibinfo{person}{Martina Hasenjäger}, \bibinfo{person}{Martin Heckmann}, {and} \bibinfo{person}{Heiko Wersing}.} \bibinfo{year}{2020}\natexlab{}.
\newblock \showarticletitle{A {Survey} of {Personalization} for {Advanced} {Driver} {Assistance} {Systems}}.
\newblock \bibinfo{journal}{\emph{IEEE Transactions on Intelligent Vehicles}} \bibinfo{volume}{5}, \bibinfo{number}{2} (\bibinfo{date}{June} \bibinfo{year}{2020}), \bibinfo{pages}{335--344}.
\newblock
\showISSN{2379-8904}
\href{https://doi.org/10.1109/TIV.2019.2955910}{doi:\nolinkurl{10.1109/TIV.2019.2955910}}
\newblock
\shownote{Conference Name: IEEE Transactions on Intelligent Vehicles}.


\bibitem[Hasenjäger and Wersing(2017)]%
        {hasenjager_personalization_2017}
\bibfield{author}{\bibinfo{person}{Martina Hasenjäger} {and} \bibinfo{person}{Heiko Wersing}.} \bibinfo{year}{2017}\natexlab{}.
\newblock \showarticletitle{Personalization in advanced driver assistance systems and autonomous vehicles: {A} review}. In \bibinfo{booktitle}{\emph{2017 {IEEE} 20th {International} {Conference} on {Intelligent} {Transportation} {Systems} ({ITSC})}}. \bibinfo{pages}{1--7}.
\newblock
\href{https://doi.org/10.1109/ITSC.2017.8317803}{doi:\nolinkurl{10.1109/ITSC.2017.8317803}}
\newblock
\shownote{ISSN: 2153-0017}.


\bibitem[Herzog et~al\mbox{.}(2019)]%
        {herzog_understanding_2019}
\bibfield{author}{\bibinfo{person}{Michael~H. Herzog}, \bibinfo{person}{Gregory Francis}, {and} \bibinfo{person}{Aaron Clarke}.} \bibinfo{year}{2019}\natexlab{}.
\newblock \bibinfo{booktitle}{\emph{Understanding {Statistics} and {Experimental} {Design}: {How} to {Not} {Lie} with {Statistics}}}.
\newblock \bibinfo{publisher}{Springer International Publishing}, \bibinfo{address}{Cham}.
\newblock
\showISBNx{978-3-030-03498-6 978-3-030-03499-3}
\href{https://doi.org/10.1007/978-3-030-03499-3}{doi:\nolinkurl{10.1007/978-3-030-03499-3}}


\bibitem[Hofbauer et~al\mbox{.}(2020)]%
        {hofbauer_measuring_2020}
\bibfield{author}{\bibinfo{person}{Markus Hofbauer}, \bibinfo{person}{Christopher~B. Kuhn}, \bibinfo{person}{Lukas Püttner}, \bibinfo{person}{Goran Petrovic}, {and} \bibinfo{person}{Eckehard Steinbach}.} \bibinfo{year}{2020}\natexlab{}.
\newblock \showarticletitle{Measuring {Driver} {Situation} {Awareness} {Using} {Region}-of-{Interest} {Prediction} and {Eye} {Tracking}}. In \bibinfo{booktitle}{\emph{2020 {IEEE} {International} {Symposium} on {Multimedia} ({ISM})}}. \bibinfo{pages}{91--95}.
\newblock
\href{https://doi.org/10.1109/ISM.2020.00022}{doi:\nolinkurl{10.1109/ISM.2020.00022}}


\bibitem[Holm et~al\mbox{.}(2009)]%
        {holm_estimating_2009}
\bibfield{author}{\bibinfo{person}{Anu Holm}, \bibinfo{person}{Kristian Lukander}, \bibinfo{person}{Jussi Korpela}, \bibinfo{person}{Mikael Sallinen}, {and} \bibinfo{person}{Kiti M.~I. Müller}.} \bibinfo{year}{2009}\natexlab{}.
\newblock \showarticletitle{Estimating {Brain} {Load} from the {EEG}}.
\newblock \bibinfo{journal}{\emph{The Scientific World Journal}} \bibinfo{volume}{9}, \bibinfo{number}{1} (\bibinfo{year}{2009}), \bibinfo{pages}{973791}.
\newblock
\showISSN{1537-744X}
\href{https://doi.org/10.1100/tsw.2009.83}{doi:\nolinkurl{10.1100/tsw.2009.83}}
\newblock
\shownote{\_eprint: https://onlinelibrary.wiley.com/doi/pdf/10.1100/tsw.2009.83}.


\bibitem[Holmqvist et~al\mbox{.}(2011)]%
        {Holmqvist_Nystrom_Andersson_Dewhurst_Jarodzka_Weijer_2011}
\bibfield{author}{\bibinfo{person}{Kenneth Holmqvist}, \bibinfo{person}{Marcus Nystrom}, \bibinfo{person}{Richard Andersson}, \bibinfo{person}{Richard Dewhurst}, \bibinfo{person}{Halszka Jarodzka}, \bibinfo{person}{Weijer}, {and} \bibinfo{person}{Joost van de}.} \bibinfo{year}{2011}\natexlab{}.
\newblock \bibinfo{booktitle}{\emph{Eye Tracking: A comprehensive guide to methods and measures}}.
\newblock \bibinfo{publisher}{Oxford University Press}, \bibinfo{address}{Oxford, New York}.
\newblock
\showISBNx{978-0-19-969708-3}


\bibitem[Horrey and Wickens(2004)]%
        {horrey_driving_2004}
\bibfield{author}{\bibinfo{person}{William~J. Horrey} {and} \bibinfo{person}{Christopher~D. Wickens}.} \bibinfo{year}{2004}\natexlab{}.
\newblock \showarticletitle{Driving and {Side} {Task} {Performance}: {The} {Effects} of {Display} {Clutter}, {Separation}, and {Modality}}.
\newblock \bibinfo{journal}{\emph{Human Factors}} \bibinfo{volume}{46}, \bibinfo{number}{4} (\bibinfo{date}{Dec.} \bibinfo{year}{2004}), \bibinfo{pages}{611--624}.
\newblock
\showISSN{0018-7208}
\href{https://doi.org/10.1518/hfes.46.4.611.56805}{doi:\nolinkurl{10.1518/hfes.46.4.611.56805}}
\newblock
\shownote{Publisher: SAGE Publications Inc}.


\bibitem[Horrey and Wickens(2007)]%
        {horrey_-vehicle_2007}
\bibfield{author}{\bibinfo{person}{William~J. Horrey} {and} \bibinfo{person}{Christopher~D. Wickens}.} \bibinfo{year}{2007}\natexlab{}.
\newblock \showarticletitle{In-{Vehicle} {Glance} {Duration}: {Distributions}, {Tails}, and {Model} of {Crash} {Risk}}.
\newblock \bibinfo{journal}{\emph{Transportation Research Record}} \bibinfo{volume}{2018}, \bibinfo{number}{1} (\bibinfo{date}{Jan.} \bibinfo{year}{2007}), \bibinfo{pages}{22--28}.
\newblock
\showISSN{0361-1981}
\href{https://doi.org/10.3141/2018-04}{doi:\nolinkurl{10.3141/2018-04}}
\newblock
\shownote{Publisher: SAGE Publications Inc}.


\bibitem[Horrey et~al\mbox{.}(2003)]%
        {horrey_effects_2003}
\bibfield{author}{\bibinfo{person}{William~J. Horrey}, \bibinfo{person}{Christopher~D. Wickens}, {and} \bibinfo{person}{Amy~L. Alexander}.} \bibinfo{year}{2003}\natexlab{}.
\newblock \showarticletitle{The {Effects} of {Head}-{Up} {Display} {Clutter} and {In}-{Vehicle} {Display} {Separation} on {Concurrent} {Driving} {Performance}}.
\newblock \bibinfo{journal}{\emph{Proceedings of the Human Factors and Ergonomics Society Annual Meeting}} \bibinfo{volume}{47}, \bibinfo{number}{16} (\bibinfo{date}{Oct.} \bibinfo{year}{2003}), \bibinfo{pages}{1880--1884}.
\newblock
\showISSN{1071-1813}
\href{https://doi.org/10.1177/154193120304701610}{doi:\nolinkurl{10.1177/154193120304701610}}
\newblock
\shownote{Publisher: SAGE Publications Inc}.


\bibitem[Kartali et~al\mbox{.}(2019)]%
        {Kartali_workload_2019}
\bibfield{author}{\bibinfo{person}{Aneta Kartali}, \bibinfo{person}{Milica~M. Janković}, \bibinfo{person}{Ivan Gligorijević}, \bibinfo{person}{Pavle Mijović}, \bibinfo{person}{Bogdan Mijović}, {and} \bibinfo{person}{Maria~Chiara Leva}.} \bibinfo{year}{2019}\natexlab{}.
\newblock \showarticletitle{Real-{Time} {Mental} {Workload} {Estimation} {Using} {EEG}}. In \bibinfo{booktitle}{\emph{Human {Mental} {Workload}: {Models} and {Applications}}}, \bibfield{editor}{\bibinfo{person}{Luca Longo} {and} \bibinfo{person}{Maria~Chiara Leva}} (Eds.). \bibinfo{publisher}{Springer International Publishing}, \bibinfo{address}{Cham}, \bibinfo{pages}{20--34}.
\newblock
\showISBNx{978-3-030-32423-0}
\href{https://doi.org/10.1007/978-3-030-32423-0_2}{doi:\nolinkurl{10.1007/978-3-030-32423-0_2}}


\bibitem[Khare and Raghavendra(2024)]%
        {khare_exploring_2024}
\bibfield{author}{\bibinfo{person}{Mayank~Deep Khare} {and} \bibinfo{person}{R. Raghavendra}.} \bibinfo{year}{2024}\natexlab{}.
\newblock \showarticletitle{Exploring {Sensor} {Technologies} and {Automation} {Levels} in {Autonomous} {Vehicles}}. In \bibinfo{booktitle}{\emph{Machine {Intelligence} for {Research} and {Innovations}}}, \bibfield{editor}{\bibinfo{person}{Om~Prakash Verma}, \bibinfo{person}{Lipo Wang}, \bibinfo{person}{Rajesh Kumar}, {and} \bibinfo{person}{Anupam Yadav}} (Eds.). \bibinfo{publisher}{Springer Nature}, \bibinfo{address}{Singapore}, \bibinfo{pages}{293--304}.
\newblock
\showISBNx{978-981-9981-35-9}
\href{https://doi.org/10.1007/978-981-99-8135-9_26}{doi:\nolinkurl{10.1007/978-981-99-8135-9_26}}


\bibitem[Kim et~al\mbox{.}(2013)]%
        {kim_exploring_2013}
\bibfield{author}{\bibinfo{person}{Hyungil Kim}, \bibinfo{person}{Xuefang Wu}, \bibinfo{person}{Joseph~L. Gabbard}, {and} \bibinfo{person}{Nicholas~F. Polys}.} \bibinfo{year}{2013}\natexlab{}.
\newblock \showarticletitle{Exploring head-up augmented reality interfaces for crash warning systems}. In \bibinfo{booktitle}{\emph{Proceedings of the 5th {International} {Conference} on {Automotive} {User} {Interfaces} and {Interactive} {Vehicular} {Applications}}} \emph{(\bibinfo{series}{{AutomotiveUI} '13})}. \bibinfo{publisher}{Association for Computing Machinery}, \bibinfo{address}{New York, NY, USA}, \bibinfo{pages}{224--227}.
\newblock
\showISBNx{978-1-4503-2478-6}
\href{https://doi.org/10.1145/2516540.2516566}{doi:\nolinkurl{10.1145/2516540.2516566}}


\bibitem[Kim et~al\mbox{.}(2017)]%
        {kim_are_2017}
\bibfield{author}{\bibinfo{person}{Naeun Kim}, \bibinfo{person}{Kwangmin Jeong}, \bibinfo{person}{Minyoung Yang}, \bibinfo{person}{Yejeon Oh}, {and} \bibinfo{person}{Jinwoo Kim}.} \bibinfo{year}{2017}\natexlab{}.
\newblock \showarticletitle{"{Are} {You} {Ready} to {Take}-over?": {An} {Exploratory} {Study} on {Visual} {Assistance} to {Enhance} {Driver} {Vigilance}}. In \bibinfo{booktitle}{\emph{Proceedings of the 2017 {CHI} {Conference} {Extended} {Abstracts} on {Human} {Factors} in {Computing} {Systems}}} \emph{(\bibinfo{series}{{CHI} {EA} '17})}. \bibinfo{publisher}{Association for Computing Machinery}, \bibinfo{address}{New York, NY, USA}, \bibinfo{pages}{1771--1778}.
\newblock
\showISBNx{978-1-4503-4656-6}
\href{https://doi.org/10.1145/3027063.3053155}{doi:\nolinkurl{10.1145/3027063.3053155}}


\bibitem[Kim et~al\mbox{.}(2022)]%
        {kim_understanding_2022}
\bibfield{author}{\bibinfo{person}{Young~Woo Kim}, \bibinfo{person}{Da~Yeong Kim}, {and} \bibinfo{person}{Sol~Hee Yoon}.} \bibinfo{year}{2022}\natexlab{}.
\newblock \showarticletitle{Understanding {Driver}'s {Situation} {Awareness} in {Highly} {Automated} {Driving}}. In \bibinfo{booktitle}{\emph{Adjunct {Proceedings} of the 14th {International} {Conference} on {Automotive} {User} {Interfaces} and {Interactive} {Vehicular} {Applications}}} \emph{(\bibinfo{series}{{AutomotiveUI} '22})}. \bibinfo{publisher}{Association for Computing Machinery}, \bibinfo{address}{New York, NY, USA}, \bibinfo{pages}{132--136}.
\newblock
\showISBNx{978-1-4503-9428-4}
\href{https://doi.org/10.1145/3544999.3552321}{doi:\nolinkurl{10.1145/3544999.3552321}}


\bibitem[Kotseruba and Tsotsos(2022)]%
        {kotseruba_attention_2022}
\bibfield{author}{\bibinfo{person}{Iuliia Kotseruba} {and} \bibinfo{person}{John~K. Tsotsos}.} \bibinfo{year}{2022}\natexlab{}.
\newblock \showarticletitle{Attention for {Vision}-{Based} {Assistive} and {Automated} {Driving}: {A} {Review} of {Algorithms} and {Datasets}}.
\newblock \bibinfo{journal}{\emph{IEEE Transactions on Intelligent Transportation Systems}} \bibinfo{volume}{23}, \bibinfo{number}{11} (\bibinfo{date}{Nov.} \bibinfo{year}{2022}), \bibinfo{pages}{19907--19928}.
\newblock
\showISSN{1558-0016}
\href{https://doi.org/10.1109/TITS.2022.3186613}{doi:\nolinkurl{10.1109/TITS.2022.3186613}}
\newblock
\shownote{Conference Name: IEEE Transactions on Intelligent Transportation Systems}.


\bibitem[Kozłowski(2016)]%
        {kozlowski_assessment_2016}
\bibfield{author}{\bibinfo{person}{Maciej Kozłowski}.} \bibinfo{year}{2016}\natexlab{}.
\newblock \showarticletitle{Assessment of safety and ride quality based on comparative studies of a new type of universal steering wheel in {3D} simulators}.
\newblock \bibinfo{journal}{\emph{Eksploatacja i Niezawodność}} \bibinfo{volume}{18}, \bibinfo{number}{4} (\bibinfo{year}{2016}), \bibinfo{pages}{481--487}.
\newblock
\showISSN{1507-2711}
\href{https://doi.org/10.17531/ein2016.4.1}{doi:\nolinkurl{10.17531/ein2016.4.1}}


\bibitem[Kutafina et~al\mbox{.}(2021)]%
        {kutafina_tracking_2021}
\bibfield{author}{\bibinfo{person}{Ekaterina Kutafina}, \bibinfo{person}{Anne Heiligers}, \bibinfo{person}{Radomir Popovic}, \bibinfo{person}{Alexander Brenner}, \bibinfo{person}{Bernd Hankammer}, \bibinfo{person}{Stephan~M. Jonas}, \bibinfo{person}{Klaus Mathiak}, {and} \bibinfo{person}{Jana Zweerings}.} \bibinfo{year}{2021}\natexlab{}.
\newblock \showarticletitle{Tracking of {Mental} {Workload} with a {Mobile} {EEG} {Sensor}}.
\newblock \bibinfo{journal}{\emph{Sensors}} \bibinfo{volume}{21}, \bibinfo{number}{15} (\bibinfo{date}{Jan.} \bibinfo{year}{2021}), \bibinfo{pages}{5205}.
\newblock
\showISSN{1424-8220}
\href{https://doi.org/10.3390/s21155205}{doi:\nolinkurl{10.3390/s21155205}}
\newblock
\shownote{Number: 15 Publisher: Multidisciplinary Digital Publishing Institute}.


\bibitem[Lavie(2005)]%
        {lavie_distracted_2005}
\bibfield{author}{\bibinfo{person}{Nilli Lavie}.} \bibinfo{year}{2005}\natexlab{}.
\newblock \showarticletitle{Distracted and confused?: {Selective} attention under load}.
\newblock \bibinfo{journal}{\emph{Trends in Cognitive Sciences}} \bibinfo{volume}{9}, \bibinfo{number}{2} (\bibinfo{date}{Feb.} \bibinfo{year}{2005}), \bibinfo{pages}{75--82}.
\newblock
\showISSN{1364-6613}
\href{https://doi.org/10.1016/j.tics.2004.12.004}{doi:\nolinkurl{10.1016/j.tics.2004.12.004}}


\bibitem[Levene(1960)]%
        {levene_robust_1960}
\bibfield{author}{\bibinfo{person}{Howard Levene}.} \bibinfo{year}{1960}\natexlab{}.
\newblock \showarticletitle{Robust {Tests} for {Equality} of {Variances}}.
\newblock In \bibinfo{booktitle}{\emph{Contributions to {Probability} and {Statistics}: {Essays} in {Honor} of {Harold} {Hotelling}}}, \bibfield{editor}{\bibinfo{person}{Ingram Olkin}, \bibinfo{person}{Sudhist~G. Ghurye}, \bibinfo{person}{Wassily Hoeffding}, \bibinfo{person}{William~G. Madow}, {and} \bibinfo{person}{Henry~B. Mann}} (Eds.). \bibinfo{publisher}{Stanford University Press}, \bibinfo{address}{Palo Alto}, \bibinfo{pages}{278--292}.
\newblock


\bibitem[Li et~al\mbox{.}(2020)]%
        {li_effects_2020}
\bibfield{author}{\bibinfo{person}{Xiaomeng Li}, \bibinfo{person}{Ronald Schroeter}, \bibinfo{person}{Andry Rakotonirainy}, \bibinfo{person}{Jonny Kuo}, {and} \bibinfo{person}{Michael~G. Lenné}.} \bibinfo{year}{2020}\natexlab{}.
\newblock \showarticletitle{Effects of different non-driving-related-task display modes on drivers’ eye-movement patterns during take-over in an automated vehicle}.
\newblock \bibinfo{journal}{\emph{Transportation Research Part F: Traffic Psychology and Behaviour}}  \bibinfo{volume}{70} (\bibinfo{date}{April} \bibinfo{year}{2020}), \bibinfo{pages}{135--148}.
\newblock
\showISSN{1369-8478}
\href{https://doi.org/10.1016/j.trf.2020.03.001}{doi:\nolinkurl{10.1016/j.trf.2020.03.001}}


\bibitem[Lilis et~al\mbox{.}(2017)]%
        {lilis_personalizing_2017}
\bibfield{author}{\bibinfo{person}{Yannis Lilis}, \bibinfo{person}{Emmanouil Zidianakis}, \bibinfo{person}{Nikolaos Partarakis}, \bibinfo{person}{Margherita Antona}, {and} \bibinfo{person}{Constantine Stephanidis}.} \bibinfo{year}{2017}\natexlab{}.
\newblock \showarticletitle{Personalizing {HMI} {Elements} in {ADAS} {Using} {Ontology} {Meta}-{Models} and {Rule} {Based} {Reasoning}}. In \bibinfo{booktitle}{\emph{Universal {Access} in {Human}–{Computer} {Interaction}. {Design} and {Development} {Approaches} and {Methods}}}, \bibfield{editor}{\bibinfo{person}{Margherita Antona} {and} \bibinfo{person}{Constantine Stephanidis}} (Eds.). \bibinfo{publisher}{Springer International Publishing}, \bibinfo{address}{Cham}, \bibinfo{pages}{383--401}.
\newblock
\showISBNx{978-3-319-58706-6}
\href{https://doi.org/10.1007/978-3-319-58706-6_31}{doi:\nolinkurl{10.1007/978-3-319-58706-6_31}}


\bibitem[Liu and Wen(2004)]%
        {liu_comparison_2004}
\bibfield{author}{\bibinfo{person}{Yung-Ching Liu} {and} \bibinfo{person}{Ming-Hui Wen}.} \bibinfo{year}{2004}\natexlab{}.
\newblock \showarticletitle{Comparison of head-up display ({HUD}) vs. head-down display ({HDD}): driving performance of commercial vehicle operators in {Taiwan}}.
\newblock \bibinfo{journal}{\emph{International Journal of Human-Computer Studies}} \bibinfo{volume}{61}, \bibinfo{number}{5} (\bibinfo{date}{Nov.} \bibinfo{year}{2004}), \bibinfo{pages}{679--697}.
\newblock
\showISSN{1071-5819}
\href{https://doi.org/10.1016/j.ijhcs.2004.06.002}{doi:\nolinkurl{10.1016/j.ijhcs.2004.06.002}}


\bibitem[Lloyd(1982)]%
        {lloyd_least_1982}
\bibfield{author}{\bibinfo{person}{S. Lloyd}.} \bibinfo{year}{1982}\natexlab{}.
\newblock \showarticletitle{Least squares quantization in {PCM}}.
\newblock \bibinfo{journal}{\emph{IEEE Transactions on Information Theory}} \bibinfo{volume}{28}, \bibinfo{number}{2} (\bibinfo{date}{March} \bibinfo{year}{1982}), \bibinfo{pages}{129--137}.
\newblock
\showISSN{1557-9654}
\href{https://doi.org/10.1109/TIT.1982.1056489}{doi:\nolinkurl{10.1109/TIT.1982.1056489}}


\bibitem[Lohani et~al\mbox{.}(2019)]%
        {lohani_review_2019}
\bibfield{author}{\bibinfo{person}{Monika Lohani}, \bibinfo{person}{Brennan~R. Payne}, {and} \bibinfo{person}{David~L. Strayer}.} \bibinfo{year}{2019}\natexlab{}.
\newblock \showarticletitle{A {Review} of {Psychophysiological} {Measures} to {Assess} {Cognitive} {States} in {Real}-{World} {Driving}}.
\newblock \bibinfo{journal}{\emph{Frontiers in Human Neuroscience}}  \bibinfo{volume}{13} (\bibinfo{date}{March} \bibinfo{year}{2019}).
\newblock
\showISSN{1662-5161}
\href{https://doi.org/10.3389/fnhum.2019.00057}{doi:\nolinkurl{10.3389/fnhum.2019.00057}}
\newblock
\shownote{Publisher: Frontiers}.


\bibitem[Lumley et~al\mbox{.}(2002)]%
        {lumley_importance_2002}
\bibfield{author}{\bibinfo{person}{Thomas Lumley}, \bibinfo{person}{Paula Diehr}, \bibinfo{person}{Scott Emerson}, {and} \bibinfo{person}{Lu Chen}.} \bibinfo{year}{2002}\natexlab{}.
\newblock \showarticletitle{The {Importance} of the {Normality} {Assumption} in {Large} {Public} {Health} {Data} {Sets}}.
\newblock \bibinfo{journal}{\emph{Annual Review of Public Health}} \bibinfo{volume}{23}, \bibinfo{number}{Volume 23, 2002} (\bibinfo{year}{2002}), \bibinfo{pages}{151--169}.
\newblock
\showISSN{1545-2093}
\href{https://doi.org/10.1146/annurev.publhealth.23.100901.140546}{doi:\nolinkurl{10.1146/annurev.publhealth.23.100901.140546}}
\newblock
\shownote{Publisher: Annual Reviews Type: Journal Article}.


\bibitem[MacQueen(1967)]%
        {macqueen_methods_1967}
\bibfield{author}{\bibinfo{person}{J.~B. MacQueen}.} \bibinfo{year}{1967}\natexlab{}.
\newblock \showarticletitle{Some {Methods} for {Classification} and {Analysis} of {MultiVariate} {Observations}}. In \bibinfo{booktitle}{\emph{Proc. of the fifth {Berkeley} {Symposium} on {Mathematical} {Statistics} and {Probability}}}, \bibfield{editor}{\bibinfo{person}{L.~M.~Le Cam} {and} \bibinfo{person}{J.~Neyman}} (Eds.), Vol.~\bibinfo{volume}{1}. \bibinfo{publisher}{University of California Press}, \bibinfo{pages}{281--297}.
\newblock


\bibitem[Mahalanobis(1936)]%
        {mahalanobis_generalised_1936}
\bibfield{author}{\bibinfo{person}{P.C. Mahalanobis}.} \bibinfo{year}{1936}\natexlab{}.
\newblock \showarticletitle{On the {Generalised} {Distance} in {Statistics}}.
\newblock \bibinfo{journal}{\emph{Proceedings of the National Academy of Sciences of India}}  \bibinfo{volume}{2} (\bibinfo{year}{1936}), \bibinfo{pages}{49--55}.
\newblock
\urldef\tempurl%
\url{https://link.springer.com/article/10.1007/s13171-019-00164-5}
\showURL{%
\tempurl}


\bibitem[Marquart et~al\mbox{.}(2015)]%
        {marquart_review_2015}
\bibfield{author}{\bibinfo{person}{Gerhard Marquart}, \bibinfo{person}{Christopher Cabrall}, {and} \bibinfo{person}{Joost de Winter}.} \bibinfo{year}{2015}\natexlab{}.
\newblock \showarticletitle{Review of {Eye}-related {Measures} of {Drivers}’ {Mental} {Workload}}.
\newblock \bibinfo{journal}{\emph{Procedia Manufacturing}}  \bibinfo{volume}{3} (\bibinfo{date}{Jan.} \bibinfo{year}{2015}), \bibinfo{pages}{2854--2861}.
\newblock
\showISSN{2351-9789}
\href{https://doi.org/10.1016/j.promfg.2015.07.783}{doi:\nolinkurl{10.1016/j.promfg.2015.07.783}}


\bibitem[Martens and Van~Winsum(2000)]%
        {martens_measuring_2000}
\bibfield{author}{\bibinfo{person}{Marieke Martens} {and} \bibinfo{person}{Wim Van~Winsum}.} \bibinfo{year}{2000}\natexlab{}.
\newblock \showarticletitle{Measuring distraction: the {Peripheral} {Detection} {Task}}.
\newblock  (\bibinfo{date}{Jan.} \bibinfo{year}{2000}).
\newblock
\urldef\tempurl%
\url{https://www-nrd.nhtsa.dot.gov/departments/Human%20Factors/driver-distraction/pdf/34.pdf}
\showURL{%
\tempurl}


\bibitem[Martin et~al\mbox{.}(2018)]%
        {martin_dynamics_2018}
\bibfield{author}{\bibinfo{person}{Sujitha Martin}, \bibinfo{person}{Sourabh Vora}, \bibinfo{person}{Kevan Yuen}, {and} \bibinfo{person}{Mohan~Manubhai Trivedi}.} \bibinfo{year}{2018}\natexlab{}.
\newblock \showarticletitle{Dynamics of {Driver}'s {Gaze}: {Explorations} in {Behavior} {Modeling} and {Maneuver} {Prediction}}.
\newblock \bibinfo{journal}{\emph{IEEE Transactions on Intelligent Vehicles}} \bibinfo{volume}{3}, \bibinfo{number}{2} (\bibinfo{date}{June} \bibinfo{year}{2018}), \bibinfo{pages}{141--150}.
\newblock
\showISSN{2379-8904}
\href{https://doi.org/10.1109/TIV.2018.2804160}{doi:\nolinkurl{10.1109/TIV.2018.2804160}}
\newblock
\shownote{Conference Name: IEEE Transactions on Intelligent Vehicles}.


\bibitem[Math et~al\mbox{.}(2013)]%
        {math_opends_2013}
\bibfield{author}{\bibinfo{person}{Rafael Math}, \bibinfo{person}{Angela Mahr}, \bibinfo{person}{Mohammad~M. Moniri}, {and} \bibinfo{person}{Christian Müller}.} \bibinfo{year}{2013}\natexlab{}.
\newblock \showarticletitle{{OpenDS}: {A} new open-source driving simulator for research}. In \bibinfo{booktitle}{\emph{{AmE}, {GMM}-{Fachtagung} {Automotive} meets {Electronics}, 4}}, Vol.~\bibinfo{volume}{75}. \bibinfo{publisher}{VDE-Verlag;}, \bibinfo{address}{Berlin, Offenbach}, \bibinfo{pages}{104--105}.
\newblock
\showISBNx{978-3-8007-3485-6}
\urldef\tempurl%
\url{https://www.tib.eu/de/suchen/id/tema%3ATEMA20140108648}
\showURL{%
\tempurl}
\newblock
\shownote{ISSN: 1432-3419}.


\bibitem[McKendrick and Cherry(2018)]%
        {mckendrick_deeper_2018}
\bibfield{author}{\bibinfo{person}{Ryan~D. McKendrick} {and} \bibinfo{person}{Erin Cherry}.} \bibinfo{year}{2018}\natexlab{}.
\newblock \showarticletitle{A {Deeper} {Look} at the {NASA} {TLX} and {Where} {It} {Falls} {Short}}.
\newblock \bibinfo{journal}{\emph{Proceedings of the Human Factors and Ergonomics Society Annual Meeting}} \bibinfo{volume}{62}, \bibinfo{number}{1} (\bibinfo{date}{Sept.} \bibinfo{year}{2018}), \bibinfo{pages}{44--48}.
\newblock
\showISSN{1071-1813}
\href{https://doi.org/10.1177/1541931218621010}{doi:\nolinkurl{10.1177/1541931218621010}}
\newblock
\shownote{Publisher: SAGE Publications Inc}.


\bibitem[Medenica et~al\mbox{.}(2011)]%
        {medenica_augmented_2011}
\bibfield{author}{\bibinfo{person}{Zeljko Medenica}, \bibinfo{person}{Andrew~L. Kun}, \bibinfo{person}{Tim Paek}, {and} \bibinfo{person}{Oskar Palinko}.} \bibinfo{year}{2011}\natexlab{}.
\newblock \showarticletitle{Augmented reality vs. street views: a driving simulator study comparing two emerging navigation aids}. In \bibinfo{booktitle}{\emph{Proceedings of the 13th {International} {Conference} on {Human} {Computer} {Interaction} with {Mobile} {Devices} and {Services}}} \emph{(\bibinfo{series}{{MobileHCI} '11})}. \bibinfo{publisher}{Association for Computing Machinery}, \bibinfo{address}{New York, NY, USA}, \bibinfo{pages}{265--274}.
\newblock
\showISBNx{978-1-4503-0541-9}
\href{https://doi.org/10.1145/2037373.2037414}{doi:\nolinkurl{10.1145/2037373.2037414}}


\bibitem[Mohamed~Selim et~al\mbox{.}(2024)]%
        {mohamed_selim_review_2024}
\bibfield{author}{\bibinfo{person}{Abdulrahman Mohamed~Selim}, \bibinfo{person}{Michael Barz}, \bibinfo{person}{Omair~Shahzad Bhatti}, \bibinfo{person}{Hasan Md~Tusfiqur Alam}, {and} \bibinfo{person}{Daniel Sonntag}.} \bibinfo{year}{2024}\natexlab{}.
\newblock \showarticletitle{A review of machine learning in scanpath analysis for passive gaze-based interaction}.
\newblock \bibinfo{journal}{\emph{Frontiers in Artificial Intelligence}}  \bibinfo{volume}{7} (\bibinfo{date}{June} \bibinfo{year}{2024}).
\newblock
\showISSN{2624-8212}
\href{https://doi.org/10.3389/frai.2024.1391745}{doi:\nolinkurl{10.3389/frai.2024.1391745}}
\newblock
\shownote{Publisher: Frontiers}.


\bibitem[Naujoks et~al\mbox{.}(2018)]%
        {naujoks_review_2018}
\bibfield{author}{\bibinfo{person}{Frederik Naujoks}, \bibinfo{person}{Dennis Befelein}, \bibinfo{person}{Katharina Wiedemann}, {and} \bibinfo{person}{Alexandra Neukum}.} \bibinfo{year}{2018}\natexlab{}.
\newblock \showarticletitle{A {Review} of {Non}-driving-related {Tasks} {Used} in {Studies} on {Automated} {Driving}}. In \bibinfo{booktitle}{\emph{Advances in {Human} {Aspects} of {Transportation}}}, \bibfield{editor}{\bibinfo{person}{Neville~A Stanton}} (Ed.). \bibinfo{publisher}{Springer International Publishing}, \bibinfo{address}{Cham}, \bibinfo{pages}{525--537}.
\newblock
\showISBNx{978-3-319-60441-1}
\href{https://doi.org/10.1007/978-3-319-60441-1_52}{doi:\nolinkurl{10.1007/978-3-319-60441-1_52}}


\bibitem[Nuwer and Coutin-Churchman(2014)]%
        {nuwer_brain_2014}
\bibfield{author}{\bibinfo{person}{M.~R. Nuwer} {and} \bibinfo{person}{P. Coutin-Churchman}.} \bibinfo{year}{2014}\natexlab{}.
\newblock \showarticletitle{Brain {Mapping} and {Quantitative} {Electroencephalogram}}.
\newblock In \bibinfo{booktitle}{\emph{Encyclopedia of the {Neurological} {Sciences} ({Second} {Edition})}}, \bibfield{editor}{\bibinfo{person}{Michael~J. Aminoff} {and} \bibinfo{person}{Robert~B. Daroff}} (Eds.). \bibinfo{publisher}{Academic Press}, \bibinfo{address}{Oxford}, \bibinfo{pages}{499--504}.
\newblock
\showISBNx{978-0-12-385158-1}
\href{https://doi.org/10.1016/B978-0-12-385157-4.00519-4}{doi:\nolinkurl{10.1016/B978-0-12-385157-4.00519-4}}


\bibitem[Olsson and Burns(2000)]%
        {olsson_measuring_2000}
\bibfield{author}{\bibinfo{person}{S. Olsson} {and} \bibinfo{person}{P.~C. Burns}.} \bibinfo{year}{2000}\natexlab{}.
\newblock \showarticletitle{Measuring {Driver} {Visual} {Distraction} with a {Peripheral} {Detection} {Task}}.
\newblock  (\bibinfo{year}{2000}).
\newblock
\urldef\tempurl%
\url{https://www-nrd.nhtsa.dot.gov/departments/Human%20Factors/driver-distraction/PDF/6.PDF}
\showURL{%
\tempurl}


\bibitem[Pakdamanian et~al\mbox{.}(2021)]%
        {pakdamanian_deeptake_2021}
\bibfield{author}{\bibinfo{person}{Erfan Pakdamanian}, \bibinfo{person}{Shili Sheng}, \bibinfo{person}{Sonia Baee}, \bibinfo{person}{Seongkook Heo}, \bibinfo{person}{Sarit Kraus}, {and} \bibinfo{person}{Lu Feng}.} \bibinfo{year}{2021}\natexlab{}.
\newblock \showarticletitle{{DeepTake}: {Prediction} of {Driver} {Takeover} {Behavior} using {Multimodal} {Data}}. In \bibinfo{booktitle}{\emph{Proceedings of the 2021 {CHI} {Conference} on {Human} {Factors} in {Computing} {Systems}}} \emph{(\bibinfo{series}{{CHI} '21})}. \bibinfo{publisher}{Association for Computing Machinery}, \bibinfo{address}{New York, NY, USA}, \bibinfo{pages}{1--14}.
\newblock
\showISBNx{978-1-4503-8096-6}
\href{https://doi.org/10.1145/3411764.3445563}{doi:\nolinkurl{10.1145/3411764.3445563}}


\bibitem[Qvarfordt(2017)]%
        {qvarfordt_gaze-informed_2017}
\bibfield{author}{\bibinfo{person}{Pernilla Qvarfordt}.} \bibinfo{year}{2017}\natexlab{}.
\newblock \showarticletitle{Gaze-informed multimodal interaction}.
\newblock In \bibinfo{booktitle}{\emph{The {Handbook} of {Multimodal}-{Multisensor} {Interfaces}: {Foundations}, {User} {Modeling}, and {Common} {Modality} {Combinations} - {Volume} 1}}. Vol.~\bibinfo{volume}{14}. \bibinfo{publisher}{Association for Computing Machinery and Morgan \& Claypool}, \bibinfo{pages}{365--402}.
\newblock
\showISBNx{978-1-970001-67-9}
\urldef\tempurl%
\url{https://doi.org/10.1145/3015783.3015794}
\showURL{%
\tempurl}


\bibitem[Radlmayr et~al\mbox{.}(2019)]%
        {radlmayr_influence_2019}
\bibfield{author}{\bibinfo{person}{Jonas Radlmayr}, \bibinfo{person}{Fabian~Marco Fischer}, {and} \bibinfo{person}{Klaus Bengler}.} \bibinfo{year}{2019}\natexlab{}.
\newblock \showarticletitle{The {Influence} of {Non}-driving {Related} {Tasks} on {Driver} {Availability} in the {Context} of {Conditionally} {Automated} {Driving}}. In \bibinfo{booktitle}{\emph{Proceedings of the 20th {Congress} of the {International} {Ergonomics} {Association} ({IEA} 2018)}}, \bibfield{editor}{\bibinfo{person}{Sebastiano Bagnara}, \bibinfo{person}{Riccardo Tartaglia}, \bibinfo{person}{Sara Albolino}, \bibinfo{person}{Thomas Alexander}, {and} \bibinfo{person}{Yushi Fujita}} (Eds.). \bibinfo{publisher}{Springer International Publishing}, \bibinfo{address}{Cham}, \bibinfo{pages}{295--304}.
\newblock
\showISBNx{978-3-319-96074-6}
\href{https://doi.org/10.1007/978-3-319-96074-6_32}{doi:\nolinkurl{10.1007/978-3-319-96074-6_32}}


\bibitem[Radlmayr et~al\mbox{.}(2014)]%
        {radlmayr_how_2014}
\bibfield{author}{\bibinfo{person}{Jonas Radlmayr}, \bibinfo{person}{Christian Gold}, \bibinfo{person}{Lutz Lorenz}, \bibinfo{person}{Mehdi Farid}, {and} \bibinfo{person}{Klaus Bengler}.} \bibinfo{year}{2014}\natexlab{}.
\newblock \showarticletitle{How {Traffic} {Situations} and {Non}-{Driving} {Related} {Tasks} {Affect} the {Take}-{Over} {Quality} in {Highly} {Automated} {Driving}}.
\newblock \bibinfo{journal}{\emph{Proceedings of the Human Factors and Ergonomics Society Annual Meeting}} \bibinfo{volume}{58}, \bibinfo{number}{1} (\bibinfo{date}{Sept.} \bibinfo{year}{2014}), \bibinfo{pages}{2063--2067}.
\newblock
\showISSN{1071-1813}
\href{https://doi.org/10.1177/1541931214581434}{doi:\nolinkurl{10.1177/1541931214581434}}
\newblock
\shownote{Publisher: SAGE Publications Inc}.


\bibitem[Raufi and Longo(2022)]%
        {raufi_evaluation_2022}
\bibfield{author}{\bibinfo{person}{Bujar Raufi} {and} \bibinfo{person}{Luca Longo}.} \bibinfo{year}{2022}\natexlab{}.
\newblock \showarticletitle{An {Evaluation} of the {EEG} {Alpha}-to-{Theta} and {Theta}-to-{Alpha} {Band} {Ratios} as {Indexes} of {Mental} {Workload}}.
\newblock \bibinfo{journal}{\emph{Frontiers in Neuroinformatics}}  \bibinfo{volume}{16} (\bibinfo{date}{May} \bibinfo{year}{2022}).
\newblock
\showISSN{1662-5196}
\href{https://doi.org/10.3389/fninf.2022.861967}{doi:\nolinkurl{10.3389/fninf.2022.861967}}
\newblock
\shownote{Publisher: Frontiers}.


\bibitem[Riener et~al\mbox{.}(2016)]%
        {riener_automotive_2016}
\bibfield{author}{\bibinfo{person}{Andreas Riener}, \bibinfo{person}{Susanne Boll}, {and} \bibinfo{person}{Andrew~L. Kun}.} \bibinfo{year}{2016}\natexlab{}.
\newblock \showarticletitle{Automotive {User} {Interfaces} in the {Age} of {Automation} ({Dagstuhl} {Seminar} 16262)}.
\newblock \bibinfo{journal}{\emph{Dagstuhl Reports}} \bibinfo{volume}{6}, \bibinfo{number}{6} (\bibinfo{year}{2016}), \bibinfo{pages}{111--157}.
\newblock
\showISSN{2192-5283}
\href{https://doi.org/10.4230/DagRep.6.6.111}{doi:\nolinkurl{10.4230/DagRep.6.6.111}}
\newblock
\shownote{Place: Dagstuhl, Germany Publisher: Schloss Dagstuhl – Leibniz-Zentrum für Informatik}.


\bibitem[Rousseeuw(1987)]%
        {rousseeuw_silhouettes_1987}
\bibfield{author}{\bibinfo{person}{Peter~J. Rousseeuw}.} \bibinfo{year}{1987}\natexlab{}.
\newblock \showarticletitle{Silhouettes: {A} graphical aid to the interpretation and validation of cluster analysis}.
\newblock \bibinfo{journal}{\emph{J. Comput. Appl. Math.}}  \bibinfo{volume}{20} (\bibinfo{date}{Nov.} \bibinfo{year}{1987}), \bibinfo{pages}{53--65}.
\newblock
\showISSN{0377-0427}
\href{https://doi.org/10.1016/0377-0427(87)90125-7}{doi:\nolinkurl{10.1016/0377-0427(87)90125-7}}


\bibitem[Sadeghian~Borojeni et~al\mbox{.}(2018)]%
        {sadeghian_borojeni_feel_2018}
\bibfield{author}{\bibinfo{person}{Shadan Sadeghian~Borojeni}, \bibinfo{person}{Susanne~C.J. Boll}, \bibinfo{person}{Wilko Heuten}, \bibinfo{person}{Heinrich~H. Bülthoff}, {and} \bibinfo{person}{Lewis Chuang}.} \bibinfo{year}{2018}\natexlab{}.
\newblock \showarticletitle{Feel the {Movement}: {Real} {Motion} {Influences} {Responses} to {Take}-over {Requests} in {Highly} {Automated} {Vehicles}}. In \bibinfo{booktitle}{\emph{Proceedings of the 2018 {CHI} {Conference} on {Human} {Factors} in {Computing} {Systems}}} \emph{(\bibinfo{series}{{CHI} '18})}. \bibinfo{publisher}{Association for Computing Machinery}, \bibinfo{address}{New York, NY, USA}, \bibinfo{pages}{1--13}.
\newblock
\showISBNx{978-1-4503-5620-6}
\href{https://doi.org/10.1145/3173574.3173820}{doi:\nolinkurl{10.1145/3173574.3173820}}


\bibitem[Salkind(2010)]%
        {salkind_encyclopedia_2010}
\bibfield{author}{\bibinfo{person}{Neil Salkind}.} \bibinfo{year}{2010}\natexlab{}.
\newblock \bibinfo{booktitle}{\emph{Encyclopedia of {Research} {Design}}}. Vol.~\bibinfo{volume}{1}.
\newblock \bibinfo{publisher}{SAGE Publications}, \bibinfo{address}{Thousand Oaks, California}.
\newblock
\href{https://doi.org/10.4135/9781412961288}{doi:\nolinkurl{10.4135/9781412961288}}


\bibitem[Salubre and Nathan-Roberts(2021)]%
        {salubre_takeover_2021}
\bibfield{author}{\bibinfo{person}{Kevin~Joel Salubre} {and} \bibinfo{person}{Dan Nathan-Roberts}.} \bibinfo{year}{2021}\natexlab{}.
\newblock \showarticletitle{Takeover {Request} {Design} in {Automated} {Driving}: {A} {Systematic} {Review}}.
\newblock \bibinfo{journal}{\emph{Proceedings of the Human Factors and Ergonomics Society Annual Meeting}} \bibinfo{volume}{65}, \bibinfo{number}{1} (\bibinfo{year}{2021}), \bibinfo{pages}{868--872}.
\newblock
\href{https://doi.org/10.1177/1071181321651296}{doi:\nolinkurl{10.1177/1071181321651296}}
\newblock
\shownote{\_eprint: https://doi.org/10.1177/1071181321651296}.


\bibitem[Schmider et~al\mbox{.}(2010)]%
        {schmider_is_2010}
\bibfield{author}{\bibinfo{person}{Emanuel Schmider}, \bibinfo{person}{Matthias Ziegler}, \bibinfo{person}{Erik Danay}, \bibinfo{person}{Luzi Beyer}, {and} \bibinfo{person}{Markus Bühner}.} \bibinfo{year}{2010}\natexlab{}.
\newblock \showarticletitle{Is it really robust? {Reinvestigating} the robustness of {ANOVA} against violations of the normal distribution assumption}.
\newblock \bibinfo{journal}{\emph{Methodology: European Journal of Research Methods for the Behavioral and Social Sciences}} \bibinfo{volume}{6}, \bibinfo{number}{4} (\bibinfo{year}{2010}), \bibinfo{pages}{147--151}.
\newblock
\showISSN{1614-2241}
\href{https://doi.org/10.1027/1614-2241/a000016}{doi:\nolinkurl{10.1027/1614-2241/a000016}}
\newblock
\shownote{Place: Germany Publisher: Hogrefe Publishing}.


\bibitem[Shapiro and Wilk(1965)]%
        {shapiro_analysis_1965}
\bibfield{author}{\bibinfo{person}{S.~S. Shapiro} {and} \bibinfo{person}{M.~B. Wilk}.} \bibinfo{year}{1965}\natexlab{}.
\newblock \showarticletitle{An analysis of variance test for normality (complete samples)†}.
\newblock \bibinfo{journal}{\emph{Biometrika}} \bibinfo{volume}{52}, \bibinfo{number}{3-4} (\bibinfo{date}{Dec.} \bibinfo{year}{1965}), \bibinfo{pages}{591--611}.
\newblock
\showISSN{0006-3444}
\href{https://doi.org/10.1093/biomet/52.3-4.591}{doi:\nolinkurl{10.1093/biomet/52.3-4.591}}
\newblock
\shownote{\_eprint: https://academic.oup.com/biomet/article-pdf/52/3-4/591/962907/52-3-4-591.pdf}.


\bibitem[Shiferaw et~al\mbox{.}(2018)]%
        {shiferaw_stationary_2018}
\bibfield{author}{\bibinfo{person}{Brook~A. Shiferaw}, \bibinfo{person}{Luke~A. Downey}, \bibinfo{person}{Justine Westlake}, \bibinfo{person}{Bronwyn Stevens}, \bibinfo{person}{Shantha M.~W. Rajaratnam}, \bibinfo{person}{David~J. Berlowitz}, \bibinfo{person}{Phillip Swann}, {and} \bibinfo{person}{Mark~E. Howard}.} \bibinfo{year}{2018}\natexlab{}.
\newblock \showarticletitle{Stationary gaze entropy predicts lane departure events in sleep-deprived drivers}.
\newblock \bibinfo{journal}{\emph{Scientific Reports}} \bibinfo{volume}{8}, \bibinfo{number}{1} (\bibinfo{date}{Feb.} \bibinfo{year}{2018}), \bibinfo{pages}{2220}.
\newblock
\showISSN{2045-2322}
\href{https://doi.org/10.1038/s41598-018-20588-7}{doi:\nolinkurl{10.1038/s41598-018-20588-7}}
\newblock
\shownote{Publisher: Nature Publishing Group}.


\bibitem[Smith et~al\mbox{.}(2016)]%
        {smith_head-up_2016}
\bibfield{author}{\bibinfo{person}{Missie Smith}, \bibinfo{person}{Joseph~L. Gabbard}, {and} \bibinfo{person}{Christian Conley}.} \bibinfo{year}{2016}\natexlab{}.
\newblock \showarticletitle{Head-{Up} vs. {Head}-{Down} {Displays}: {Examining} {Traditional} {Methods} of {Display} {Assessment} {While} {Driving}}. In \bibinfo{booktitle}{\emph{Proceedings of the 8th {International} {Conference} on {Automotive} {User} {Interfaces} and {Interactive} {Vehicular} {Applications}}} \emph{(\bibinfo{series}{Automotive'{UI} 16})}. \bibinfo{publisher}{Association for Computing Machinery}, \bibinfo{address}{New York, NY, USA}, \bibinfo{pages}{185--192}.
\newblock
\showISBNx{978-1-4503-4533-0}
\href{https://doi.org/10.1145/3003715.3005419}{doi:\nolinkurl{10.1145/3003715.3005419}}


\bibitem[Smith et~al\mbox{.}(2015)]%
        {smith_visual_2015}
\bibfield{author}{\bibinfo{person}{Missie Smith}, \bibinfo{person}{Jillian Streeter}, \bibinfo{person}{Gary Burnett}, {and} \bibinfo{person}{Joseph~L. Gabbard}.} \bibinfo{year}{2015}\natexlab{}.
\newblock \showarticletitle{Visual search tasks: the effects of head-up displays on driving and task performance}. In \bibinfo{booktitle}{\emph{Proceedings of the 7th {International} {Conference} on {Automotive} {User} {Interfaces} and {Interactive} {Vehicular} {Applications}}} \emph{(\bibinfo{series}{{AutomotiveUI} '15})}. \bibinfo{publisher}{Association for Computing Machinery}, \bibinfo{address}{New York, NY, USA}, \bibinfo{pages}{80--87}.
\newblock
\showISBNx{978-1-4503-3736-6}
\href{https://doi.org/10.1145/2799250.2799291}{doi:\nolinkurl{10.1145/2799250.2799291}}


\bibitem[Stokes and Wickens(1988)]%
        {stokes_aviation_1988}
\bibfield{author}{\bibinfo{person}{Alan~F. Stokes} {and} \bibinfo{person}{Christopher~D. Wickens}.} \bibinfo{year}{1988}\natexlab{}.
\newblock \showarticletitle{Aviation displays}.
\newblock In \bibinfo{booktitle}{\emph{Human factors in aviation}}. \bibinfo{publisher}{Academic Press}, \bibinfo{address}{San Diego, CA, US}, \bibinfo{pages}{387--431}.
\newblock
\showISBNx{978-0-12-750030-0}


\bibitem[Sugiono et~al\mbox{.}(2017)]%
        {sugiono_investigating_2017}
\bibfield{author}{\bibinfo{person}{Sugiono Sugiono}, \bibinfo{person}{Denny Widhayanuriyawan}, {and} \bibinfo{person}{Debrina~P. Andriani}.} \bibinfo{year}{2017}\natexlab{}.
\newblock \showarticletitle{Investigating the {Impact} of {Road} {Condition} {Complexity} on {Driving} {Workload} {Based} on {Subjective} {Measurement} using {NASA} {TLX}}.
\newblock \bibinfo{journal}{\emph{MATEC Web of Conferences}}  \bibinfo{volume}{136} (\bibinfo{year}{2017}), \bibinfo{pages}{02007}.
\newblock
\showISSN{2261-236X}
\href{https://doi.org/10.1051/matecconf/201713602007}{doi:\nolinkurl{10.1051/matecconf/201713602007}}
\newblock
\shownote{Publisher: EDP Sciences}.


\bibitem[Tsang and Vidulich(2006)]%
        {tsang_mental_2006}
\bibfield{author}{\bibinfo{person}{Pamela~S. Tsang} {and} \bibinfo{person}{Michael~A. Vidulich}.} \bibinfo{year}{2006}\natexlab{}.
\newblock \showarticletitle{Mental {Workload} and {Situation} {Awareness}}.
\newblock In \bibinfo{booktitle}{\emph{Handbook of {Human} {Factors} and {Ergonomics}}}. \bibinfo{publisher}{John Wiley \& Sons, Ltd}, \bibinfo{pages}{243--268}.
\newblock
\showISBNx{978-0-470-04820-7}
\href{https://doi.org/10.1002/0470048204.ch9}{doi:\nolinkurl{10.1002/0470048204.ch9}}
\newblock
\shownote{Section: 9 \_eprint: https://onlinelibrary.wiley.com/doi/pdf/10.1002/0470048204.ch9}.


\bibitem[Vogelpohl et~al\mbox{.}(2018)]%
        {vogelpohl_transitioning_2018}
\bibfield{author}{\bibinfo{person}{Tobias Vogelpohl}, \bibinfo{person}{Matthias Kühn}, \bibinfo{person}{Thomas Hummel}, \bibinfo{person}{Tina Gehlert}, {and} \bibinfo{person}{Mark Vollrath}.} \bibinfo{year}{2018}\natexlab{}.
\newblock \showarticletitle{Transitioning to manual driving requires additional time after automation deactivation}.
\newblock \bibinfo{journal}{\emph{Transportation Research Part F: Traffic Psychology and Behaviour}}  \bibinfo{volume}{55} (\bibinfo{date}{May} \bibinfo{year}{2018}), \bibinfo{pages}{464--482}.
\newblock
\showISSN{1369-8478}
\href{https://doi.org/10.1016/j.trf.2018.03.019}{doi:\nolinkurl{10.1016/j.trf.2018.03.019}}


\bibitem[Wang et~al\mbox{.}(2023)]%
        {wang_multi-modal_2023}
\bibfield{author}{\bibinfo{person}{Yingjie Wang}, \bibinfo{person}{Qiuyu Mao}, \bibinfo{person}{Hanqi Zhu}, \bibinfo{person}{Jiajun Deng}, \bibinfo{person}{Yu Zhang}, \bibinfo{person}{Jianmin Ji}, \bibinfo{person}{Houqiang Li}, {and} \bibinfo{person}{Yanyong Zhang}.} \bibinfo{year}{2023}\natexlab{}.
\newblock \showarticletitle{Multi-{Modal} {3D} {Object} {Detection} in {Autonomous} {Driving}: {A} {Survey}}.
\newblock \bibinfo{journal}{\emph{International Journal of Computer Vision}} \bibinfo{volume}{131}, \bibinfo{number}{8} (\bibinfo{date}{Aug.} \bibinfo{year}{2023}), \bibinfo{pages}{2122--2152}.
\newblock
\showISSN{1573-1405}
\href{https://doi.org/10.1007/s11263-023-01784-z}{doi:\nolinkurl{10.1007/s11263-023-01784-z}}


\bibitem[Wei et~al\mbox{.}(2023)]%
        {wei_state_2023}
\bibfield{author}{\bibinfo{person}{Sijie Wei}, \bibinfo{person}{Peter~E. Pfeffer}, {and} \bibinfo{person}{Johannes Edelmann}.} \bibinfo{year}{2023}\natexlab{}.
\newblock \showarticletitle{State of the {Art}: {Ongoing} {Research} in {Assessment} {Methods} for {Lane} {Keeping} {Assistance} {Systems}}.
\newblock \bibinfo{journal}{\emph{IEEE Transactions on Intelligent Vehicles}} (\bibinfo{year}{2023}), \bibinfo{pages}{1--28}.
\newblock
\showISSN{2379-8904}
\href{https://doi.org/10.1109/TIV.2023.3269156}{doi:\nolinkurl{10.1109/TIV.2023.3269156}}


\bibitem[Zeeb et~al\mbox{.}(2017)]%
        {zeeb_why_2017}
\bibfield{author}{\bibinfo{person}{Kathrin Zeeb}, \bibinfo{person}{Manuela Härtel}, \bibinfo{person}{Axel Buchner}, {and} \bibinfo{person}{Michael Schrauf}.} \bibinfo{year}{2017}\natexlab{}.
\newblock \showarticletitle{Why is steering not the same as braking? {The} impact of non-driving related tasks on lateral and longitudinal driver interventions during conditionally automated driving}.
\newblock \bibinfo{journal}{\emph{Transportation Research Part F: Traffic Psychology and Behaviour}}  \bibinfo{volume}{50} (\bibinfo{date}{Oct.} \bibinfo{year}{2017}), \bibinfo{pages}{65--79}.
\newblock
\showISSN{1369-8478}
\href{https://doi.org/10.1016/j.trf.2017.07.008}{doi:\nolinkurl{10.1016/j.trf.2017.07.008}}


\bibitem[Zhang et~al\mbox{.}(2019)]%
        {zhang_determinants_2019}
\bibfield{author}{\bibinfo{person}{Bo Zhang}, \bibinfo{person}{Joost de Winter}, \bibinfo{person}{Silvia Varotto}, \bibinfo{person}{Riender Happee}, {and} \bibinfo{person}{Marieke Martens}.} \bibinfo{year}{2019}\natexlab{}.
\newblock \showarticletitle{Determinants of take-over time from automated driving: {A} meta-analysis of 129 studies}.
\newblock \bibinfo{journal}{\emph{Transportation Research Part F: Traffic Psychology and Behaviour}}  \bibinfo{volume}{64} (\bibinfo{date}{July} \bibinfo{year}{2019}), \bibinfo{pages}{285--307}.
\newblock
\showISSN{1369-8478}
\href{https://doi.org/10.1016/j.trf.2019.04.020}{doi:\nolinkurl{10.1016/j.trf.2019.04.020}}


\bibitem[Zhou et~al\mbox{.}(2022)]%
        {zhou_using_2022}
\bibfield{author}{\bibinfo{person}{Feng Zhou}, \bibinfo{person}{X.~Jessie Yang}, {and} \bibinfo{person}{Joost C.~F. de Winter}.} \bibinfo{year}{2022}\natexlab{}.
\newblock \showarticletitle{Using {Eye}-{Tracking} {Data} to {Predict} {Situation} {Awareness} in {Real} {Time} {During} {Takeover} {Transitions} in {Conditionally} {Automated} {Driving}}.
\newblock \bibinfo{journal}{\emph{IEEE Transactions on Intelligent Transportation Systems}} \bibinfo{volume}{23}, \bibinfo{number}{3} (\bibinfo{date}{March} \bibinfo{year}{2022}), \bibinfo{pages}{2284--2295}.
\newblock
\showISSN{1558-0016}
\href{https://doi.org/10.1109/TITS.2021.3069776}{doi:\nolinkurl{10.1109/TITS.2021.3069776}}
\newblock
\shownote{Conference Name: IEEE Transactions on Intelligent Transportation Systems}.


\end{thebibliography}



\end{document}